%% file: paper.tex
\documentclass[a4paper,fleqn,usenatbib]{mnras}

\usepackage[T1]{fontenc}
\usepackage{ae,aecompl}

\usepackage{graphicx}	
\usepackage{amsmath}	
\usepackage{amssymb}	

\usepackage[outdir=figures/maps/]{epstopdf}
\usepackage{booktabs}
\usepackage{amsmath,amssymb}
\usepackage{gensymb}
\usepackage{subcaption}
\newcommand{\mps}{m s$^{-1}$} 

\newcommand*\subtxt[1]{_{\textnormal{#1}}} 
\DeclareRobustCommand\_{\ifmmode\expandafter\subtxt\else\textunderscore\fi}

\title[Impact of unresolved magnetic spots on RV]{The impact of unresolved magnetic spots on high precision radial velocity measurements}

\author[M. Lisogorskyi et al.]{
	M. Lisogorskyi\thanks{E-mail: m.lisogorskyi@herts.ac.uk}$^{1}$,
	S. Boro Saikia$^{2}$,
	S. V. Jeffers$^{3}$,
	H. R. A. Jones$^{1}$,
	J. Morin$^{4}$,
	\newauthor
	M. Mengel$^{5}$,
	A. Reiners$^{3}$,
	A. A. Vidotto$^{6}$,
	and P. Petit$^{7}$
\\
\\
$^{1}$Centre for Astrophysics Research, University of Hertfordshire, College Lane, AL10 9AB, Hatfield, UK\\
$^{2}$University of Vienna, Department of Astrophysics, Turkenschanzstrasse 17, 1180 Vienna, Austria\\
$^{3}$Institut f{\"u}r Astrophysik, Universit{\"a}t G{\"o}ttingen, Friedrich Hund Platz 1, 37077 G{\"o}ttingen, Germany\\
$^{4}$Laboratoire Univers et Particules de Montpellier (LUPM), Universit{\'e} de Montpellier, CNRS, 34095 Montpellier, France\\
$^{5}$University of Southern Queensland, Centre for Astrophysics, Toowoomba 4350, Australia\\
$^{6}$Trinity College Dublin, the University of Dublin, College Green, Dublin, D-2, Ireland\\
$^{7}$Institut de Recherche en Astrophysique et Plan\'etologie, Universit\'e de Toulouse, CNRS, CNES, 31400 Toulouse, France\\
}

\date{Accepted 2020 July 22. Received 2020 July 22; in original form 2020 January 24}

\pubyear{2020}

\begin{document}
\label{firstpage}
\pagerange{\pageref{firstpage}--\pageref{lastpage}}
\maketitle

\begin{abstract}
	The Doppler method of exoplanet detection has been extremely successful, but suffers from contaminating noise from stellar activity.
	In this work a model of a rotating star with a magnetic field based on the geometry of the K2 star $\epsilon$ Eridani is presented and used to estimate its effect on simulated radial velocity measurements.
	A number of different distributions of unresolved magnetic spots were simulated on top of the observed large-scale magnetic maps obtained from eight years of spectropolarimetric observations.
	The radial velocity signals due to the magnetic spots have amplitudes of up to 10 m s$^{-1}$, high enough to prevent the detection of planets under 20 Earth masses in temperate zones of solar type stars.
	We show that the radial velocity depends heavily on spot distribution.
	Our results emphasize that understanding stellar magnetic activity and spot distribution is crucial for detection of Earth analogues.
\end{abstract}



\section{Introduction}

The Doppler method is one of the most important methods of exoplanet detection that led to the discovery or confirmation of a wide range of exoplanets.
It measures the reflect motion of the star, as the planet orbits it, by measuring small doppler shifts in narrow spectral absorption features of the star.
It is very reliable for exoplanets that are large and close to the host star, but gets more challenging towards lower planet masses and higher separations.
This affects our capability to detect Earth mass planets in temperate zones of their host stars.
To detect those systems, extremely precise radial velocity (RV) measurements are required ($\sim$0.09 m s$^{-1}$ to detect Earth around the Sun, for instance).
Current Doppler velocitimeters are very stable high resolution spectrographs (HARPS, CARMENES, SOPHIE, ESPRESSO etc), and are getting close to this precision (e.g. \citealt{2010SPIE.7735E..0FP, 2018haex.bookE.157G}).
However, rotation and magnetic activity of the host star can hide an existing planet or mimic a planetary signal.
As a result, even though some studies focus on young stars \citep[e.g.][]{2013A&A...559A..83L, 2020A&A...633A..44G}, the majority of studies are directed towards older and less active stars.
One example of a young star that has been extensively observed with the radial velocity technique is $\epsilon$ Eridani.
$\epsilon$ Eri (HD 22049) is a young \citep[440 Myr,][]{2007ApJ...669.1167B} Sun-like star \citep[K2V,][]{1989ApJS...71..245K}, at a distance of 3.212 pc from the Sun.
It is more active than the Sun \citep[log $R'\_{HK}=-4.455$,][]{2007ApJ...669.1167B}, with a rotation period of 11.68 days \citep{1997AA...318..429R}.
Observations of its chromospheric activity using the Ca II H\&K lines indicate a strong and highly variable magnetic activity.
Unlike the quasi periodic activity cycle of the Sun, $\epsilon$ Eridani has two co-existing activity cycles of 3 and 13 years \citep{2013ApJ...763L..26M}.

The first observations of $\epsilon$ Eridani's variable radial velocity were reported by \cite{1988ApJ...331..902C}.
Later, a signal with a period of approximately seven years was detected in RV data \citep{1999ApJ...526..890C} and then interpreted as planetary signal \citep{2000ApJ...544L.145H} corresponding to a Jovian-mass exoplanet with an exceptionally high eccentricity of 0.6, an orbital period of 2500 days and semi-amplitude of 19 \mps.
The existence of the planet has been debated for a long time, as the orbital parameters obtained using additional observations differed significantly from the previous solutions \citep{2012ApJS..200...15A}, and such a high eccentricity is incompatible with the debris disk around the star \citep{2009A&A...499L..13B}.
After 20 years of debates about possible confusion with stellar noise and joint Bayesian analysis of state-of-the art direct imaging observations and 30 years worth of radial velocity data, it is asserted as a confirmed planet \citep{2019AJ....157...33M} with a close to circular orbit ($e = {0.07}_{-0.05}^{+0.06}$) of $2691.8 \pm 25.6$ days and an amplitude of $11.49\pm0.66$ \mps.
To find smaller planets or planets with longer periods around this or similar stars, we would need a better understanding of stellar activity.

The impact of dark spots and bright plages on RV data has been extensively studied in the literature \citep[e.g.][]{1997ApJ...485..319S, 2002AN....323..392H, 2010A&A...512A..38L, 2014MNRAS.438.2717J, 2017MNRAS.466.1733B, 2019A&A...624A..83K}, but only few studies investigated effects of the magnetic field in starspots via Zeeman broadening
\citep{2013A&A...552A.103R, 2014MNRAS.443.2599H, 2016MNRAS.461.1465H, 2016csss.confE.134M, 2017MNRAS.465.3343D, 2020arXiv200513386H}, showing that magnetic activity is a good tracer of activity component of RV.

In this work we aim to quantify the contribution of resolved and unresolved magnetic spots (small scale magnetic regions) on radial velocity measurements of the star and, subsequently, detection of exoplanets.
Other effects such as brightness contrast or the convective blueshift have previously been investigated by e.g. \citet{2017A&A...607A.124M}, and are not included in our model at this stage.
Our aim in this paper is to quantify the impact of both the large magnetic features and small unresolved magnetic spots on the star's RV precision.
We use observations of the magnetic field of $\epsilon$ Eri, recovered using Zeeman-Doppler Imaging \citep[ZDI,][]{2008MNRAS.388...80P,2014AA...569A..79J,2017MNRAS.471L..96J}.
ZDI enables reconstruction of the geometry of the star's large-scale magnetic field.
The small scale features like magnetic spots are not resolved with this technique \citep{2014MNRAS.439.2122L}, so we model them in addition to the ZDI maps.

This paper is structured as follows: the observations of the large-scale magnetic field and magnetic spot modelling are described in Sections \ref{sec:modelling:field}, and Section \ref{sec:modelling:spots}, and the radial velocity measurements in Section \ref{sec:modelling:rv}.
The results are shown in Section \ref{sec:results}.
The radial velocity impact of the magnetic field is shown in Section \ref{sec:results:field} and planet detectability in Section \ref{sec:results:detectability}.
Our conclusions are discussed in Section \ref{sec:summary}.

\section{Modelling}
\label{sec:modelling}

The model presented here only considers the impact of the Zeeman effect on the radial velocity measurements and magnetic spots are modelled as small areas with high magnetic field strength.

\subsection{Magnetic field}
\label{sec:modelling:field}

We use eight epochs of magnetic maps from \citet{2014AA...569A..79J} and \citet{2017MNRAS.471L..96J} that span nearly eight years and cover almost three S-index cycles \citep{2013ApJ...763L..26M}: 2007.08, 2008.09, 2010.04, 2011.81, 2012.82, 2013.75, 2014.84, and 2015.01.
The observations were secured with the \'echelle spectropolarimeter NARVAL ($R\sim65000$).
The data analysis techniques used to reconstruct the large-scale magnetic field of $\epsilon$ Eri is described by \citet{2014AA...569A..79J}.
Maps of the radial component of the magnetic field are shown in the appendix (Figure \ref{fig:case_i_maps}).
The large-scale magnetic field topology changes quite substantially during the period of observations.
In epoch 2007.08 a very distinct dipolar structure is observed, which is not present in 2008.09 or 2010.04, where the main feature is a polar region of negative polarity.
This region changes to a positive polarity and back, finally showing a dipolar structure again in 2013.75, but with reversed polarity.

\subsection{Simulated magnetic spots}
\label{sec:modelling:spots}

While ZDI can recover the large-scale geometry of surface magnetic field, small features, such as magnetic spots, remain undetected.
Even though presence of magnetic spots was determined for some stars, the sizes and distribution of these spots remain unknown and might differ for young low-mass stars compared to the Sun \citep{2005LRSP....2....8B, 2009ARA&A..47..333D, 2009A&ARv..17..251S}.
We account for this in our models by using multiple distributions of the small and unresolved magnetic spots. 

For low-mass stars, the averaged surface magnetic field measured from \textit{Stokes I} is at least 10 times stronger than that measured from \textit{Stokes V} \citep{2000MNRAS.313..823W, 2012LRSP....9....1R, Lehmann2018, 2020A&A...635A.142K}, because it is not cancelled out like \textit{Stokes V}.
If there are magnetic spots (even of opposite polarity) on the surface, they add together to result in the \textit{Stokes I} field.
For an active star we can assume that the \textit{Stokes I} field is entirely coming from star spots.
If that is the case then we can define a simple relation between magnetic spot numbers and the \textit{Stokes I} field. 
Our simulated stellar surface is made up of $N$ elements and we assume a single spot is equal to one surface element. 
We calculate the number of spots based on the following equation:

\begin{equation}
	N\_{spots} = \frac{N B\_{I}}{B\_{spot}} \approx \frac{10 N B\_{V}}{B\_{spot}},
\end{equation}

where $B\_{I}$ is magnetic field from \textit{Stokes I}, $B\_{V}$ is magnetic field from \textit{Stokes V} and $B_{spot}$ is the field strength of a single magnetic spot.
In this paper we use a grid of 5000 elements between latitudes of $70\degree$ and $-30\degree$ 
(due to the observed inclination of the star, not an intrinsic property of the magnetic field distribution).
Maximum magnetic field strength $B_{V}$  measured by \citet{2014AA...569A..79J} is 42 G, maximum field strength of a spot $B_{spot}$ that we used is 3 kG, resulting in 700 spots in total.
The size of a single surface element is 180 $\mu$Hem, which is comparable to a small sunspot \citep{2016ApJ...830L..33M}.
As the spots are randomly distributed across the surface elements, some of them will be next to each other, thus creating a bigger spot.

The magnetic spot distribution cases considered in this paper are as follows:

\begin{enumerate}

	\item[\bf Case (i)] Only the large scale magnetic field measured using ZDI, up to 42 G, as measured by \citet{2014AA...569A..79J} and \citet{2017MNRAS.471L..96J} without any additional simulated magnetic spots. 
	\item[\bf Case (ii)] Positive and negative spots of equal strength (1 kG) are randomly distributed across all rotation phases and effectively all latitudes.
	\item[\bf Case (iii)] Randomly distributed magnetic spots of both positive and negative polarity that have at least a 3 kG magnetic field. It is a strong field but not unreasonable \citep[e.g.][]{1990IAUS..138..427S, 2020A&A...635A.202L} 
	\item[\bf Case (iv)] Same number of spots as in the previous case, but the positive spots have field strengths of 4 kG and the negative spots have field strength of 2 kG. 
	\item[\bf Case (v)] The spots are only simulated in areas where large magnetic regions are present.
	In the regions of strong positive field we simulate stronger positive spots (3 kG) and weaker negative spots (1 kG), and the opposite for the negative regions.
	The distribution is still random but just localized to certain phase ranges. 
	\item[\bf Case (vi)] Artificial star with the same stellar parameters as $\epsilon$ Eri (inclination, $v$ sin$i$, rotation period etc) with no large-scale field.
	Both positive and negative spots are 3 kG.
	Only one epoch of observations with a random distribution of spots is produced. 

\end{enumerate}

The resulting magnetic maps can be found in the appendix (Figures \ref{fig:case_i_maps}--\ref{fig:case_vi_maps}).
The reconstruction of the large-scale magnetic field at epoch of observation takes into account stellar effects such as differential rotation and evolution of the magnetic field.
The lifetime of the small unresolved spots is not included in the model and is effectively the same in all cases.
The lifetimes of the small spots is longer than the timeframe used to reconstruct each ZDI map.
The only difference between the cases is the distribution and field strength of the magnetic spots.

To compute synthetic line  profiles, we use the simulated magnetic field maps divided into a grid of pixels, each being associated with a local \textit{Stokes I} profile, using the method from \citet{2009A&A...508L...9P}.
Profiles between the observed epochs are interpolated.

\subsection{Simulation of radial velocity measurements}
\label{sec:modelling:rv}

To simulate radial velocity observations, we use the LSD profiles generated for a set of stellar rotation phases.
Each profile, centred at 5500\AA, is chosen depending on the rotational phase of the star at the time of the simulated observation and the epoch.
If we inject a planetary signal into the simulation (Section \ref{sec:results:detectability}), the line profile is Doppler-shifted according to the Keplerian signal.
To simulate instrumental noise of approximately 10 cm s$^{-1}$ in radial velocity \citep[achievable with ESPRESSO,][]{2010SPIE.7735E..0FP}, Gaussian noise is added to the profile.
All the profiles in the simulated observation set are averaged to produce a template, that is cross-correlated with every profile to measure the radial velocity.

We have approximately one magnetic map per year, but it was shown that the magnetic field of $\epsilon$ Eri changes drastically on a time-scale of only months \citep{2017MNRAS.471L..96J}.
To account for this, we include two maps spaced only 2 months apart in November 2014 and January 2015.
In this paper, the shape of each LSD profile (before adding Doppler shift and noise) was interpolated between the epochs, according to the time of observation, thus creating a smooth shape transition from one magnetic map to another.
This approach does not provide any information about shorter time-scales due to the data sampling, but still provides good constraints on long term effects caused by the stellar magnetic cycle and the amplitude of the radial velocity noise.

We compute a Lomb-Scargle periodogram \citep{1976Ap&SS..39..447L, 1982ApJ...263..835S} of the resulting RV curve and fit a Keplerian orbit using a non-linear least squares method to retrieve the planetary signal.
The code was developed in \textsc{python} and is available on \textsc{github}\footnote{\url{https://github.com/timberhill/radiant}}.

\section{Results}
\label{sec:results}

Based of the set up described above we investigate the effect of magnetic field on radial velocity measurements with and without magnetic spots and the detectability of planets in presence of magnetic spots.

The main results are presented in Figures \ref{fig:rv_rot} and \ref{fig:longbase}, which show the simulated radial velocity measurements for each of the considered magnetic spot distribution cases (top to bottom) and for each epoch of the observations (left to right), and the whole time series, respectively.
The detectability of a range of different planets in presence of this noise is shown on Figure \ref{fig:detectability} for each of the cases and for two observational strategies: randomly spaced observations (left column) and clusters of observations (right column).

\subsection{Effect of the magnetic field}
\label{sec:results:field}

\begin{figure*}
	\hspace{0.085\textwidth} 2007.08 \hfill 2008.09 \hfill 2010.04 \hfill 2011.81 \hfill 2012.82 \hfill 2013.75 \hfill 2014.84 \hfill 2015.01 \hspace{0.05\textwidth}
\par
	\centering
	\begin{subfigure}{\textwidth}
		\centering
		\includegraphics[width=.95\linewidth]{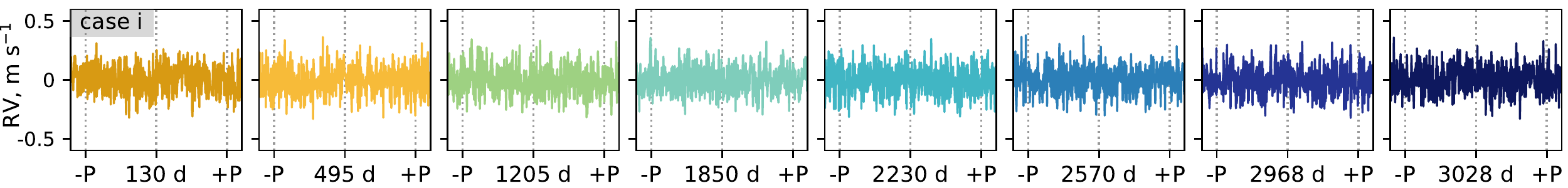}
		\label{fig:case_i}
	\end{subfigure}
	\begin{subfigure}{\textwidth}
		\centering
		\includegraphics[width=.95\linewidth]{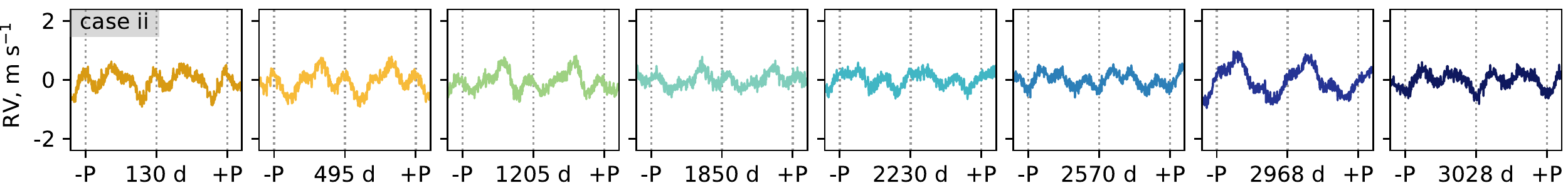}
		\label{fig:case_ii}
	\end{subfigure}
	\begin{subfigure}{\textwidth}
		\centering
		\includegraphics[width=.95\linewidth]{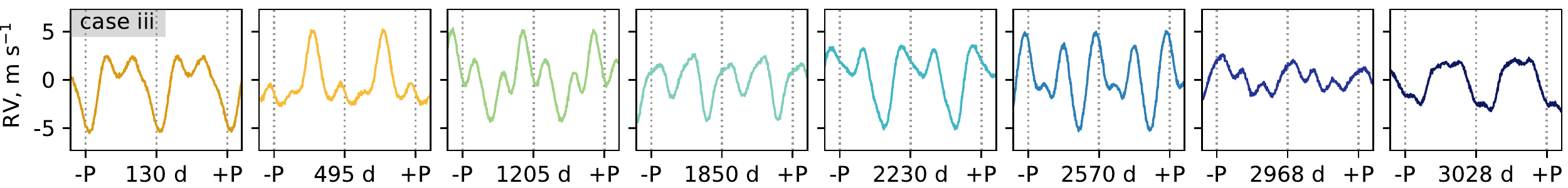}
		\label{fig:case_iii}
	\end{subfigure}
	\begin{subfigure}{\textwidth}
		\centering
		\includegraphics[width=.95\linewidth]{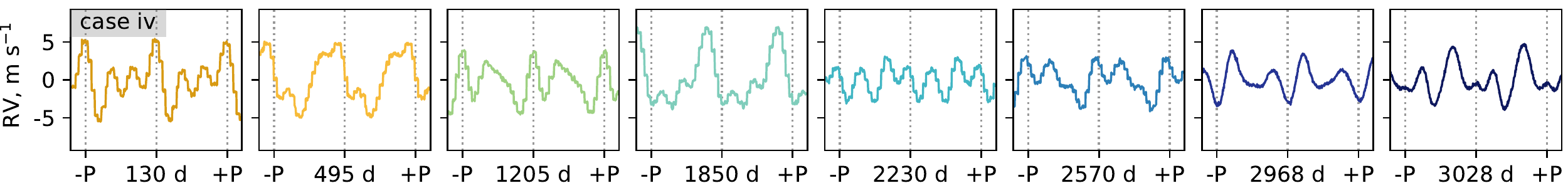}
		\label{fig:case_iv}
	\end{subfigure}
	\begin{subfigure}{\textwidth}
		\centering
		\includegraphics[width=.95\linewidth]{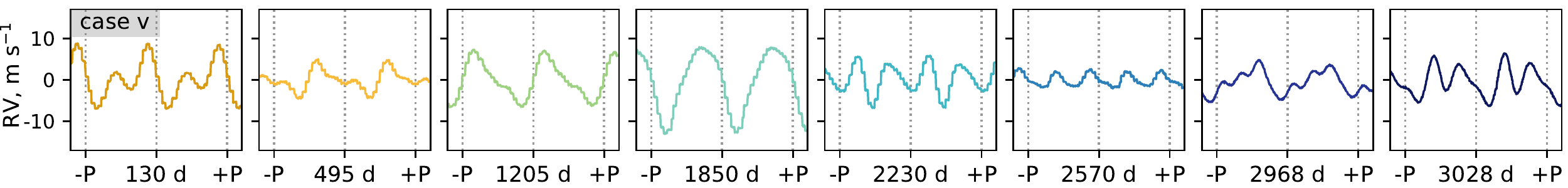}
		\label{fig:case_v}
	\end{subfigure}
	\begin{subfigure}{\textwidth}
		\centering
		\includegraphics[width=.95\linewidth]{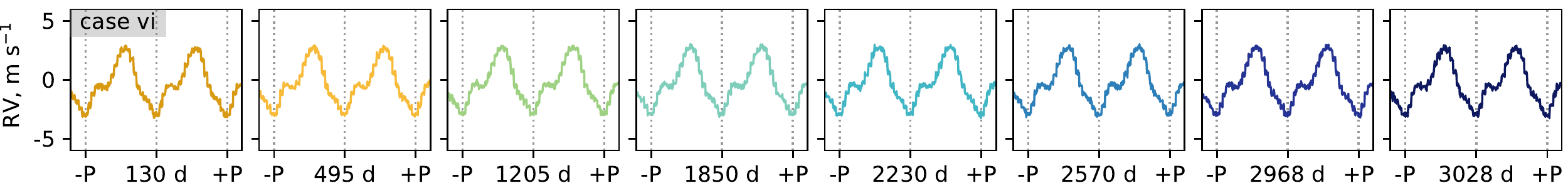}
		\label{fig:case_vi}
	\end{subfigure}
	\caption{
		Radial velocity curves, computed using magnetic maps and the simulated spots.
		The simulations using different magnetic maps are shown top to bottom (see Section \ref{sec:modelling:spots} and the appendix for the spots distribution).
		The observing epochs are shown left to right for each case.
		Two periods of stellar rotation (11.68 days) are plotted and indicated by the gray vertical lines.
		The colours represent different epochs and are consistent with Figure \ref{fig:longbase}.
	}
	\label{fig:rv_rot}

	\centering
	\includegraphics[width=0.95\textwidth]{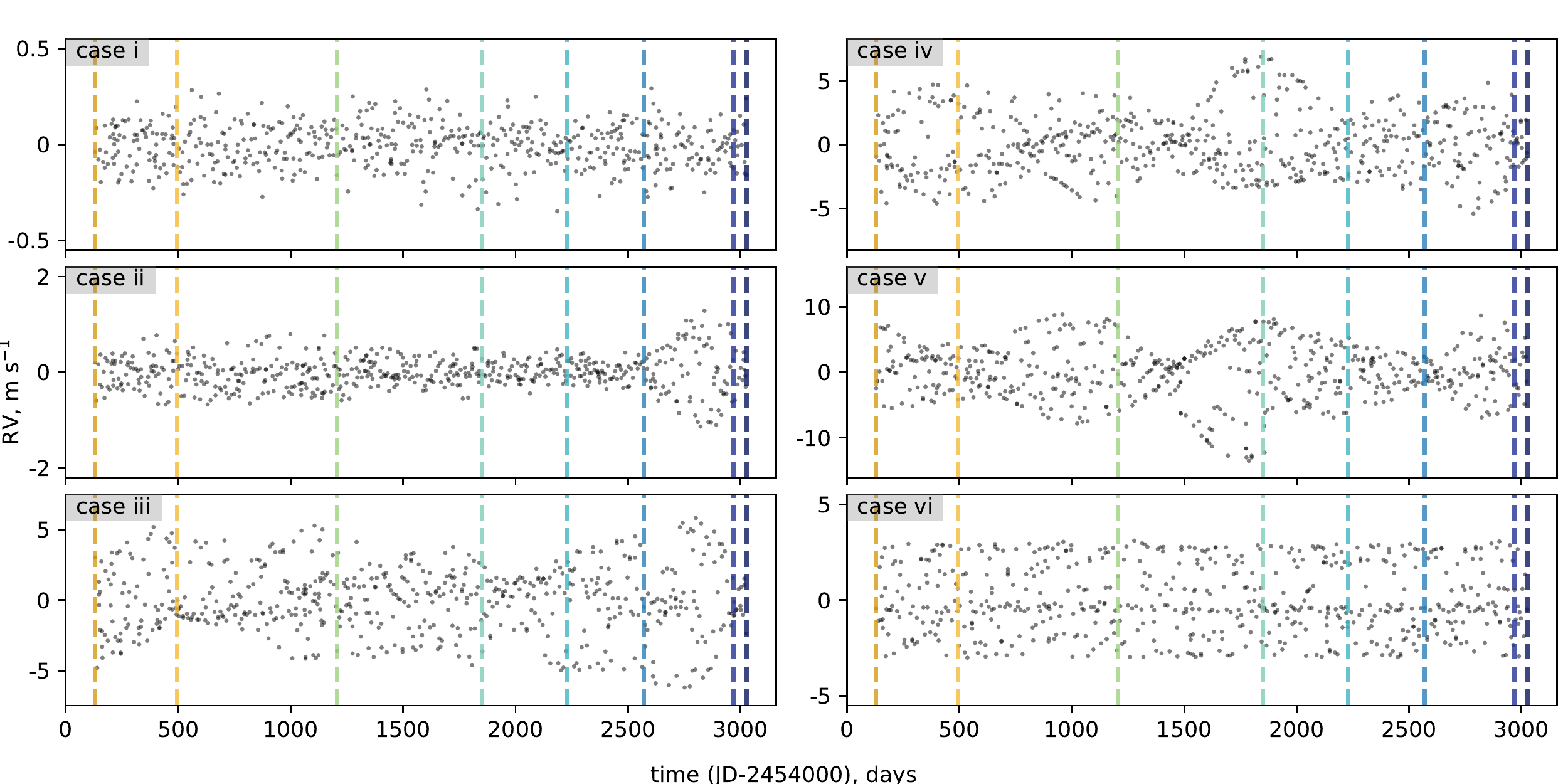}
	\caption{
		Radial velocity curves, computed using magnetic maps and the simulated spots over the whole span of observations.
		The eight simulations use different magnetic maps (see Section \ref{sec:modelling:spots} and the appendix for the spots distribution).
		Vertical dashed lines show observation epochs, colours are consistent with Figure \ref{fig:longbase}.
		This figure shows the same results as Figure \ref{fig:rv_rot}, but on a longer time-scale.
		}
	\label{fig:longbase}
\end{figure*}

In this section only effects of the magnetic field and instrumental noise are introduced, without any planetary signals.
First we consider the radial velocity produced by the magnetic field with spots at the stellar rotation period.
The simulated radial velocity curves around each observational epoch for each of the magnetic maps are shown on Figure \ref{fig:rv_rot}.
The different spot distribution cases (described in Section \ref{sec:modelling:spots}) are shown from top to bottom (marked on the left), and the eight epochs are plotted from left to right.
The measured line shape variation due to the large scale field ({\it case i}) has a radial velocity effect of the order of 2 cm s$^{-1}$ and is almost completely hidden behind the noise.
Including the magnetic spots created a significant radial velocity effect with amplitudes ranging from $\sim1$ m s$^{-1}$ ({\it case ii}) to 10 m s$^{-1}$ ({\it case v}).
This also shows how significantly the distribution of the spots changes the radial velocity of the star.

As one would expect, the Zeeman effect can introduce RV curves with various shapes, frequently having primary and secondary peaks.
A similar radial velocity pattern is present in in the CHIRON (Cerro Tololo Inter-American Observatory High Resolution Spectrometer) observations of $\epsilon$ Eri in September 2014 \citep{2016ApJ...824..150G}, close to the 2014.84 epoch, but with a higher amplitude.

Figure \ref{fig:longbase} shows the effect of the magnetic field on radial velocity measurements for each of the considered cases.
It shows the same result as Figure \ref{fig:rv_rot}, but on a longer time-scale.
The observed epochs are indicated with vertical lines and the simulated observations between them are interpolated as described in Section \ref{sec:modelling:rv}.
The observations were randomly distributed and no signals beyond the stellar rotation period are present.

This plot illustrates the dramatic changes of the RMS from one epoch to another, even using the same approach to the spot distribution.
The radial velocity effect of the large scale magnetic field, which does not exceed 42 G across all considered epochs, is quite small and is below the precision of current instruments ($\sim2$ cm s$^{-1}$) \citep{2018haex.bookE.157G}.
The small unresolved magnetic spots, on the other hand, produce enough noise to mask Earth mass planets via the Zeeman effect alone.

The seemingly smooth curves originate from the way profiles in between observations are interpolated and the actual data is eight epochs, marked with vertical lines on Figure \ref{fig:longbase}.
Even though there is no sign of the three year activity cycle, the simulated radial velocity measurements do show some structure, especially prominent in {\it case v} (centre right panel on Figure \ref{fig:longbase}), where spots are only simulated in areas with large magnetic regions.
If we consider the probability of measuring a given radial velocity without prior knowledge about the phase of the star and magnetic field geometry, then the probability of measuring 2 m s$^{-1}$ around $t=1500$ is higher than measuring 0 m s$^{-1}$.
The same applies to the cluster of points around day $t=500$ and $t=2000$ of {\it case iii}.
This may create a bias in long baseline RV measurements depending on the observing strategy if they don't cover a full rotation of the star \citep[e.g.][]{2016Natur.536..437A}.

\subsection{Detectability of planets}
\label{sec:results:detectability}

To investigate whether the magnetic field effect on radial velocity influences planetary detection, we simulated a grid of planets with masses of 1, 2, 5, 10, 20, 50, 100, 159, 318, 636, 1589 $M_\oplus$ and semi-major axes of 0.01, 0.02, 0.05, 0.1, 0.2, 0.5, 1, 2, 5 au.
All simulated planets have circular orbits.

Each orbit is simulated and fitted 10 times in the different parts of the available observations span of 8 years.
The fitted planetary mass errors are then calculated and averaged for each of the cases.
Two observational strategies were compared -- singular observations with random step and clustered (observations made in batches).

We use the relative errors in mass and period as detectability criteria, calculated as $|M\_{fit} - M| / M$, and $|P\_{fit} - P| / P$, respectively, where $M\_{fit}$ and $P\_{fit}$ are the derived mass and period of a planet, $M$ and $P$ are the true mass and period of a planet.
Some of the fits, especially of the long-period and low-mass planets, result in a mass error of more than 100 per cent and they are listed as $>1$ as it is a natural cut-off.
For periods, we consider 50 per cent error to be a cut-off.

\begin{figure*}
	\centering
	\includegraphics[width=0.49\textwidth,keepaspectratio,trim={0.2cm 0 0.4cm 0},clip]{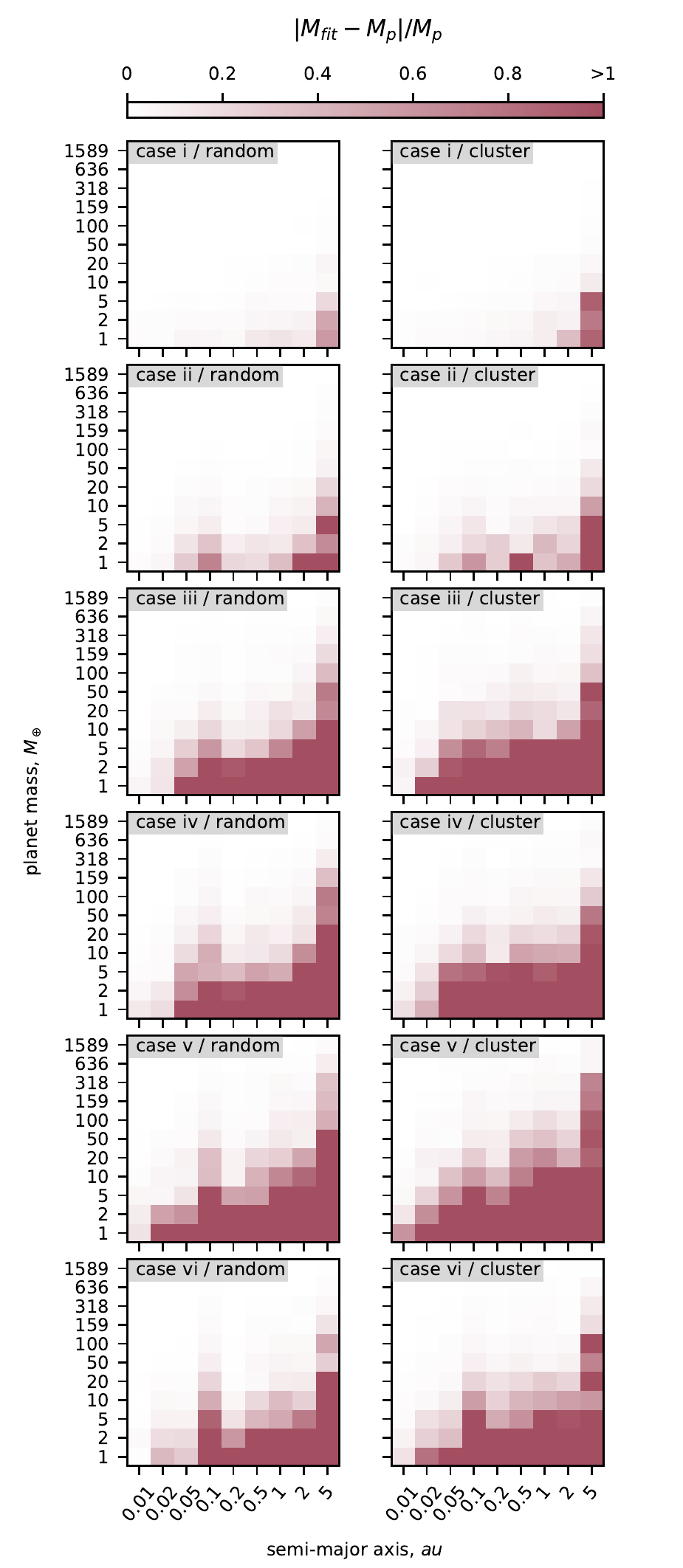}
	\includegraphics[width=0.49\textwidth,keepaspectratio,trim={0.2cm 0 0.4cm 0},clip]{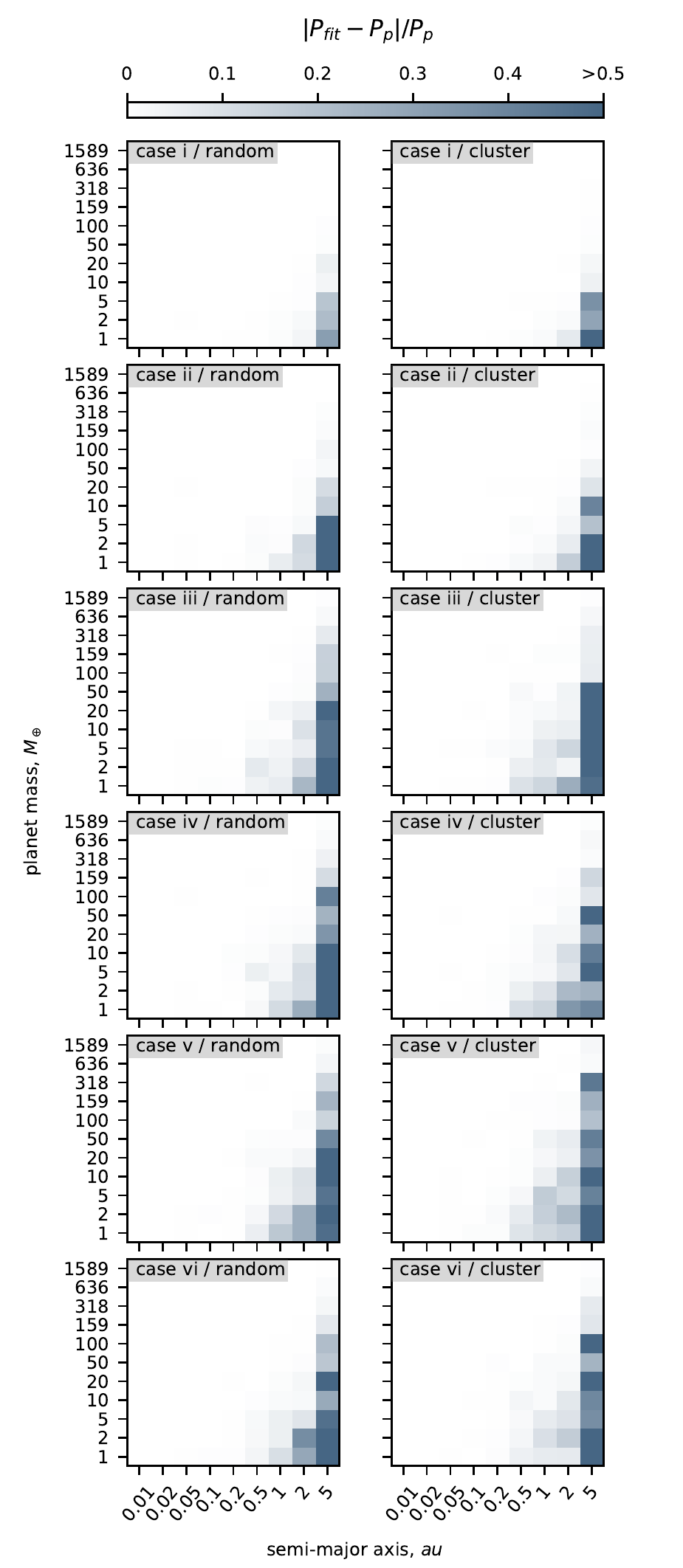}

	\caption{
		Planet detection simulations that show a mass and period fit errors for grids of simulated planets for each of the spot distribution cases (section \ref{sec:modelling:spots}) and observational strategies.
		\textit{Top to bottom:} different spot distributions, as described in Section \ref{sec:modelling:spots}.
		\textit{Left to right:} mass error using randomly spaced observations, mass error using clustered observations, period error using randomly spaced observations, period error using clustered observations.
		Red colour indicates a relative error in mass fit over 100 per cent and the blue colour indicates a relative error in period determination over 50 per cent.
		White denotes a small mass or period error and a precise retrieval of the planet (typically within 1 per cent).
	}
	\label{fig:detectability}
\end{figure*}

In both cases the three orbital periods were covered with ~30 observations per period.
The fits at 5 au have orbital period longer than the available observation span and therefore are not reliable but are included for completeness.
Figure \ref{fig:detectability} shows the detectability of planets for each of the corresponding cases and observational strategies.
The left panel shows relative mass error (in red) and the right panel shows relative error in period determination (in blue).
White squares mean the planet's mass or period were recovered precisely (typically within 1 per cent, darkest squares indicate a mass error of $>100$ per cent, or a period error of $>50$ per cent.
The panels have two columns, showing the results computed using randomly spaced observations and clustered observations.
The fits in general were of comparable quality for both approaches.

{\it Case i} (top panels on Figure \ref{fig:detectability}) can be used as a benchmark as the large scale magnetic field effect is very small and the instrumental noise is dominant.
In all cases, high-mass objects are nearly always recovered and so appear as white whereas lower mass planets (e.g., 1 $M_\oplus$) are only recovered at small semi-major axes or where there are no spots ({\it case i}).

Majority of the initial planetary periods are recovered within 1 per cent, except for the planets with amplitudes well below the noise.
The masses are less well recovered which is explained by the fact that we are not introducing signals apart from the rotation of the star.
The masses of the planets, however, are much larger due to the noise introduced.
In addition, there is an increase of detectable planets mass at $a = 0.1$ au, where orbital period is very close to stellar rotation period.
For heavier planets the fits get better as the planetary signal starts to dominate in amplitude.
This includes the proposed $\epsilon$ Eri b, which has a mass of 247.9 $M_\oplus$ and the semi-major axis of $3.48$ au, resulting in the semi-amplitude of the radial velocity signal of 11.8 m s$^{-1}$ \citep{2019AJ....157...33M}.

We adopted a simple approach of planet recovery that is consistent and illustrates the detectability between different magnetic field configurations.
The rotational signal could be subtracted to improve the fits, but it would still be hiding planets close to the rotation period or its harmonics.
Furthermore, the magnetic field noise can be also be decorrelated by measuring widths of Zeeman-sensitive lines in the spectrum, but this is outside the scope of this paper.

\section{Summary and discussion}
\label{sec:summary}

In this work we presented a simple model for estimating radial velocity effects of magnetic spots using \textit{Stokes V} observations of $\epsilon$ Eridani.
We quantified the radial velocity impact of the measured large scale magnetic field of the star, as well as unobserved magnetic spots.

The observed large scale magnetic fields have a very small impact on the radial velocity -- about 2 cm s$^{-1}$ -- which below the detection limit even of the most precise instruments like ESPRESSO.
The unresolved magnetic spots, on the other hand, might create a strong radial velocity signature up to 10 m s$^{-1}$, which is consistent with observation of the Sun as a star.
A signal of this amplitude can hide or mimic planets under 20 Earth masses in a temperate zone of a Sun-like star.
This level of noise is introduced by the simulated rotation of the star and long-term variability present in the magnetic maps that span almost three S-index cycles.

The radial velocity amplitude also depends heavily on the distribution of the magnetic spots.
Using the same approach to the spot distribution, the radial velocity effect can change drastically from one observing season to another.
In the future we will apply this approach to other stars with a range of spectral types.
For instance, the unresolved magnetic field can be measured with an indication of its complexity using methods such as those developed by \citep{2019A&A...626A..86S}. 
A better understanding of the relationship between the magnetic field of a star and its measured radial velocity is essential to understanding stellar activity and detection of Earth-sized planets around solar type stars.

\section*{Acknowledgements}

We would like to thank Guillem Anglada-Escud\'e for fruiful discussions that led to this paper.
ML acknowledges financial support from Astromundus programme for MSc at the University of G\"ottingen and a University of Hertfordshire PhD studentship.
SJ acknowledges the support of the German Science Foundation (DFG) Research Unit FOR2544 `Blue Planets around Red Stars', project JE 701/3-1 and DFG priority program SPP 1992 `Exploring the Diversity of Extrasolar Planets' (JE 701/5-1).
SBS acknowledges funding via the Austrian Space Application Programme (ASAP) of the Austrian Research Promotion Agency (FFG) within ASAP11, the FWF NFN project S11601-N16, and the sub-project S11604-N16. 
HJ acknowledges support from the UK Science and Technology Facilities Council [ST/R006598/1].
AAV has received funding from the European Research Council (ERC) under the European Union's Horizon 2020 research and innovation programme (grant agreement No 817540, ASTROFLOW).

This research made use of \textsc{numpy} \citep{van2011numpy}, \textsc{astropy}, a community-developed core Python package for Astronomy \citep{2013A&A...558A..33A}, \textsc{PyAstronomy} (\url{https://github.com/sczesla/PyAstronomy}), \textsc{scipy} \citep{jones_scipy_2001}, \textsc{scikit-learn} \citep{mckinney}, and \textsc{matplotlib}, a Python library for publication quality graphics \citep{Hunter:2007}.

\section*{Data availability statement}

The data underlying this article are available on \textsc{Github}, at \url{https://github.com/timberhill/radiant} along with the code used to generate it.



\bibliographystyle{mnras}
\bibliography{paper}



\appendix
\input{appendix.tex}


\bsp	
\label{lastpage}
\end{document}

%% file: appendix.tex
\begin{figure*}
    \centering
    
    \begin{subfigure}{0.49\textwidth}
        \vspace{0.5cm}
        \caption{2007.08}
        \centering
        \includegraphics[width=.95\linewidth, trim={1.8cm 0cm 1.7cm 1.5cm},clip]{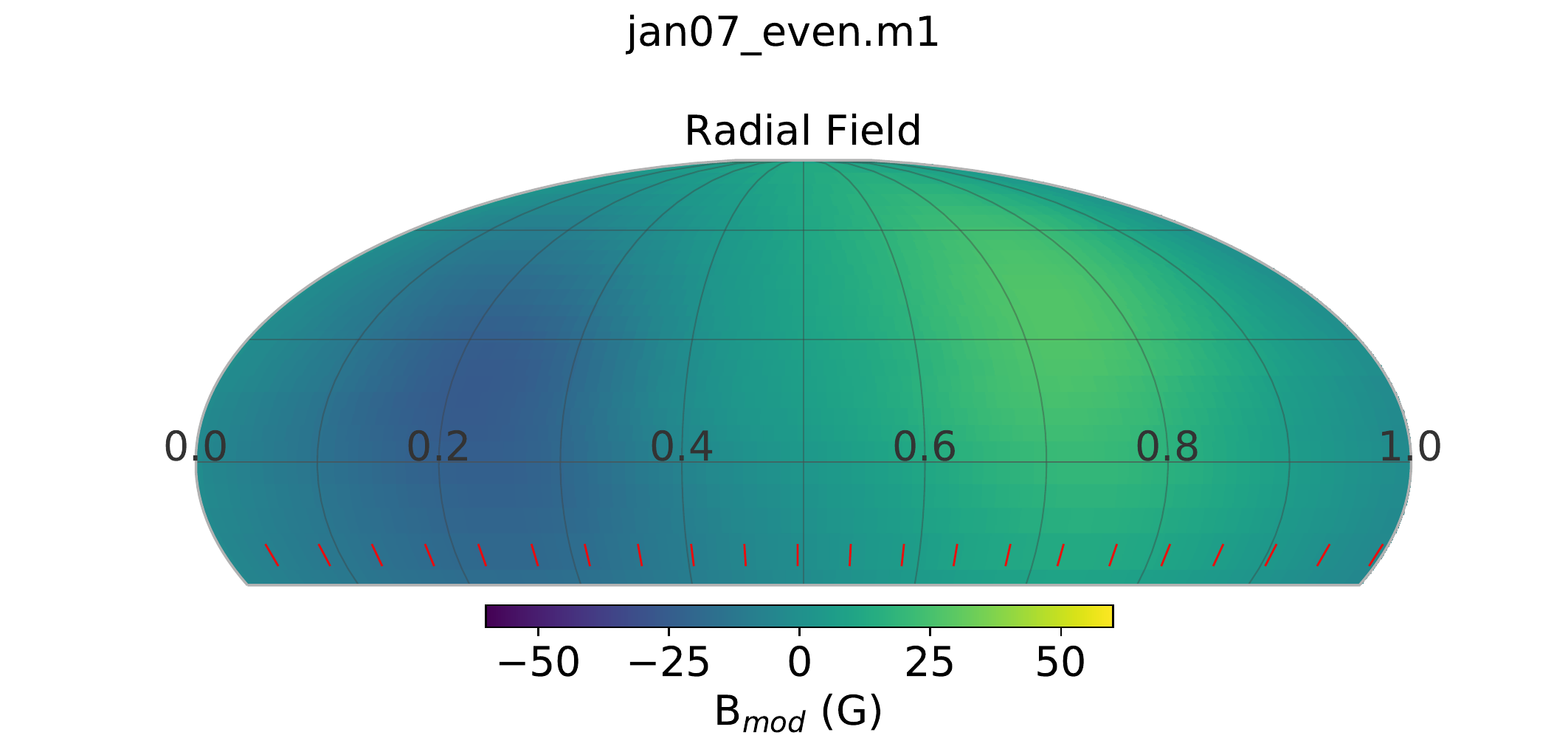}
        \label{fig:case_i_jan07}
        \vspace{0.5cm}
    \end{subfigure}
    \begin{subfigure}{0.49\textwidth}
        \vspace{0.5cm}
        \caption{2008.09}
        \centering
        \includegraphics[width=.95\linewidth, trim={1.8cm 0cm 1.7cm 1.5cm},clip]{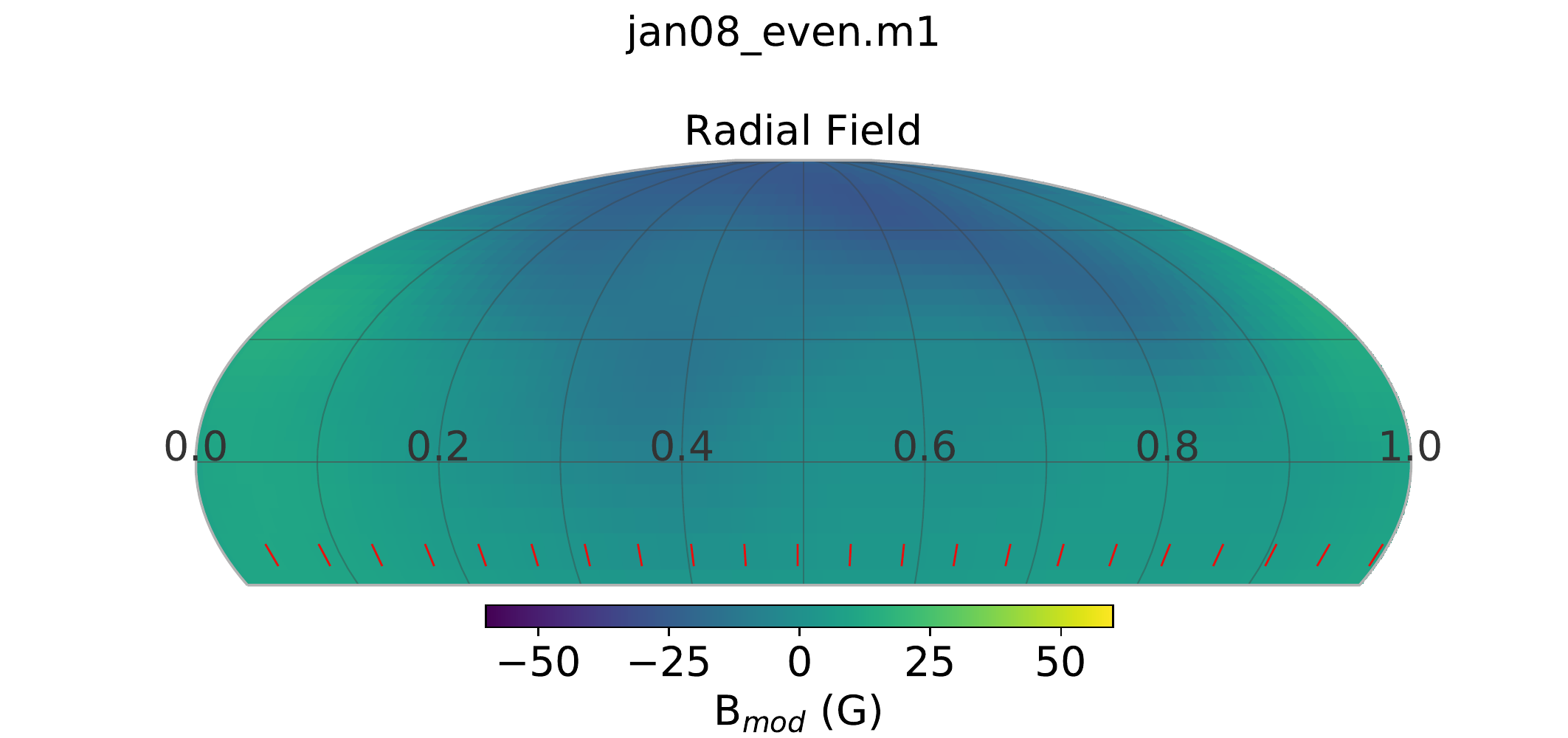}
        \label{fig:case_i_jan08}
        \vspace{0.5cm}
    \end{subfigure}
    \begin{subfigure}{0.49\textwidth}
        \vspace{0.5cm}
        \caption{2010.04}
        \centering
        \includegraphics[width=.95\linewidth, trim={1.8cm 0cm 1.7cm 1.5cm},clip]{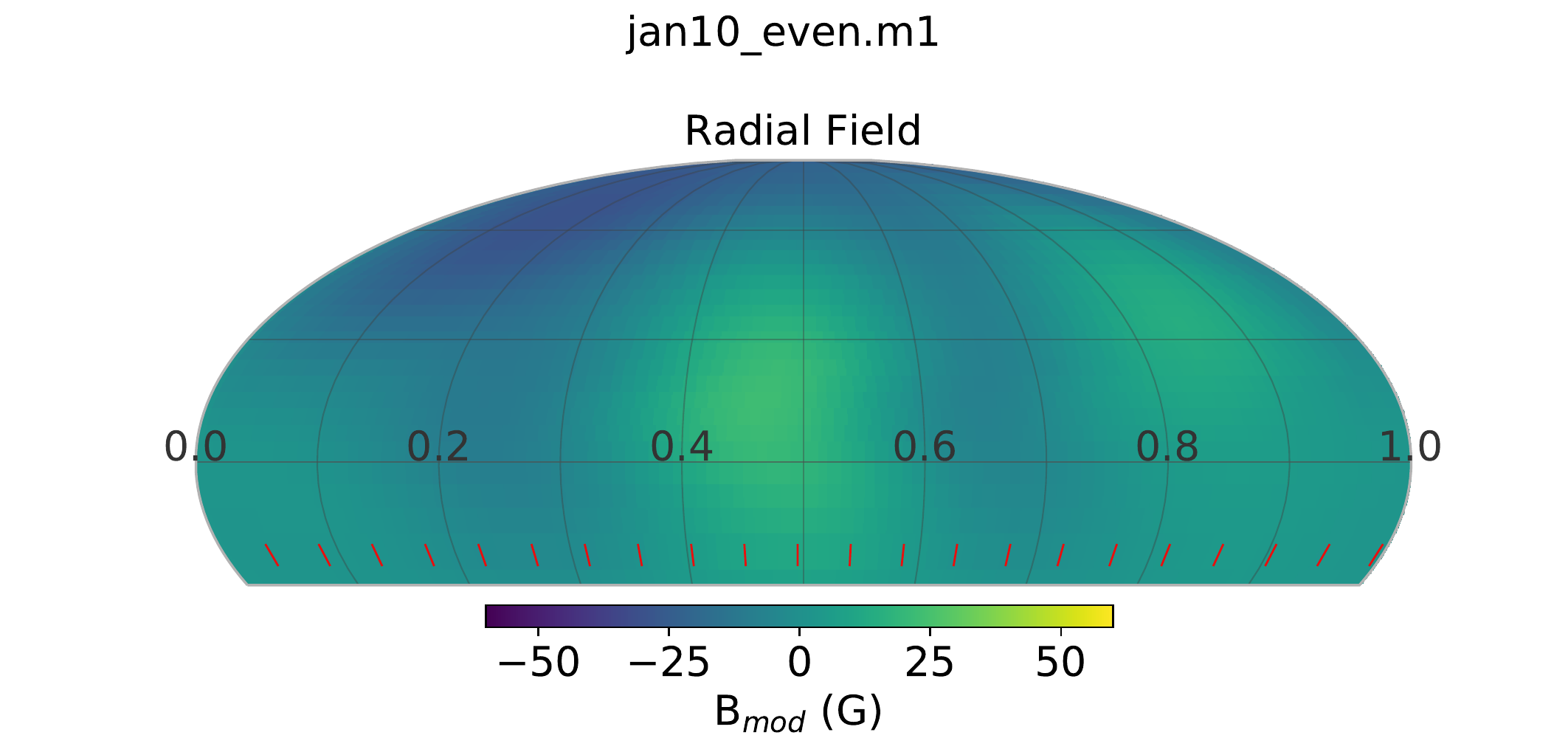}
        \label{fig:case_i_jan10}
        \vspace{0.5cm}
    \end{subfigure}
    \begin{subfigure}{0.49\textwidth}
        \vspace{0.5cm}
        \caption{2011.81}
        \centering
        \includegraphics[width=.95\linewidth, trim={1.8cm 0cm 1.7cm 1.5cm},clip]{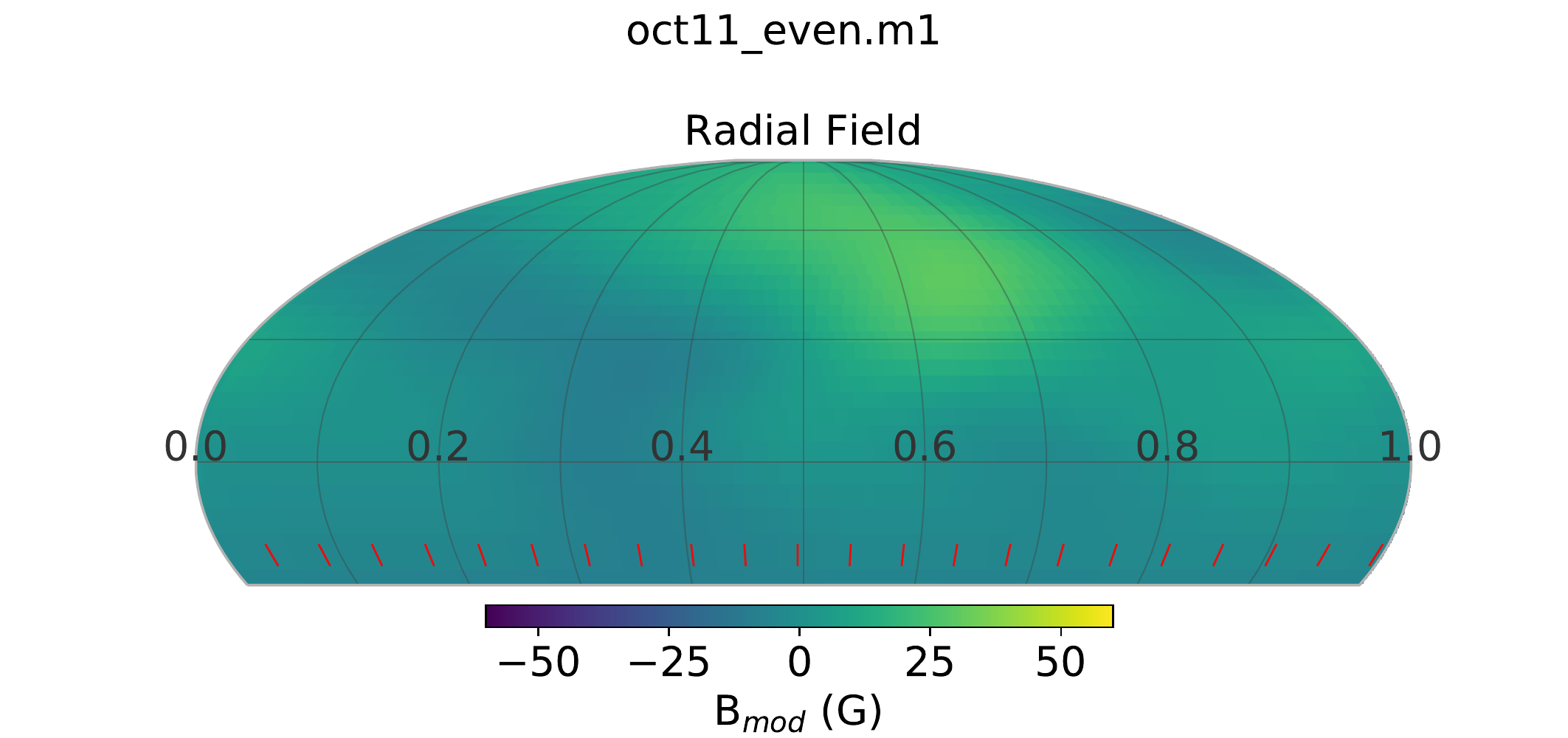}
        \label{fig:case_i_oct11}
        \vspace{0.5cm}
    \end{subfigure}
    \begin{subfigure}{0.49\textwidth}
        \vspace{0.5cm}
        \caption{2012.82}
        \centering
        \includegraphics[width=.95\linewidth, trim={1.8cm 0cm 1.7cm 1.5cm},clip]{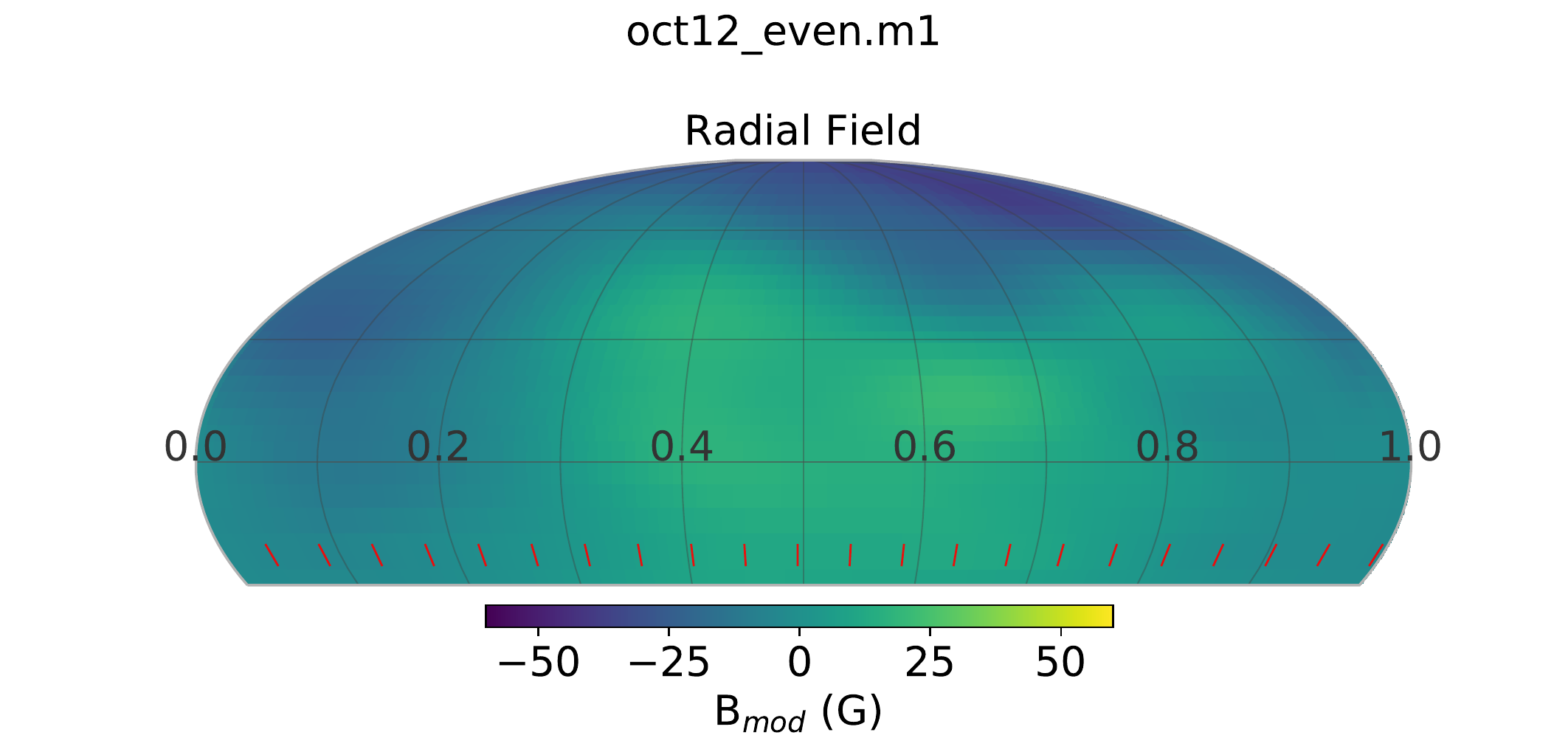}
        \label{fig:case_i_oct12}
        \vspace{0.5cm}
    \end{subfigure}
    \begin{subfigure}{0.49\textwidth}
        \vspace{0.5cm}
        \caption{2013.75}
        \centering
        \includegraphics[width=.95\linewidth, trim={1.8cm 0cm 1.7cm 1.5cm},clip]{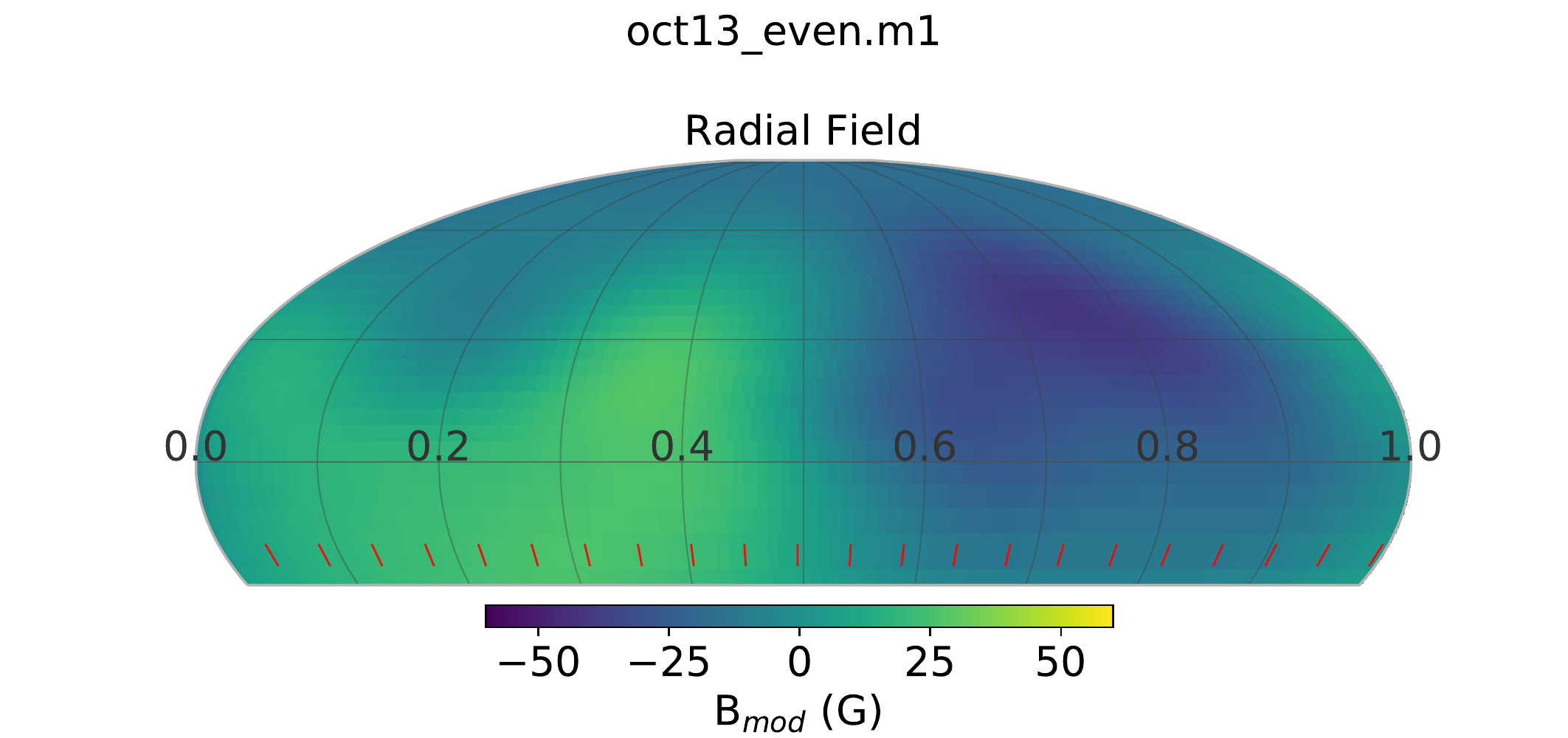}
        \label{fig:case_i_oct13}
        \vspace{0.5cm}
    \end{subfigure}
    \begin{subfigure}{0.49\textwidth}
        \vspace{0.5cm}
        \caption{2014.84}
        \centering
        \includegraphics[width=.95\linewidth, trim={1.8cm 0cm 1.7cm 1.5cm},clip]{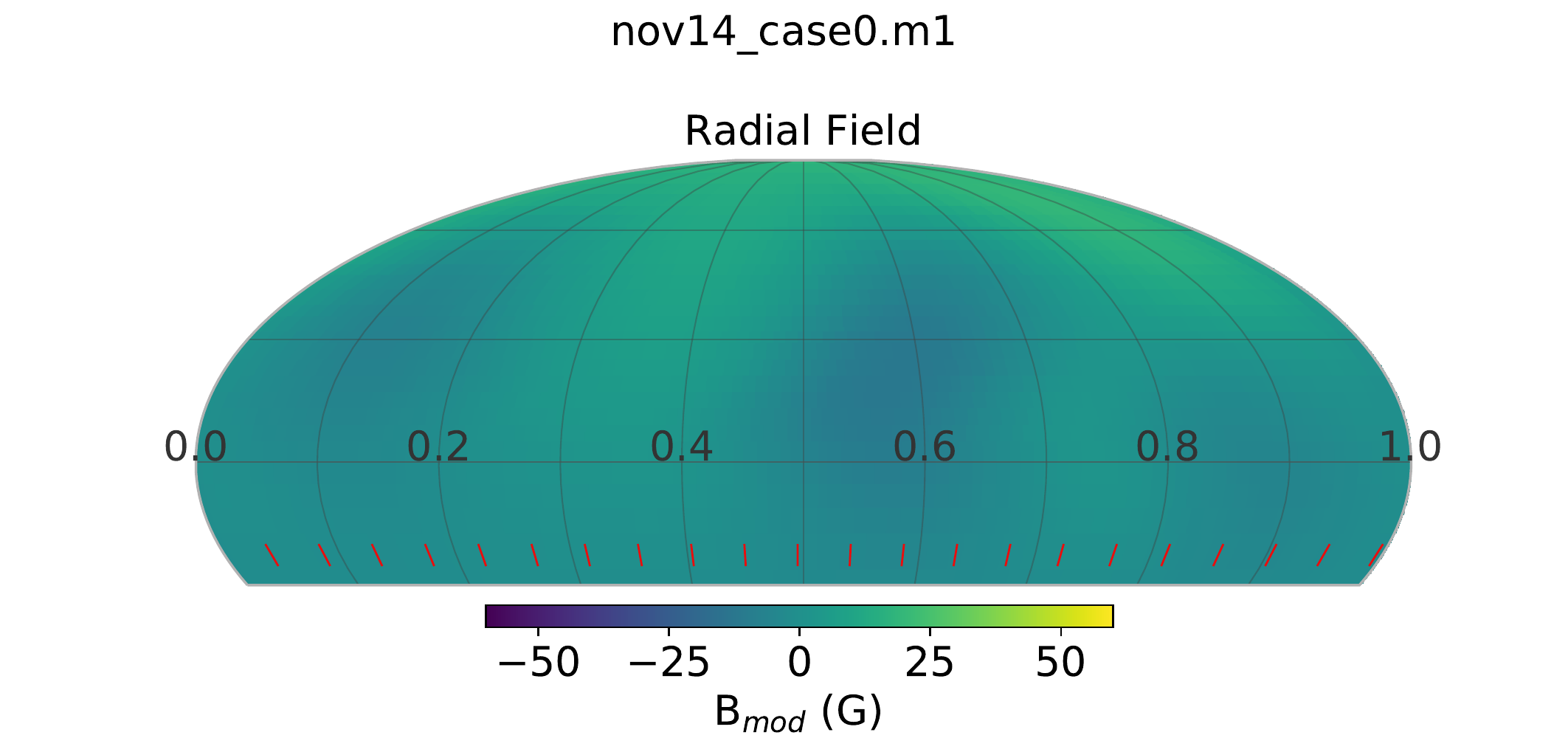}
        \label{fig:case_i_nov14}
        \vspace{0.5cm}
    \end{subfigure}
    \begin{subfigure}{0.49\textwidth}
        \vspace{0.5cm}
        \caption{2015.01}
        \centering
        \includegraphics[width=.95\linewidth, trim={1.8cm 0cm 1.7cm 1.5cm},clip]{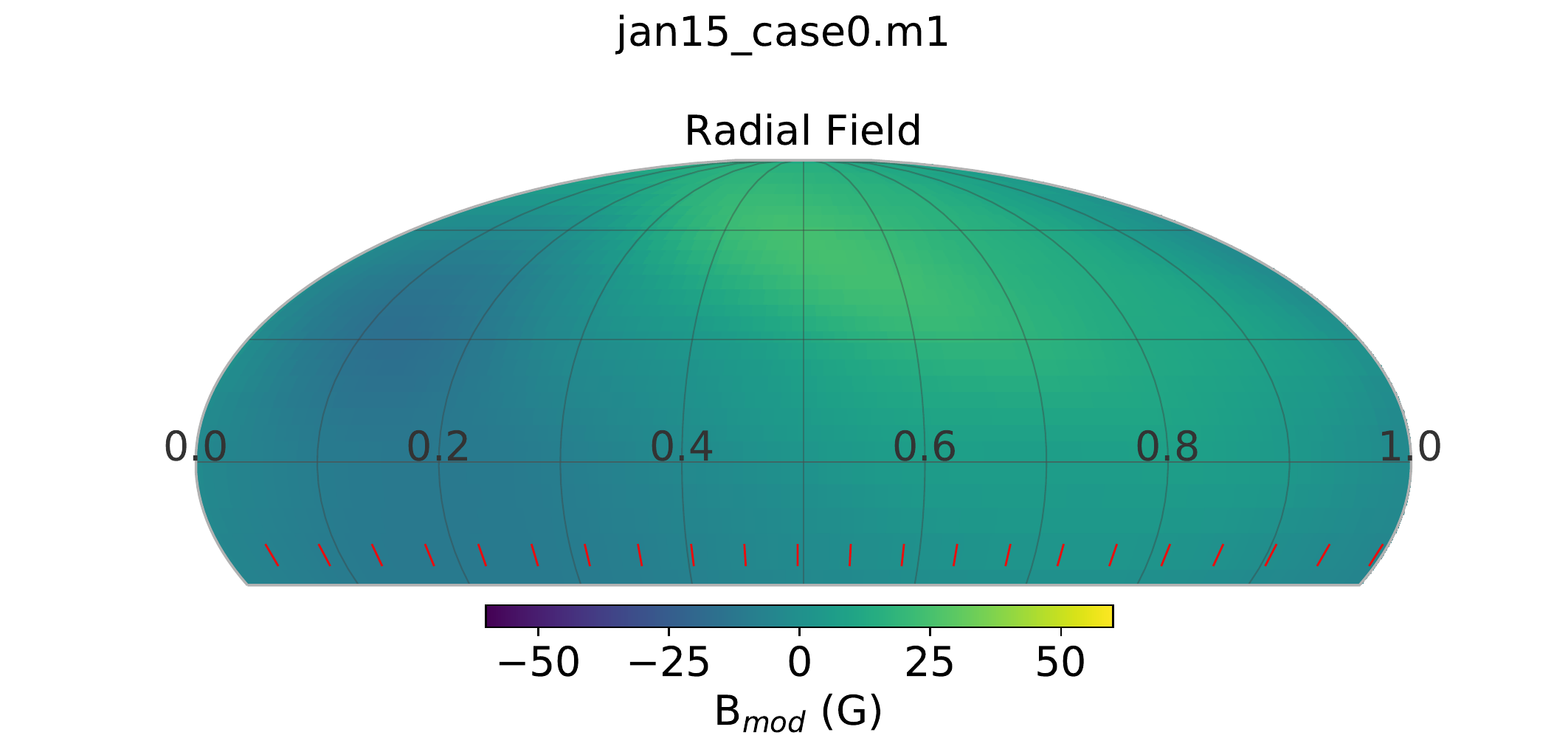}
        \label{fig:case_i_jan15}
        \vspace{0.5cm}
    \end{subfigure}
    \caption{
        Magnetic field maps of $\epsilon$ Eridani for eight epochs of observations (see sub-captions), according to {\it case i}.
        Bright areas indicate positive polarity and dark -- negative polarity.
    }
    \label{fig:case_i_maps}
\end{figure*}

\begin{figure*}
    \centering
    
    \begin{subfigure}{0.49\textwidth}
        \vspace{0.5cm}
        \caption{2007.08}
        \centering
        \includegraphics[width=.95\linewidth, trim={1.8cm 0cm 1.7cm 1.5cm},clip]{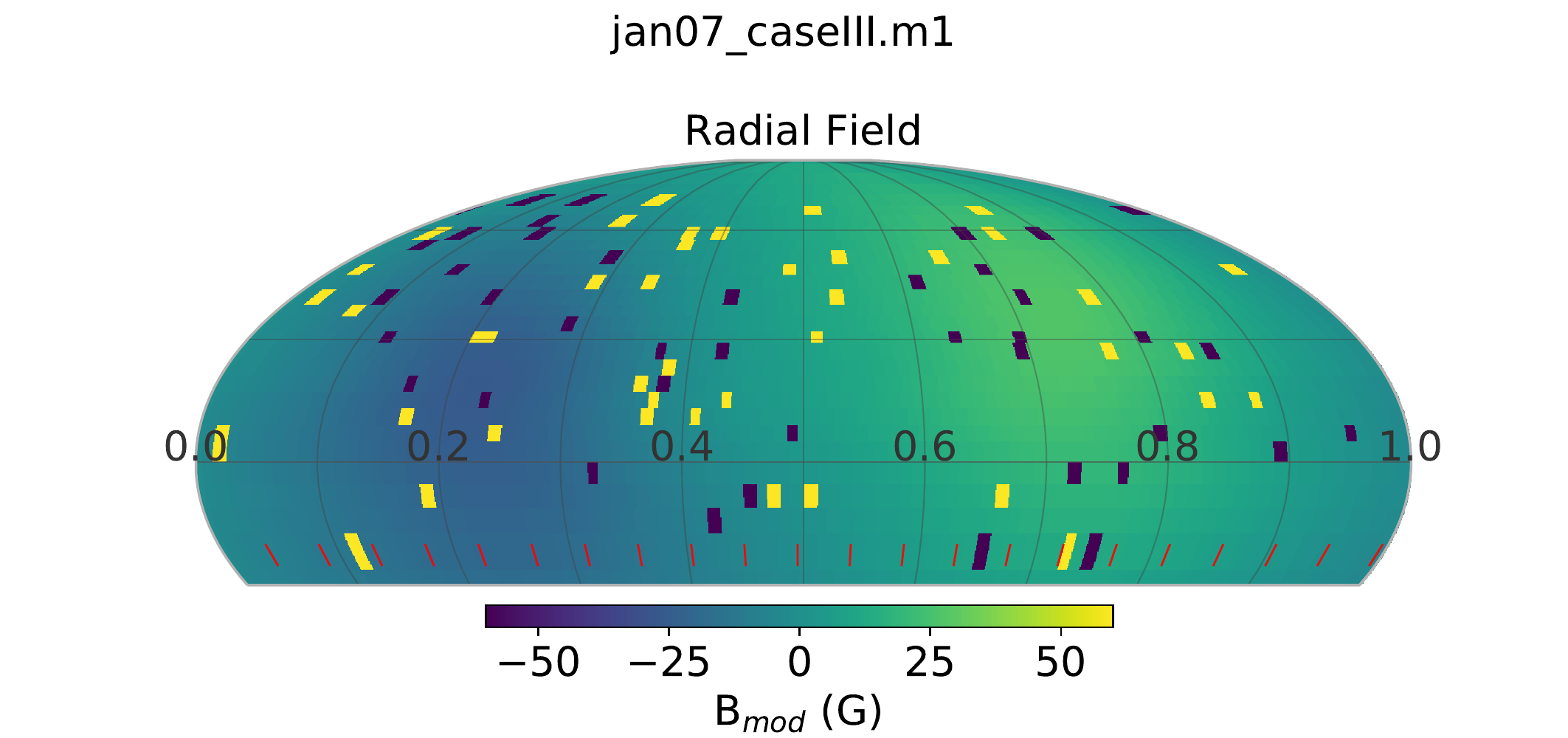}
        \label{fig:case_ii_jan07}
        \vspace{0.5cm}
    \end{subfigure}
    \begin{subfigure}{0.49\textwidth}
        \vspace{0.5cm}
        \caption{2008.09}
        \centering
        \includegraphics[width=.95\linewidth, trim={1.8cm 0cm 1.7cm 1.5cm},clip]{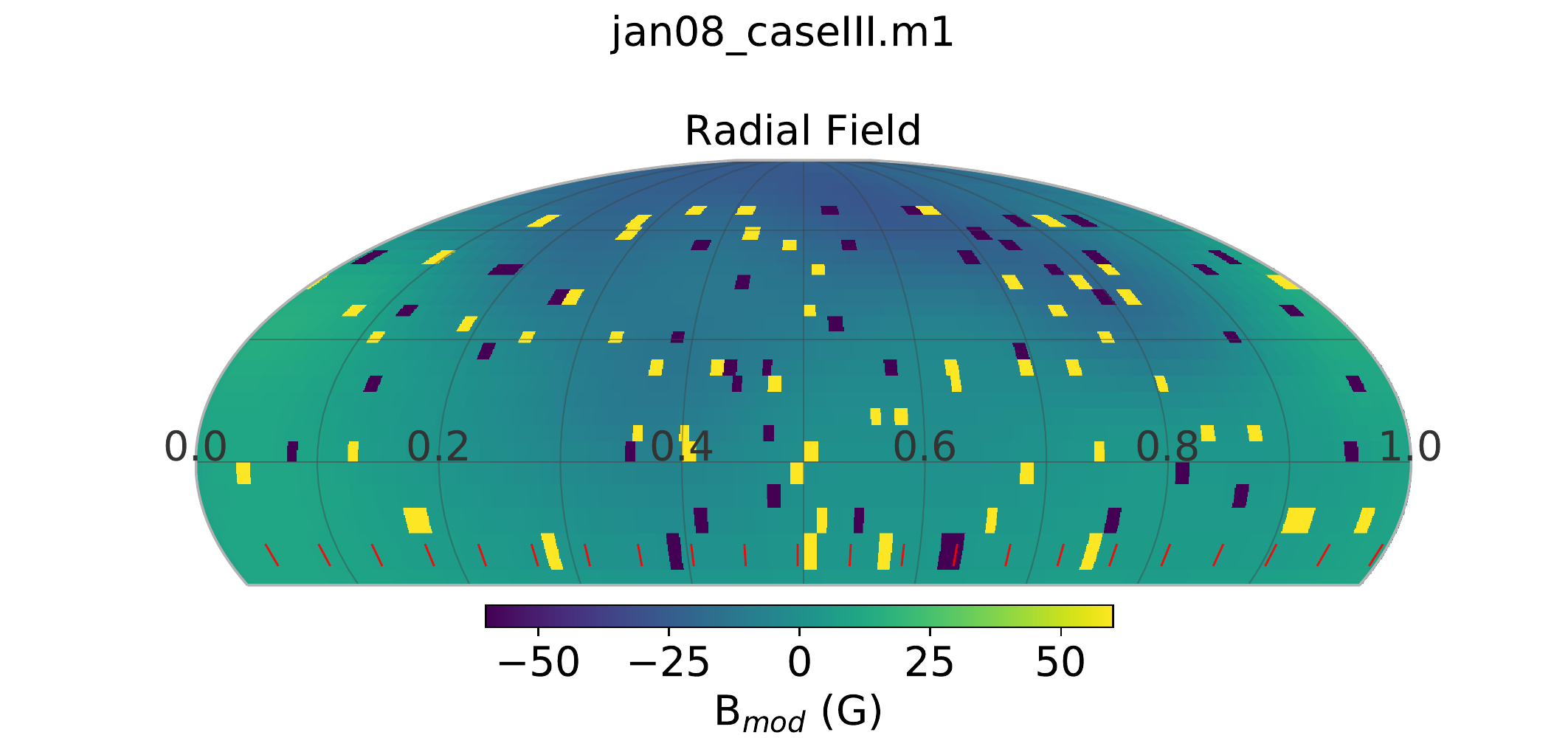}
        \label{fig:case_ii_jan08}
        \vspace{0.5cm}
    \end{subfigure}
    \begin{subfigure}{0.49\textwidth}
        \vspace{0.5cm}
        \caption{2010.04}
        \centering
        \includegraphics[width=.95\linewidth, trim={1.8cm 0cm 1.7cm 1.5cm},clip]{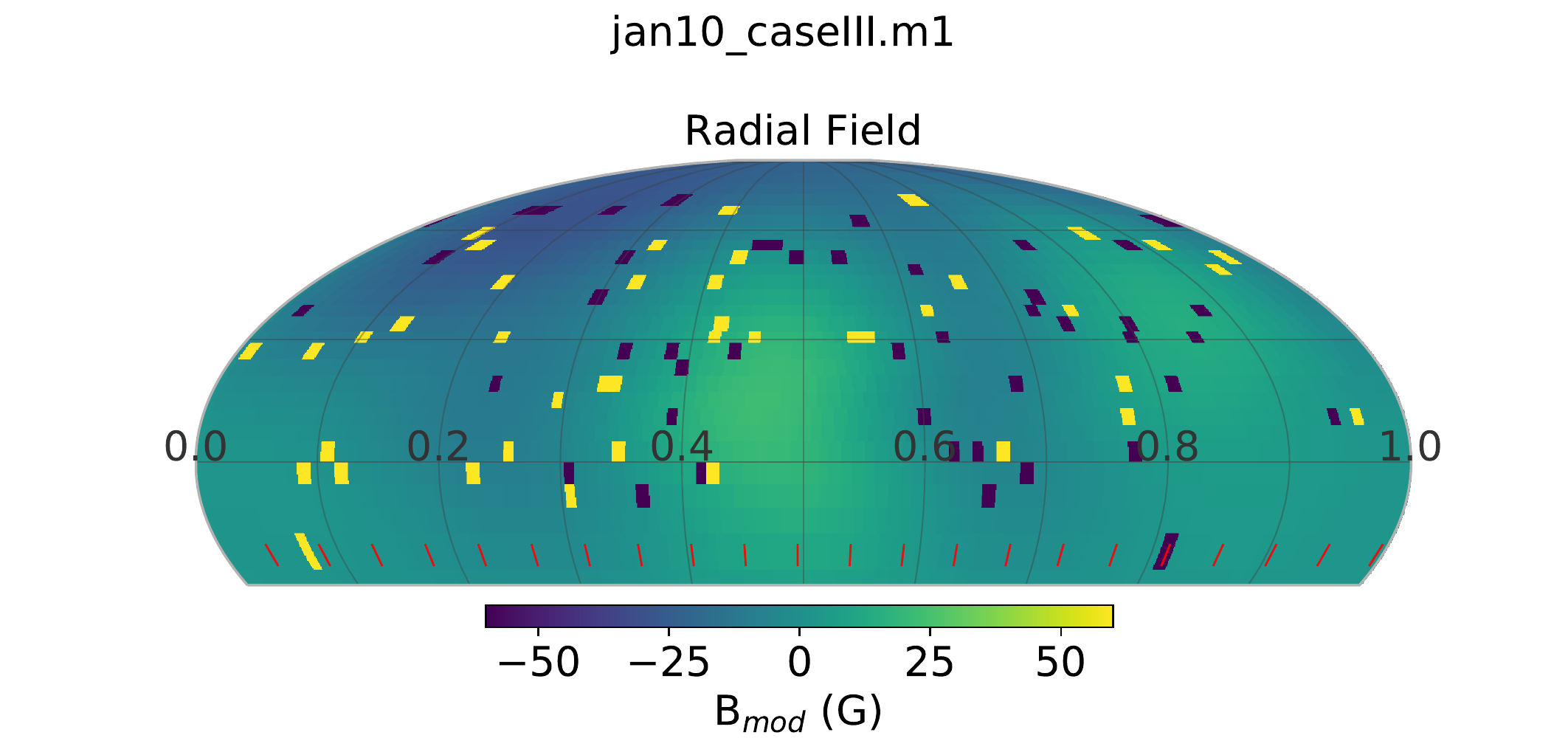}
        \label{fig:case_ii_jan10}
        \vspace{0.5cm}
    \end{subfigure}
    \begin{subfigure}{0.49\textwidth}
        \vspace{0.5cm}
        \caption{2011.81}
        \centering
        \includegraphics[width=.95\linewidth, trim={1.8cm 0cm 1.7cm 1.5cm},clip]{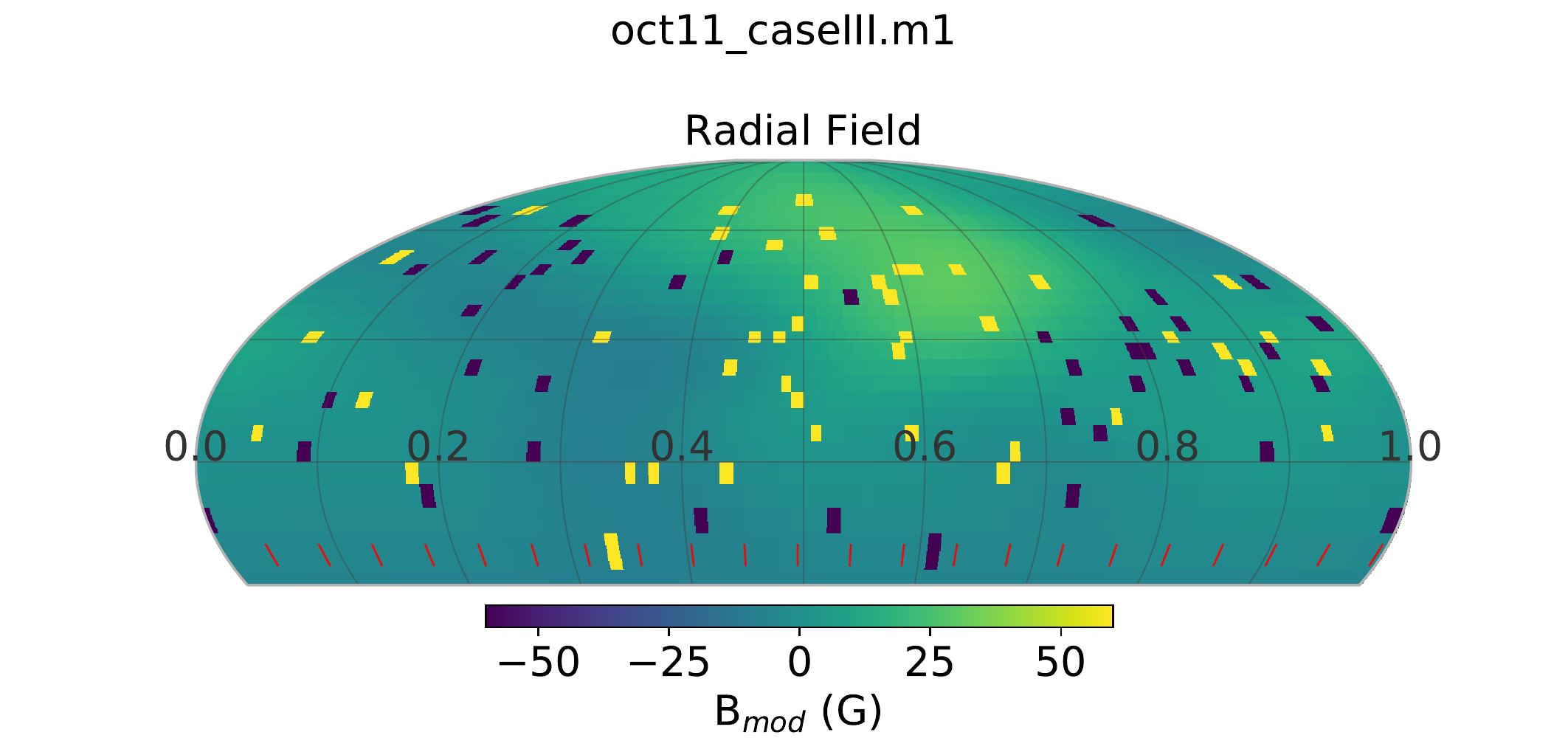}
        \label{fig:case_ii_oct11}
        \vspace{0.5cm}
    \end{subfigure}
    \begin{subfigure}{0.49\textwidth}
        \vspace{0.5cm}
        \caption{2012.82}
        \centering
        \includegraphics[width=.95\linewidth, trim={1.8cm 0cm 1.7cm 1.5cm},clip]{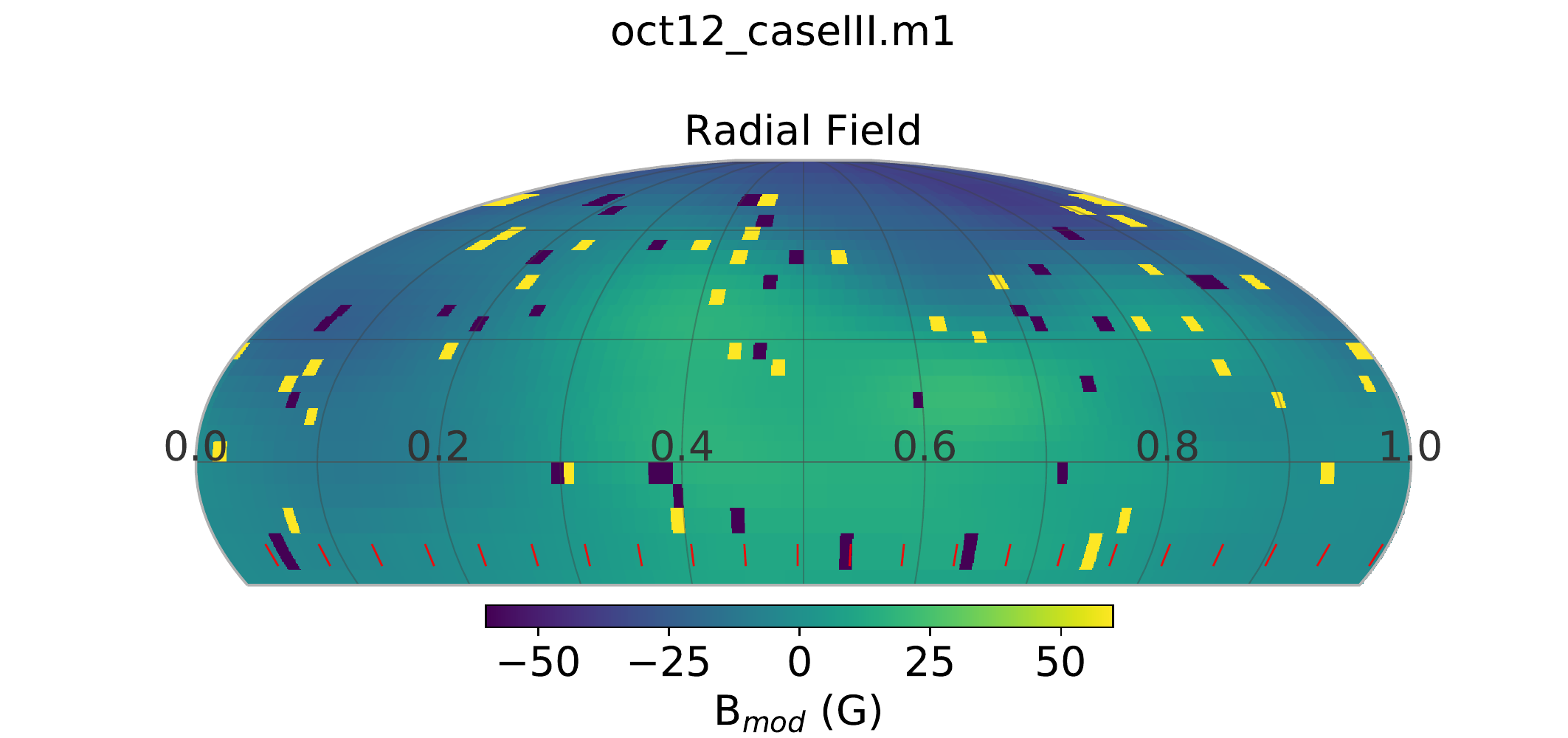}
        \label{fig:case_ii_oct12}
        \vspace{0.5cm}
    \end{subfigure}
    \begin{subfigure}{0.49\textwidth}
        \vspace{0.5cm}
        \caption{2013.75}
        \centering
        \includegraphics[width=.95\linewidth, trim={1.8cm 0cm 1.7cm 1.5cm},clip]{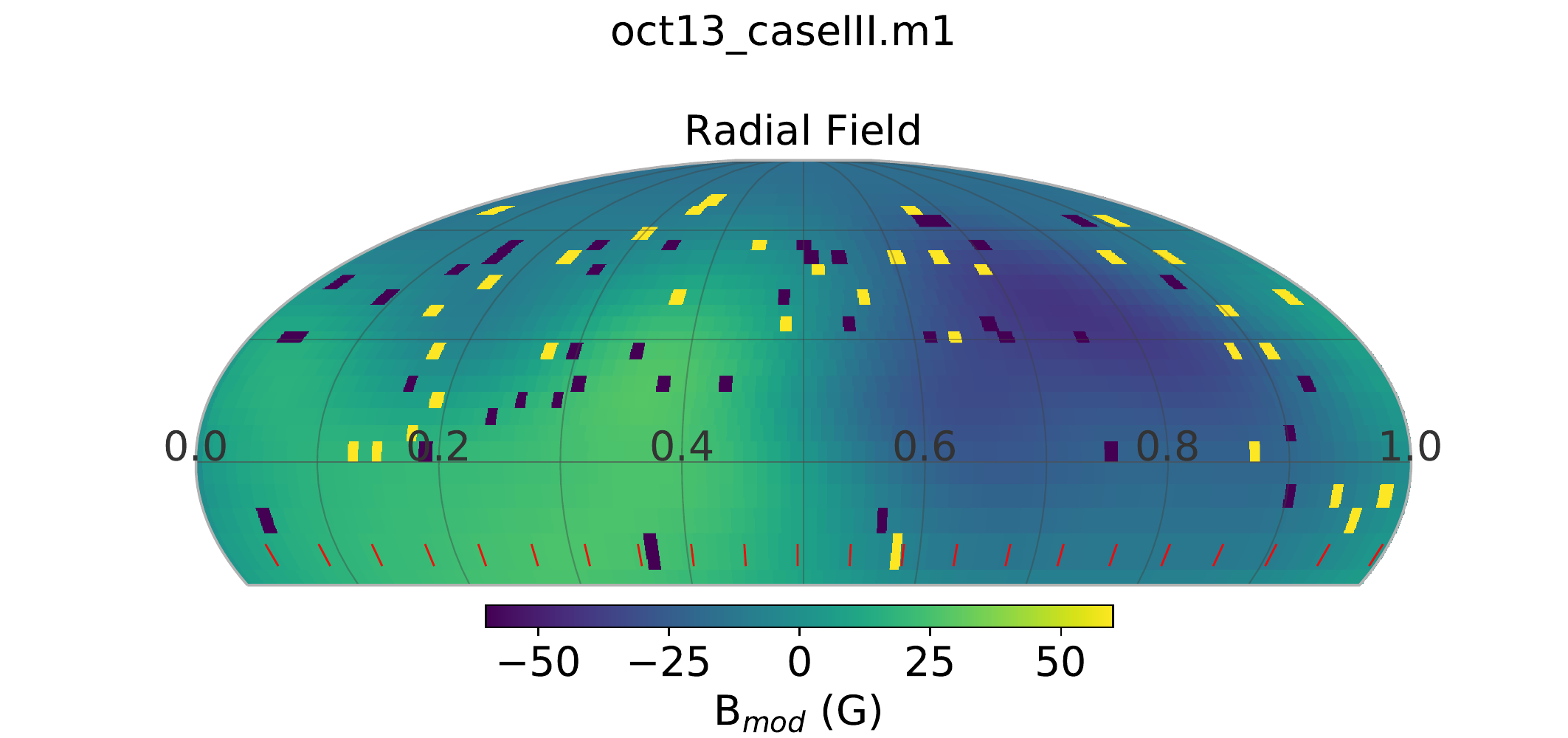}
        \label{fig:case_ii_oct13}
        \vspace{0.5cm}
    \end{subfigure}
    \begin{subfigure}{0.49\textwidth}
        \vspace{0.5cm}
        \caption{2014.84}
        \centering
        \includegraphics[width=.95\linewidth, trim={1.8cm 0cm 1.7cm 1.5cm},clip]{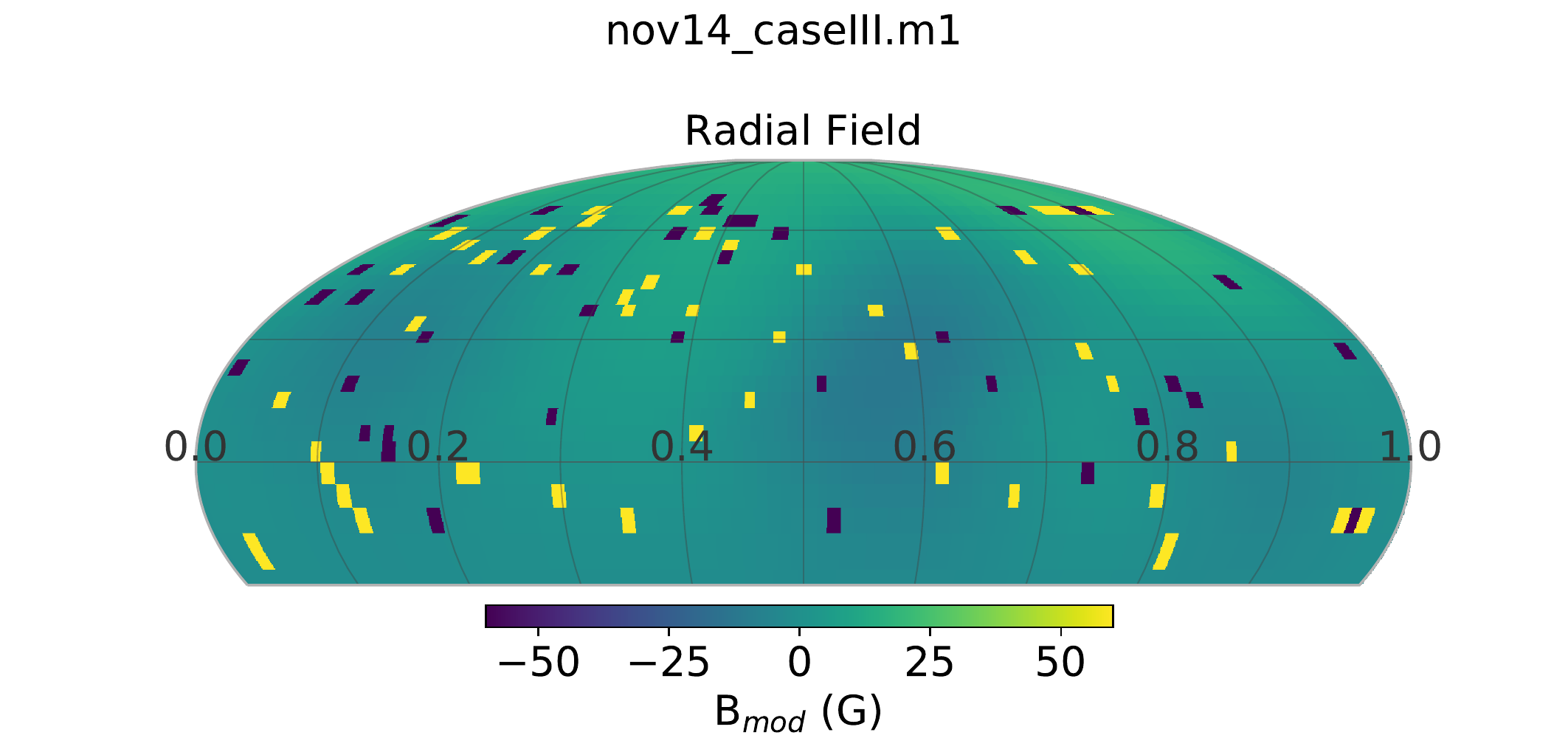}
        \label{fig:case_ii_nov14}
        \vspace{0.5cm}
    \end{subfigure}
    \begin{subfigure}{0.49\textwidth}
        \vspace{0.5cm}
        \caption{2015.01}
        \centering
        \includegraphics[width=.95\linewidth, trim={1.8cm 0cm 1.7cm 1.5cm},clip]{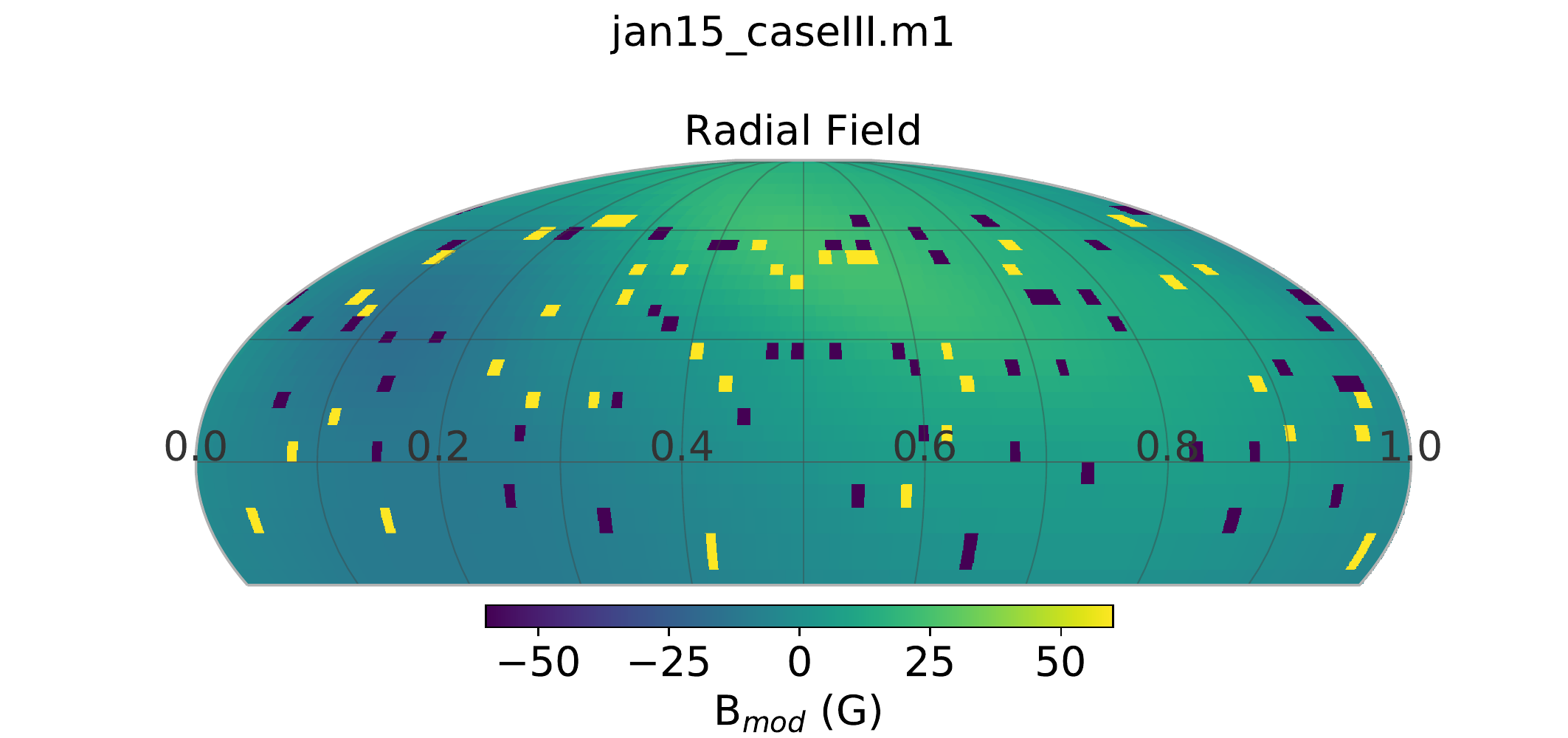}
        \label{fig:case_ii_jan15}
        \vspace{0.5cm}
    \end{subfigure}
    \caption{
        Magnetic field maps of $\epsilon$ Eridani for eight epochs of observations (see sub-captions) and simulated magnetic spots, according to {\it case ii}.
        Bright areas indicate positive polarity and dark -- negative polarity.
    }
    \label{fig:case_ii_maps}
\end{figure*}

\begin{figure*}
    \centering
    
    \begin{subfigure}{0.49\textwidth}
        \vspace{0.5cm}
        \caption{2007.08}
        \centering
        \includegraphics[width=.95\linewidth, trim={1.8cm 0cm 1.7cm 1.5cm},clip]{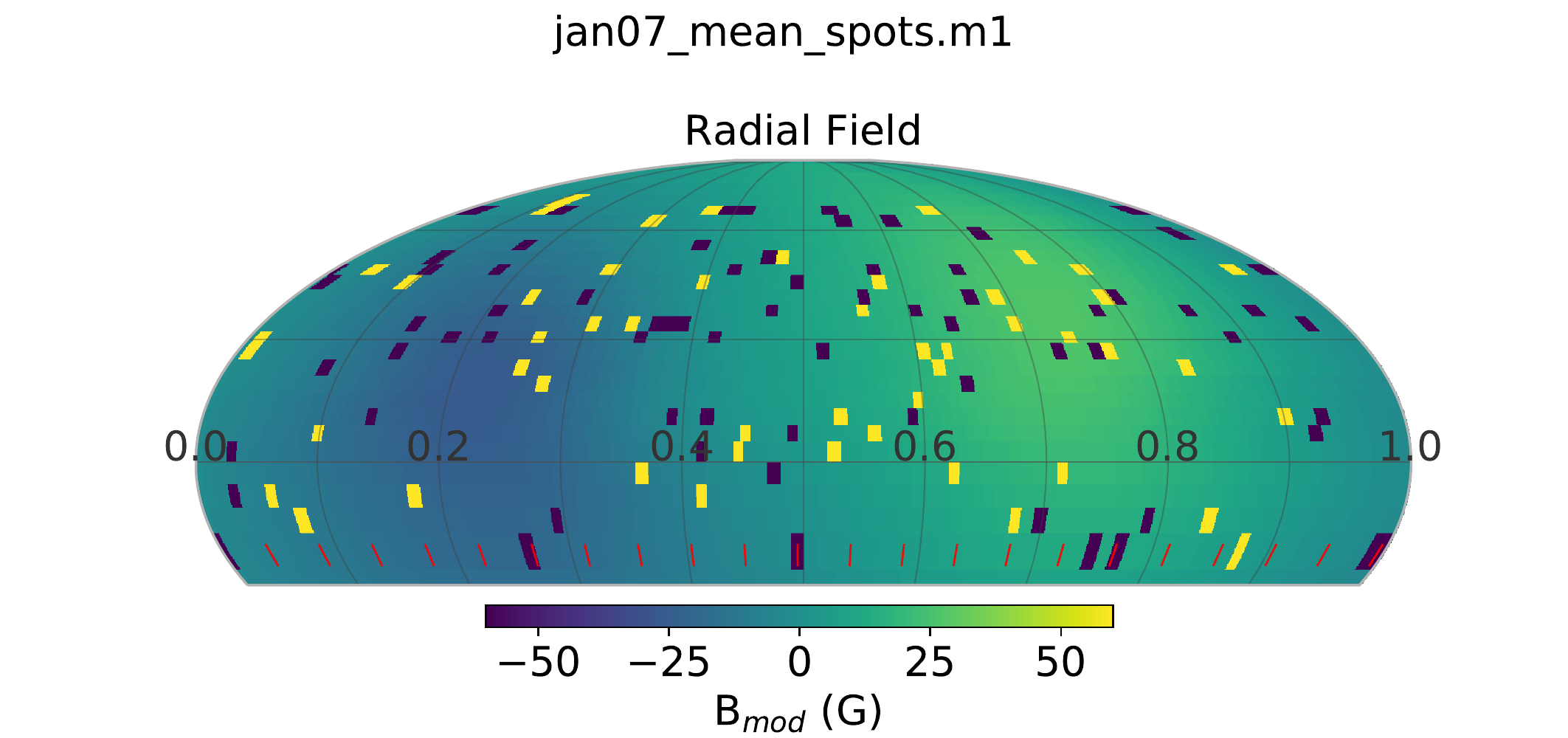}
        \label{fig:case_iii_jan07}
        \vspace{0.5cm}
    \end{subfigure}
    \begin{subfigure}{0.49\textwidth}
        \vspace{0.5cm}
        \caption{2008.09}
        \centering
        \includegraphics[width=.95\linewidth, trim={1.8cm 0cm 1.7cm 1.5cm},clip]{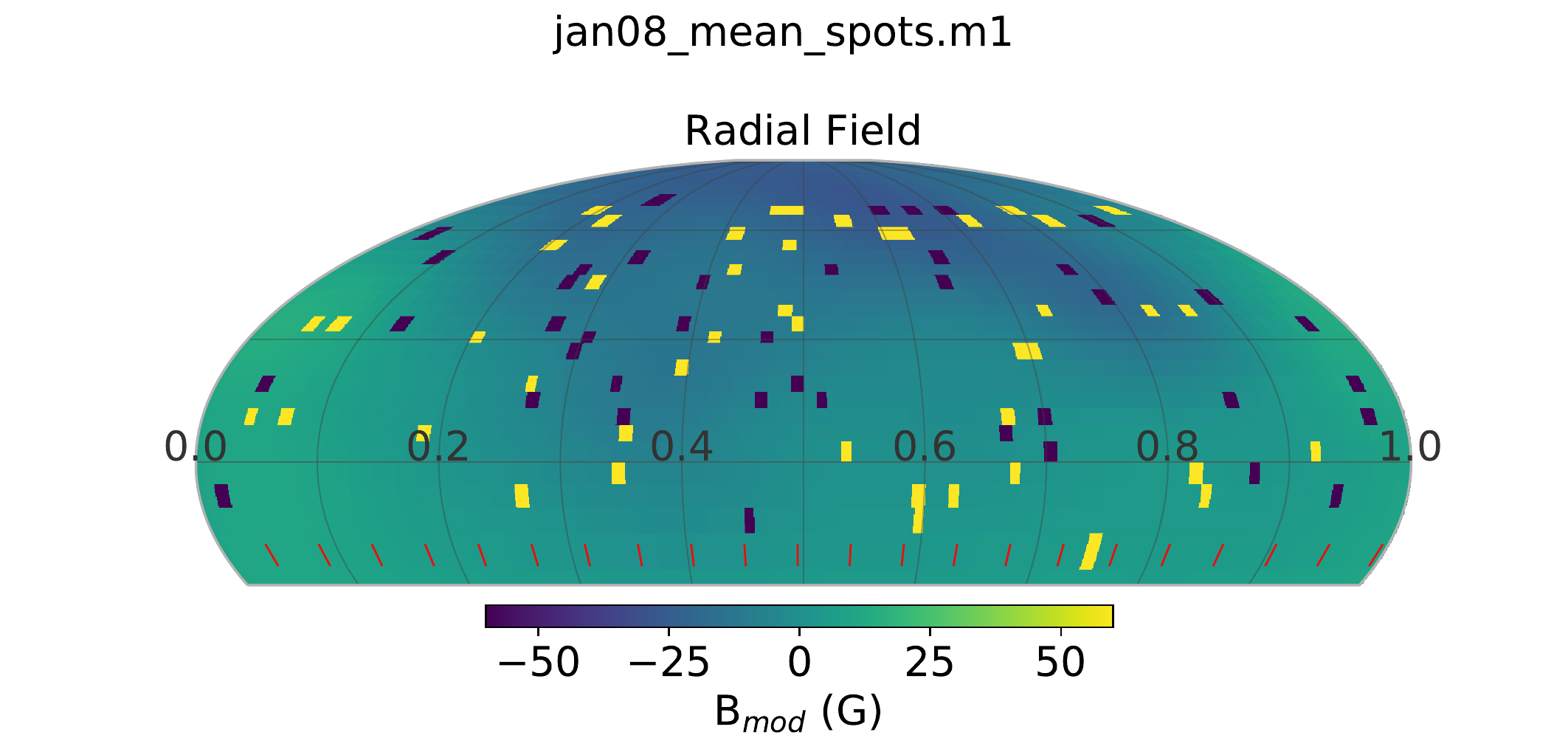}
        \label{fig:case_iii_jan08}
        \vspace{0.5cm}
    \end{subfigure}
    \begin{subfigure}{0.49\textwidth}
        \vspace{0.5cm}
        \caption{2010.04}
        \centering
        \includegraphics[width=.95\linewidth, trim={1.8cm 0cm 1.7cm 1.5cm},clip]{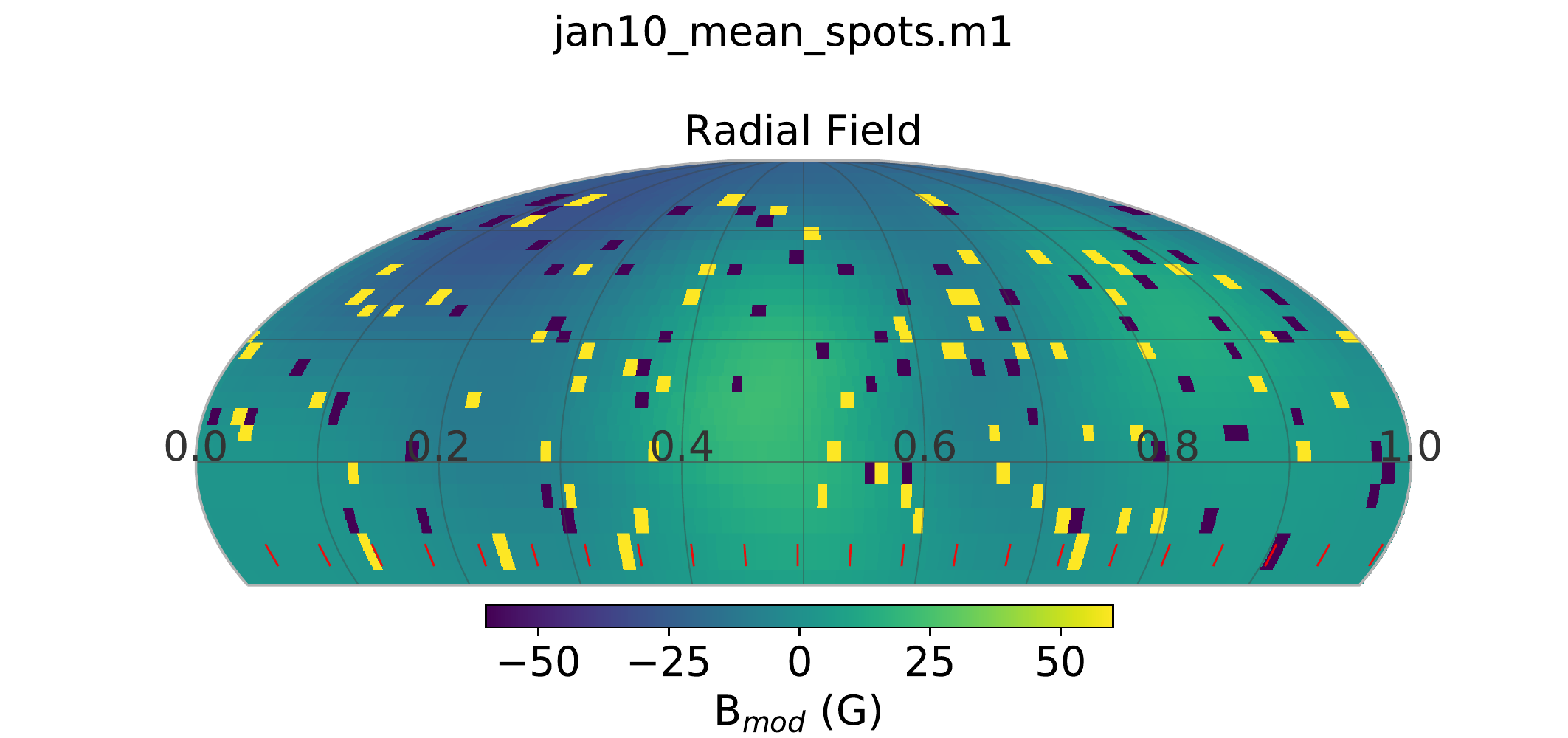}
        \label{fig:case_iii_jan10}
        \vspace{0.5cm}
    \end{subfigure}
    \begin{subfigure}{0.49\textwidth}
        \vspace{0.5cm}
        \caption{2011.81}
        \centering
        \includegraphics[width=.95\linewidth, trim={1.8cm 0cm 1.7cm 1.5cm},clip]{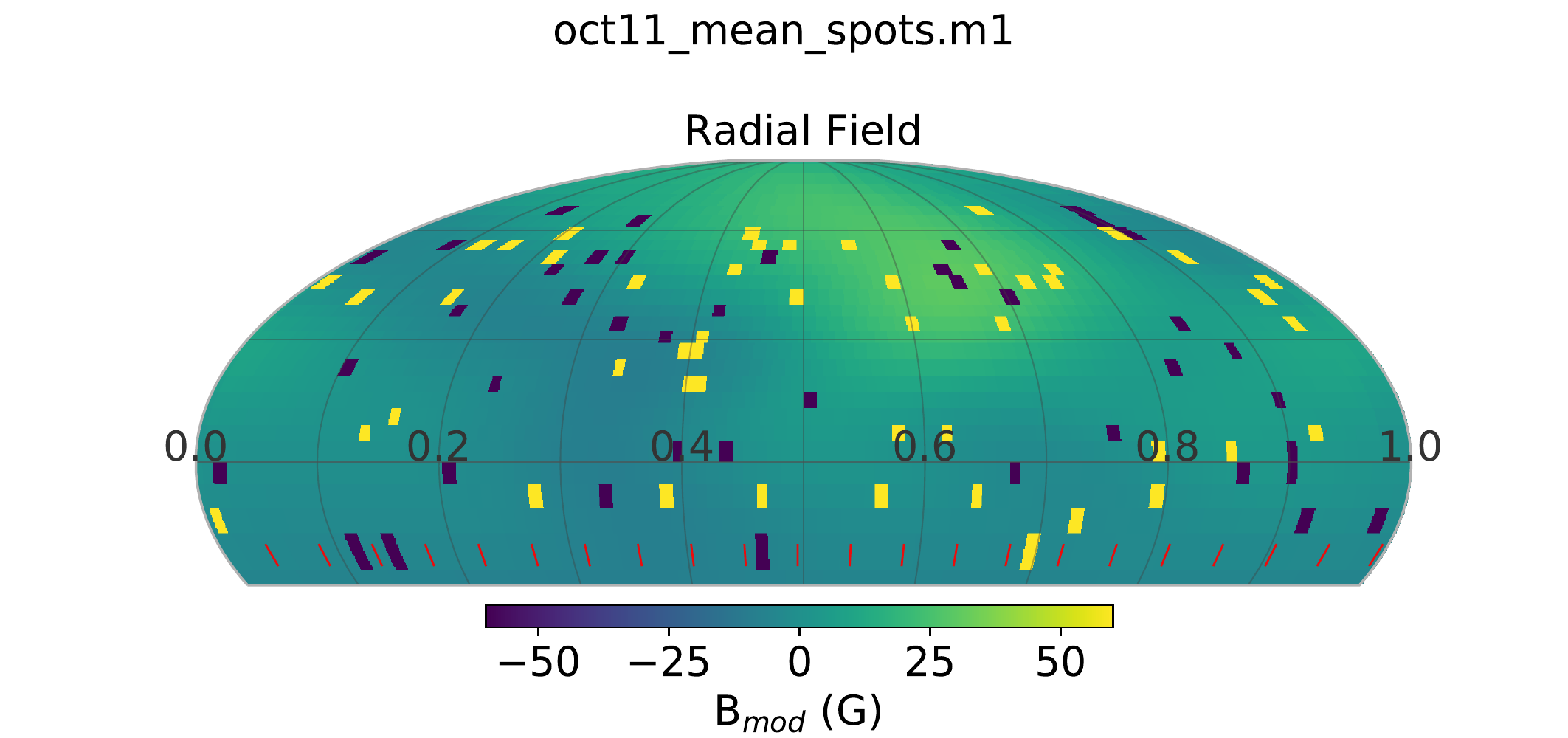}
        \label{fig:case_iii_oct11}
        \vspace{0.5cm}
    \end{subfigure}
    \begin{subfigure}{0.49\textwidth}
        \vspace{0.5cm}
        \caption{2012.82}
        \centering
        \includegraphics[width=.95\linewidth, trim={1.8cm 0cm 1.7cm 1.5cm},clip]{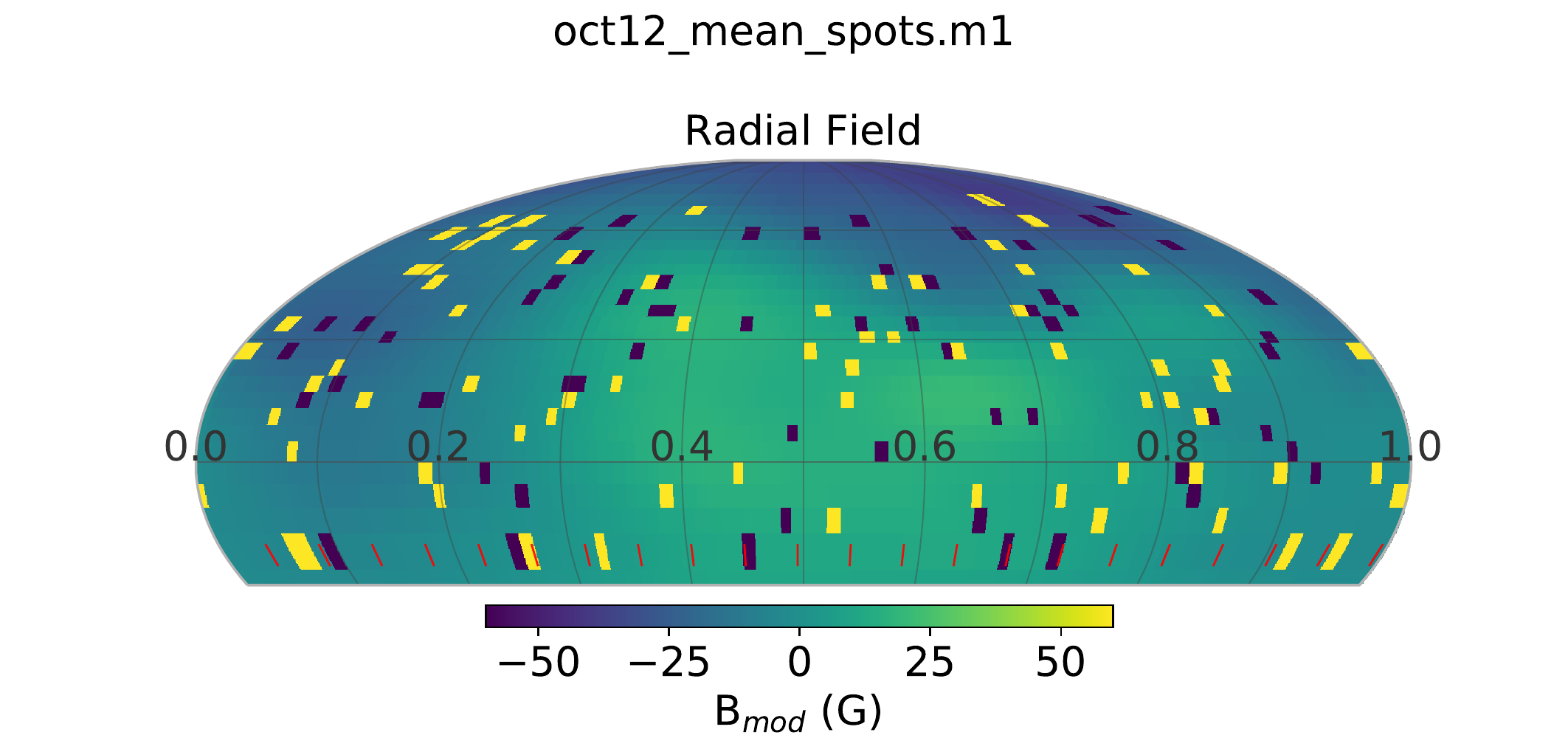}
        \label{fig:case_iii_oct12}
        \vspace{0.5cm}
    \end{subfigure}
    \begin{subfigure}{0.49\textwidth}
        \vspace{0.5cm}
        \caption{2013.75}
        \centering
        \includegraphics[width=.95\linewidth, trim={1.8cm 0cm 1.7cm 1.5cm},clip]{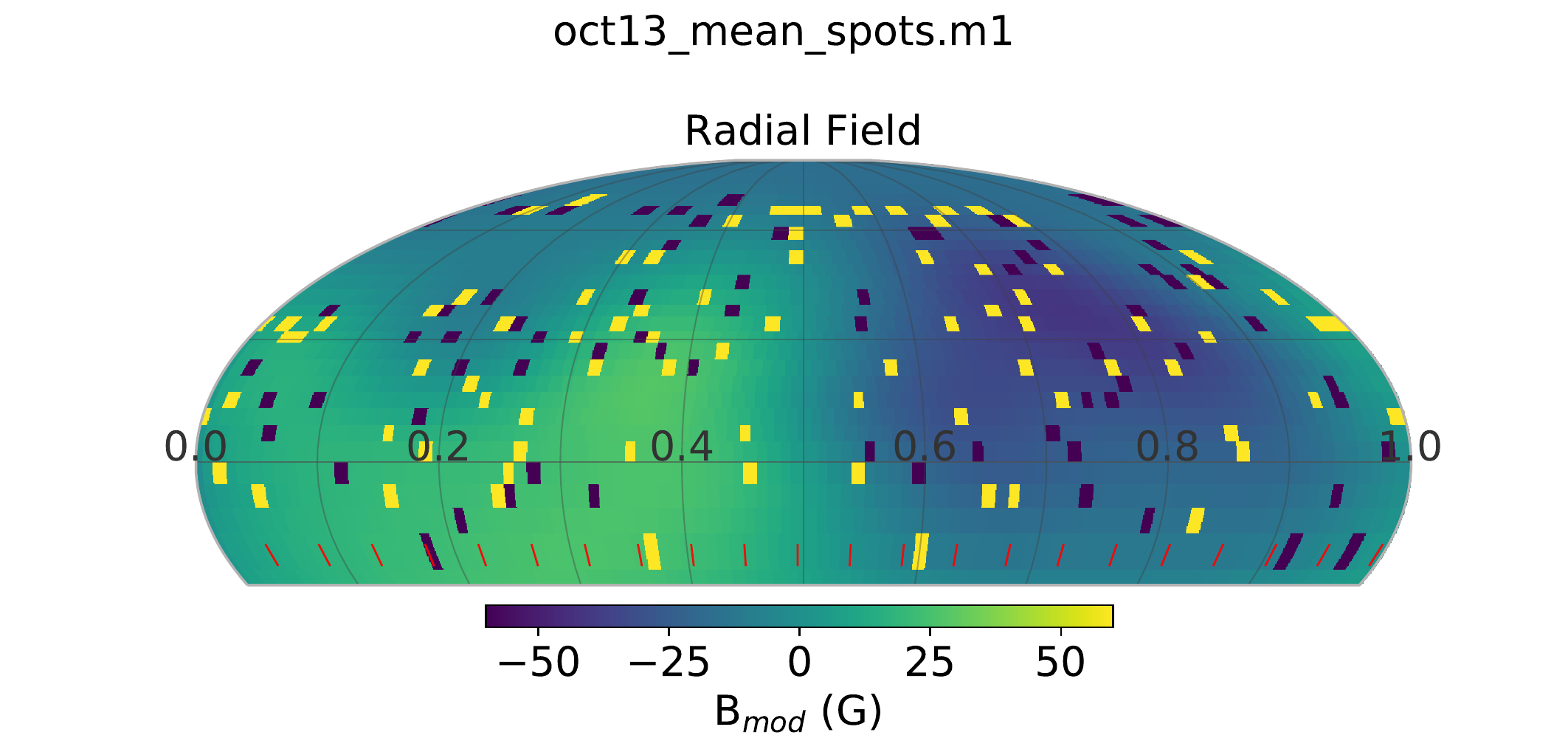}
        \label{fig:case_iii_oct13}
        \vspace{0.5cm}
    \end{subfigure}
    \begin{subfigure}{0.49\textwidth}
        \vspace{0.5cm}
        \caption{2014.84}
        \centering
        \includegraphics[width=.95\linewidth, trim={1.8cm 0cm 1.7cm 1.5cm},clip]{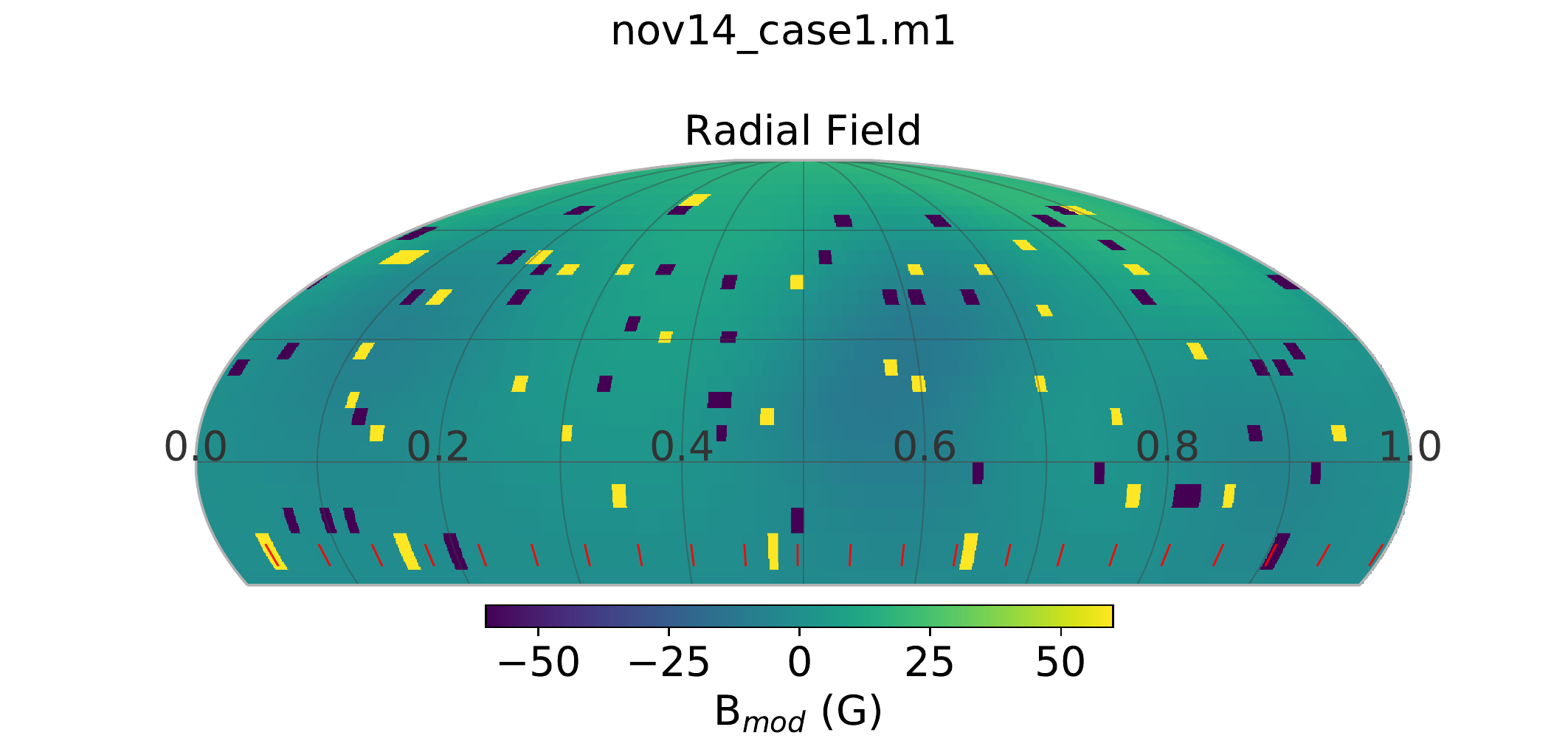}
        \label{fig:case_iii_nov14}
        \vspace{0.5cm}
    \end{subfigure}
    \begin{subfigure}{0.49\textwidth}
        \vspace{0.5cm}
        \caption{2015.01}
        \centering
        \includegraphics[width=.95\linewidth, trim={1.8cm 0cm 1.7cm 1.5cm},clip]{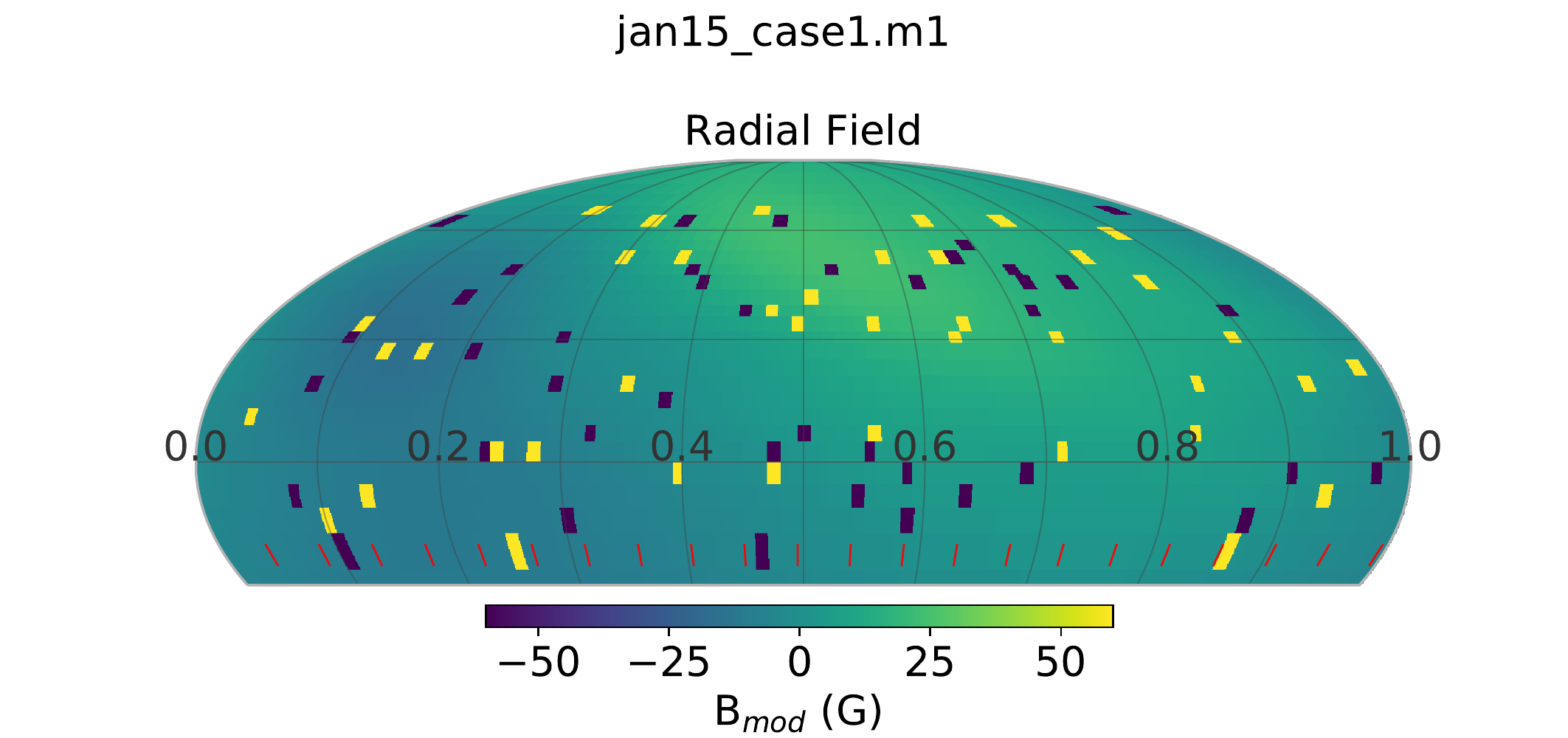}
        \label{fig:case_iii_jan15}
        \vspace{0.5cm}
    \end{subfigure}
    \caption{
        Magnetic field maps of $\epsilon$ Eridani for eight epochs of observations (see sub-captions) and simulated magnetic spots, according to {\it case iii}.
        Bright areas indicate positive polarity and dark -- negative polarity.
    }
    \label{fig:case_iii_maps}
\end{figure*}

\begin{figure*}
    \centering
    
    \begin{subfigure}{0.49\textwidth}
        \vspace{0.5cm}
        \caption{2007.08}
        \centering
        \includegraphics[width=.95\linewidth, trim={1.8cm 0cm 1.7cm 1.5cm},clip]{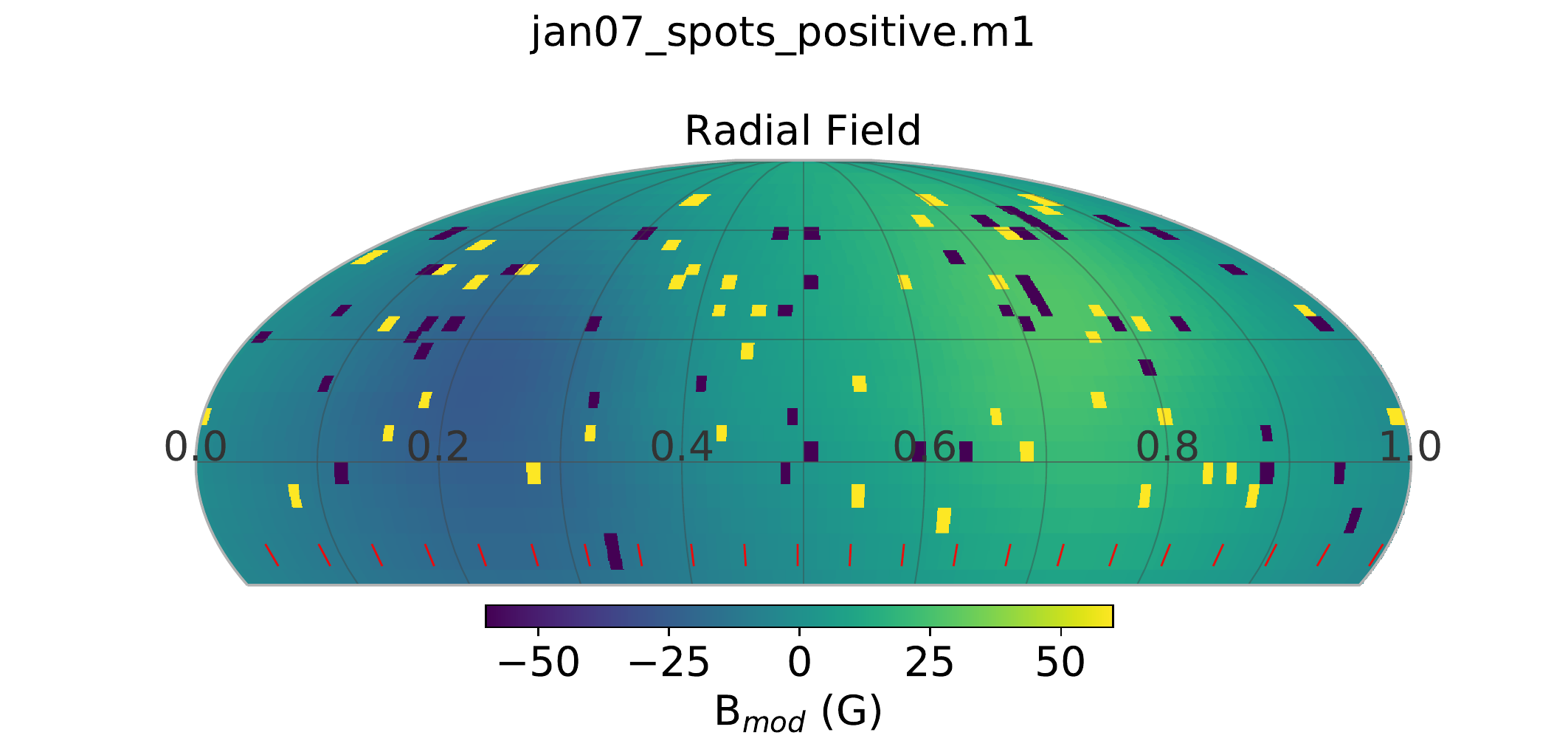}
        \label{fig:case_iv_jan07}
        \vspace{0.5cm}
    \end{subfigure}
    \begin{subfigure}{0.49\textwidth}
        \vspace{0.5cm}
        \caption{2008.09}
        \centering
        \includegraphics[width=.95\linewidth, trim={1.8cm 0cm 1.7cm 1.5cm},clip]{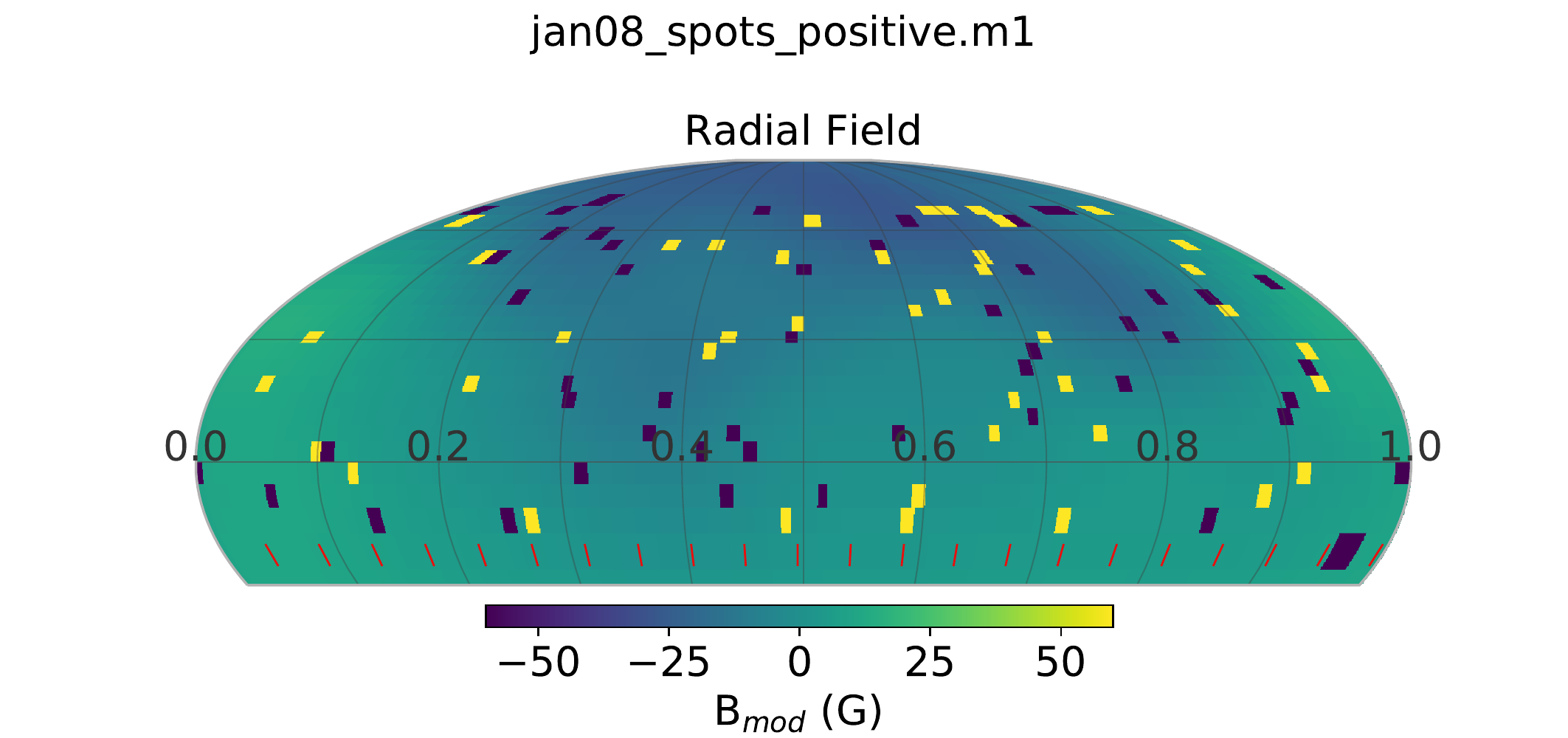}
        \label{fig:case_iv_jan08}
        \vspace{0.5cm}
    \end{subfigure}
    \begin{subfigure}{0.49\textwidth}
        \vspace{0.5cm}
        \caption{2010.04}
        \centering
        \includegraphics[width=.95\linewidth, trim={1.8cm 0cm 1.7cm 1.5cm},clip]{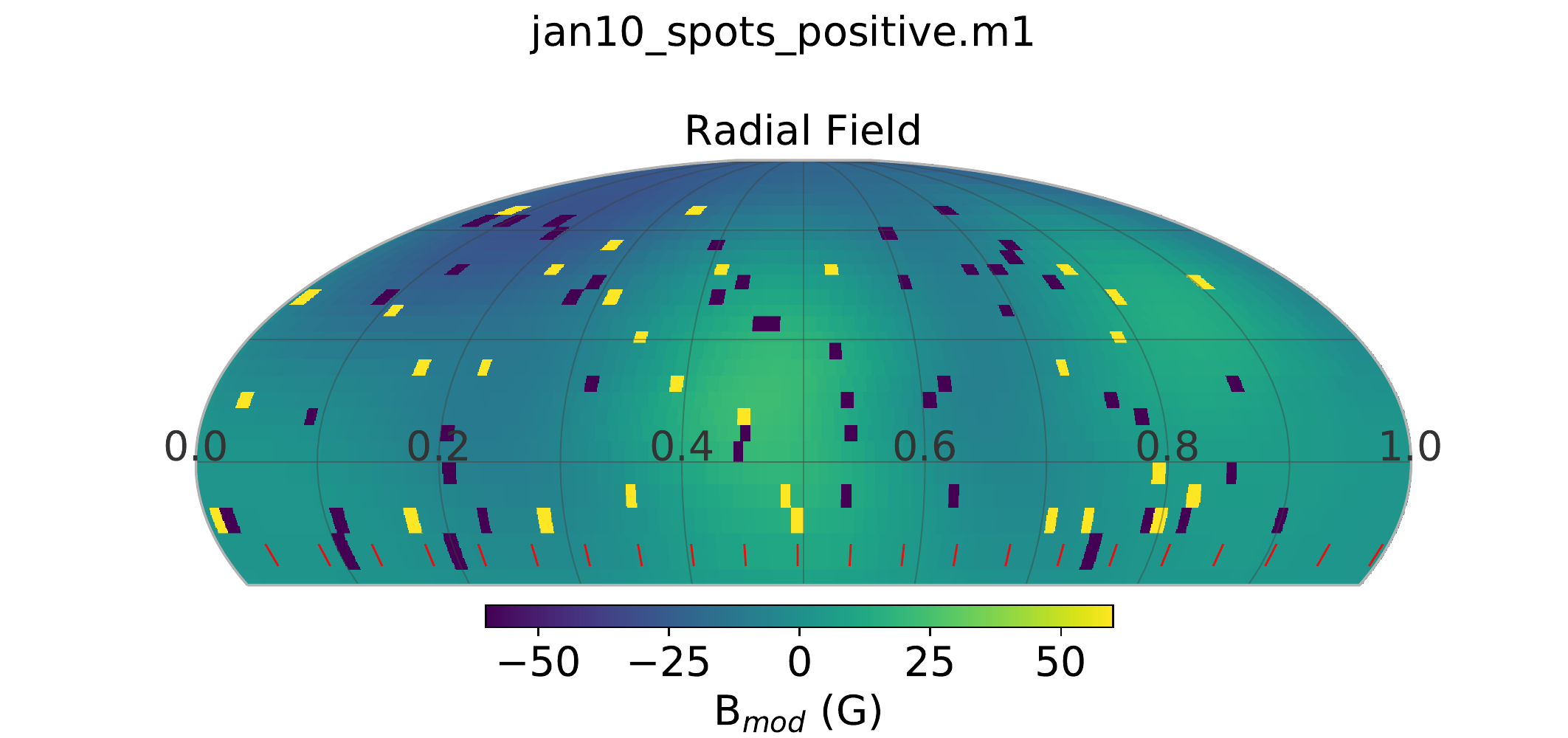}
        \label{fig:case_iv_jan10}
        \vspace{0.5cm}
    \end{subfigure}
    \begin{subfigure}{0.49\textwidth}
        \vspace{0.5cm}
        \caption{2011.81}
        \centering
        \includegraphics[width=.95\linewidth, trim={1.8cm 0cm 1.7cm 1.5cm},clip]{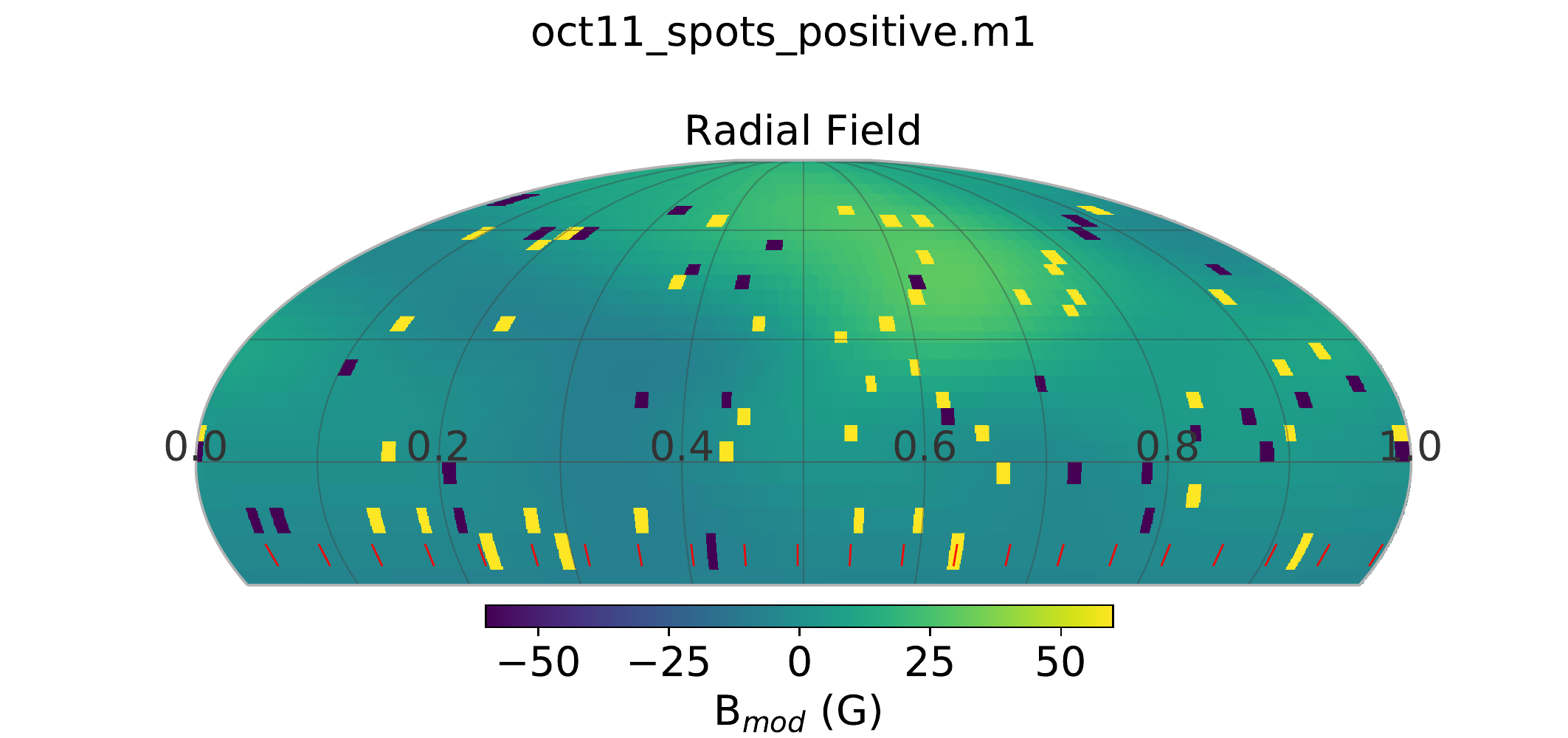}
        \label{fig:case_iv_oct11}
        \vspace{0.5cm}
    \end{subfigure}
    \begin{subfigure}{0.49\textwidth}
        \vspace{0.5cm}
        \caption{2012.82}
        \centering
        \includegraphics[width=.95\linewidth, trim={1.8cm 0cm 1.7cm 1.5cm},clip]{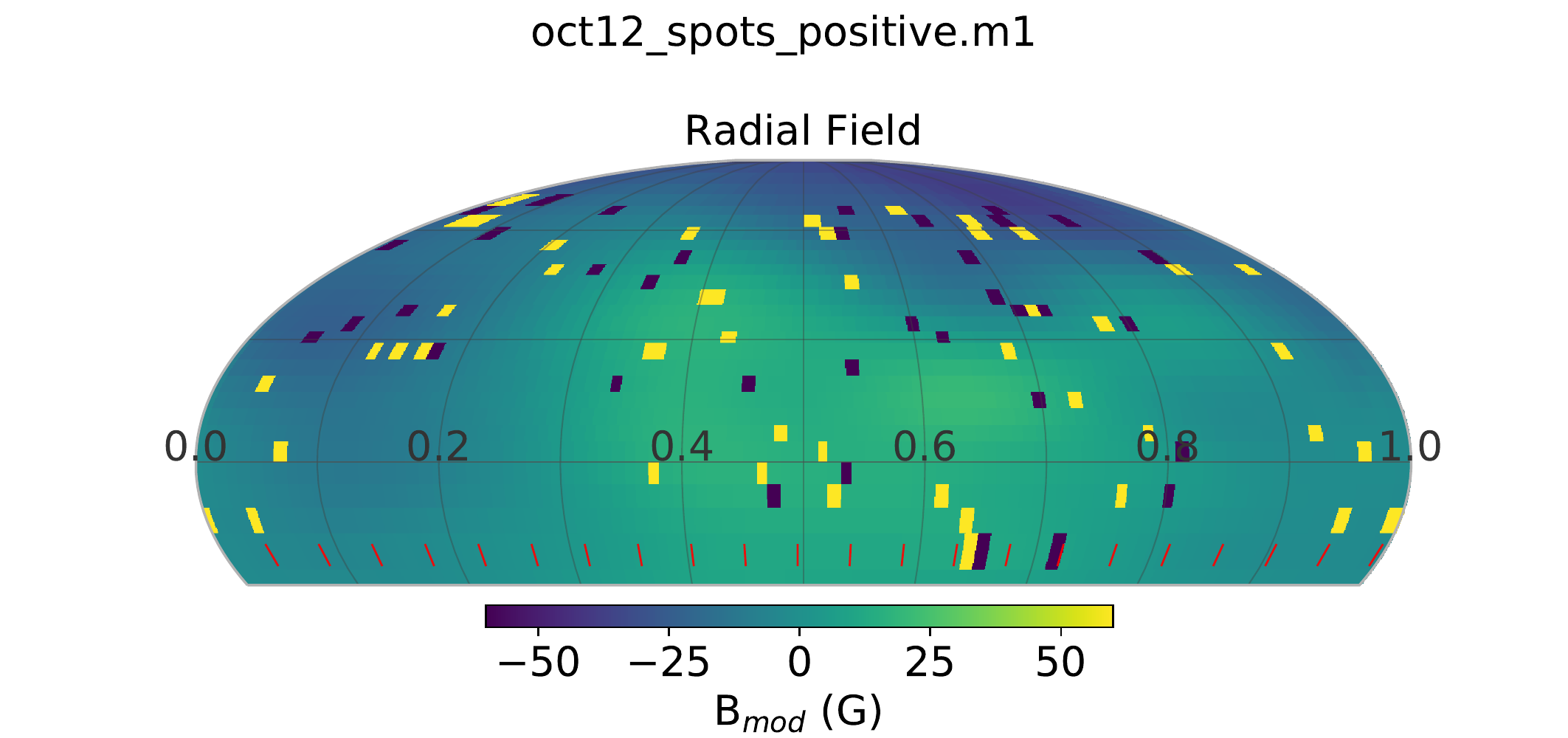}
        \label{fig:case_iv_oct12}
        \vspace{0.5cm}
    \end{subfigure}
    \begin{subfigure}{0.49\textwidth}
        \vspace{0.5cm}
        \caption{2013.75}
        \centering
        \includegraphics[width=.95\linewidth, trim={1.8cm 0cm 1.7cm 1.5cm},clip]{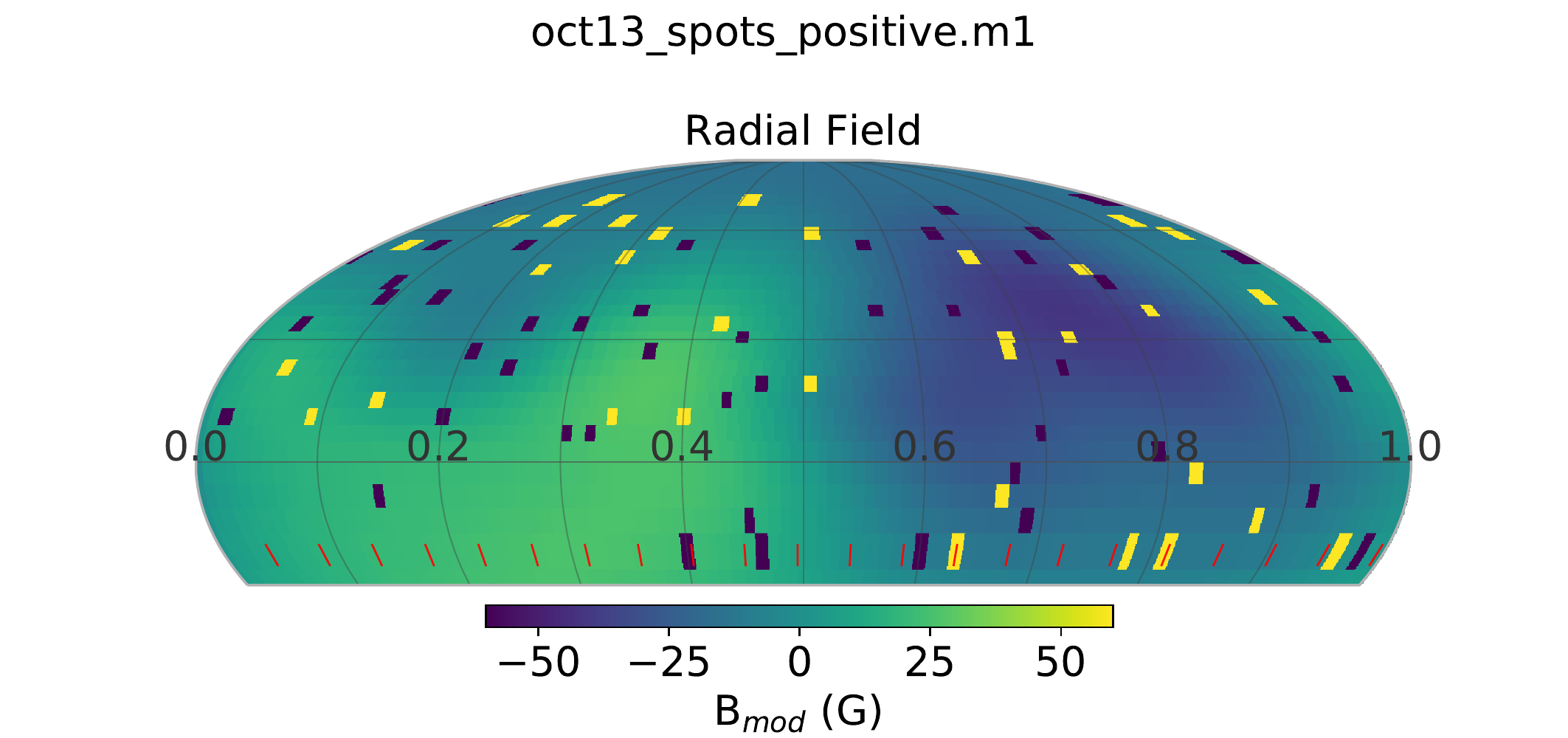}
        \label{fig:case_iv_oct13}
        \vspace{0.5cm}
    \end{subfigure}
    \begin{subfigure}{0.49\textwidth}
        \vspace{0.5cm}
        \caption{2014.84}
        \centering
        \includegraphics[width=.95\linewidth, trim={1.8cm 0cm 1.7cm 1.5cm},clip]{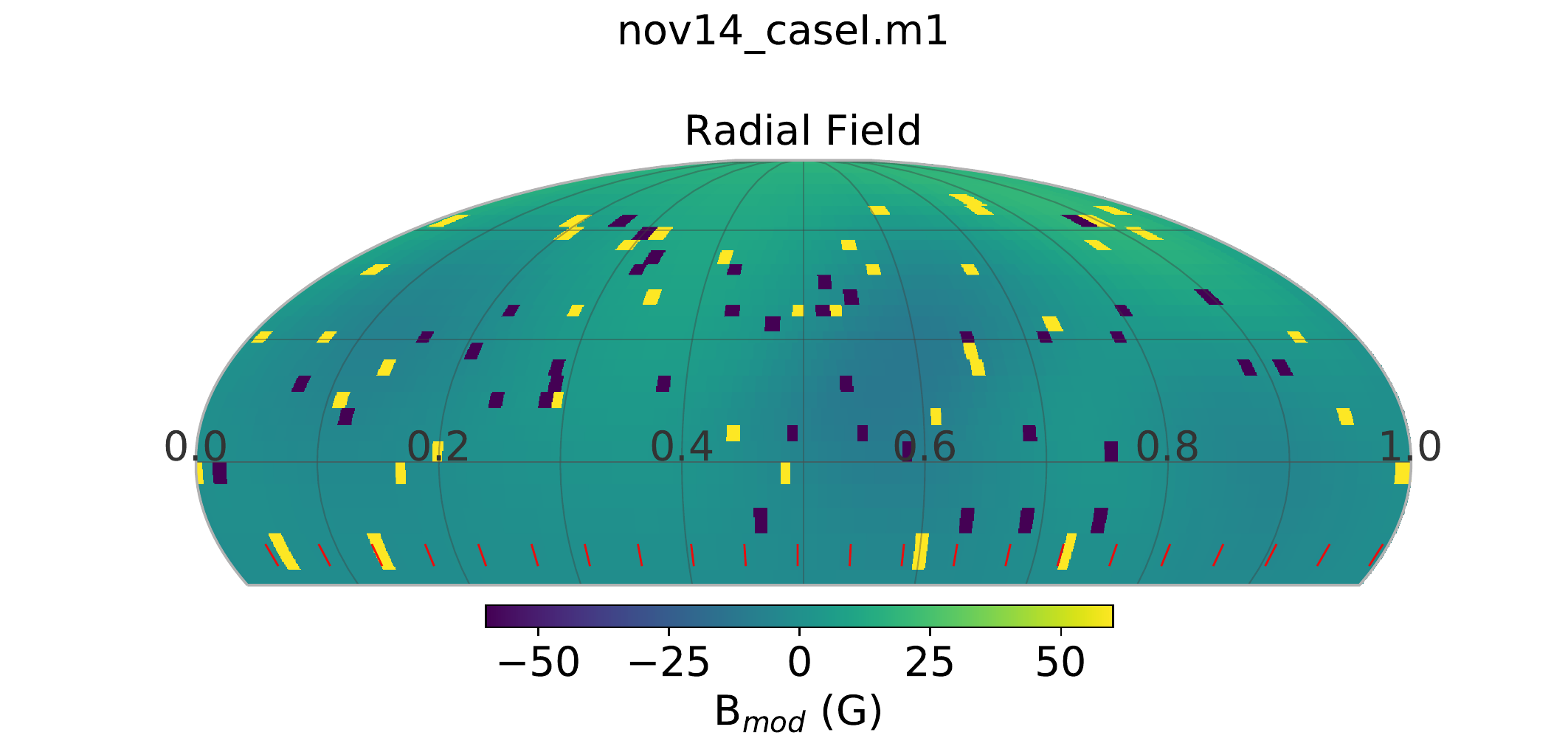}
        \label{fig:case_iv_nov14}
        \vspace{0.5cm}
    \end{subfigure}
    \begin{subfigure}{0.49\textwidth}
        \vspace{0.5cm}
        \caption{2015.01}
        \centering
        \includegraphics[width=.95\linewidth, trim={1.8cm 0cm 1.7cm 1.5cm},clip]{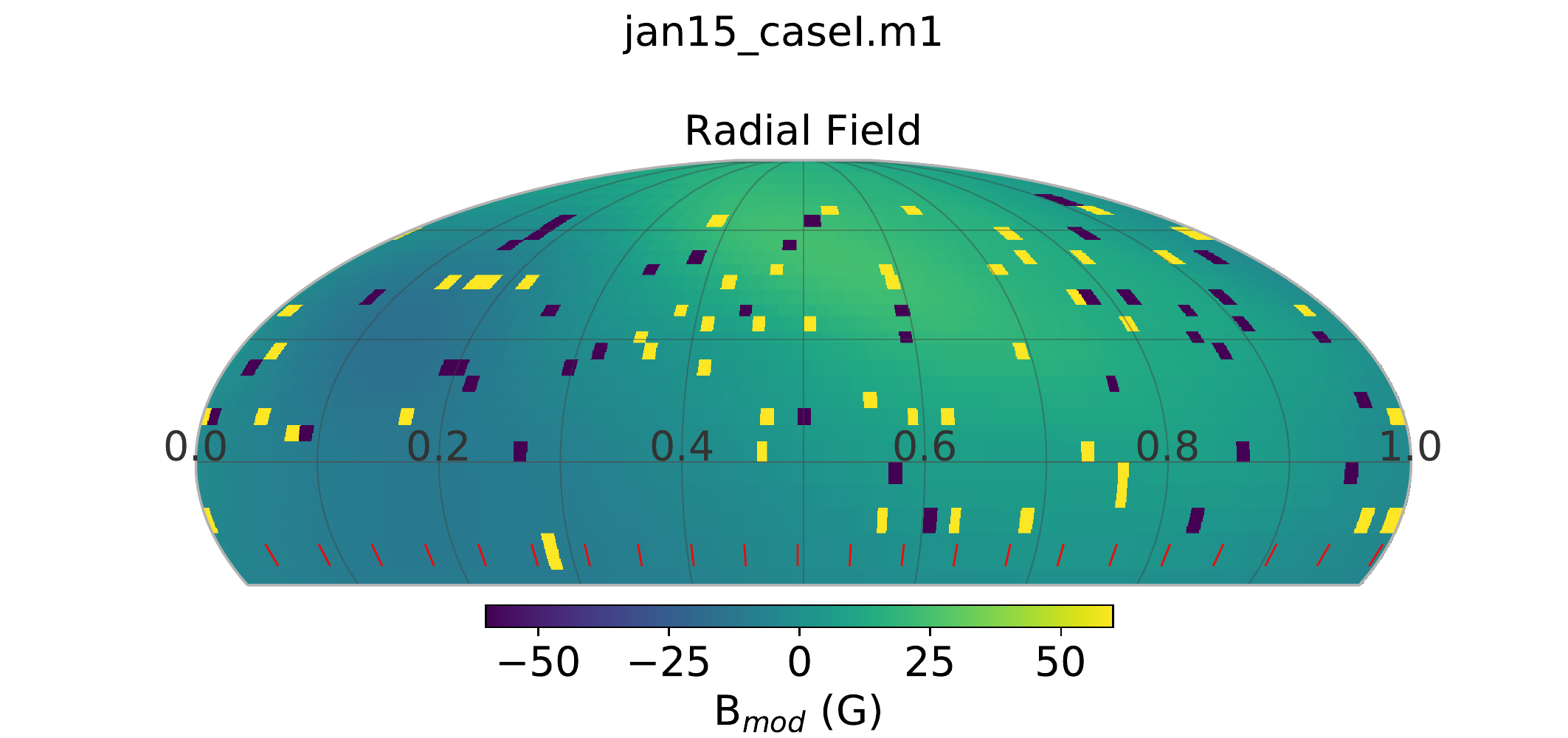}
        \label{fig:case_iv_jan15}
        \vspace{0.5cm}
    \end{subfigure}
    \caption{
        Magnetic field maps of $\epsilon$ Eridani for eight epochs of observations (see sub-captions) and simulated magnetic spots, according to {\it case iv}.
        Bright areas indicate positive polarity and dark -- negative polarity.
    }
    \label{fig:case_iv_maps}
\end{figure*}

\begin{figure*}
    \centering
    
    \begin{subfigure}{0.49\textwidth}
        \vspace{0.5cm}
        \caption{2007.08}
        \centering
        \includegraphics[width=.95\linewidth, trim={1.8cm 0cm 1.7cm 1.5cm},clip]{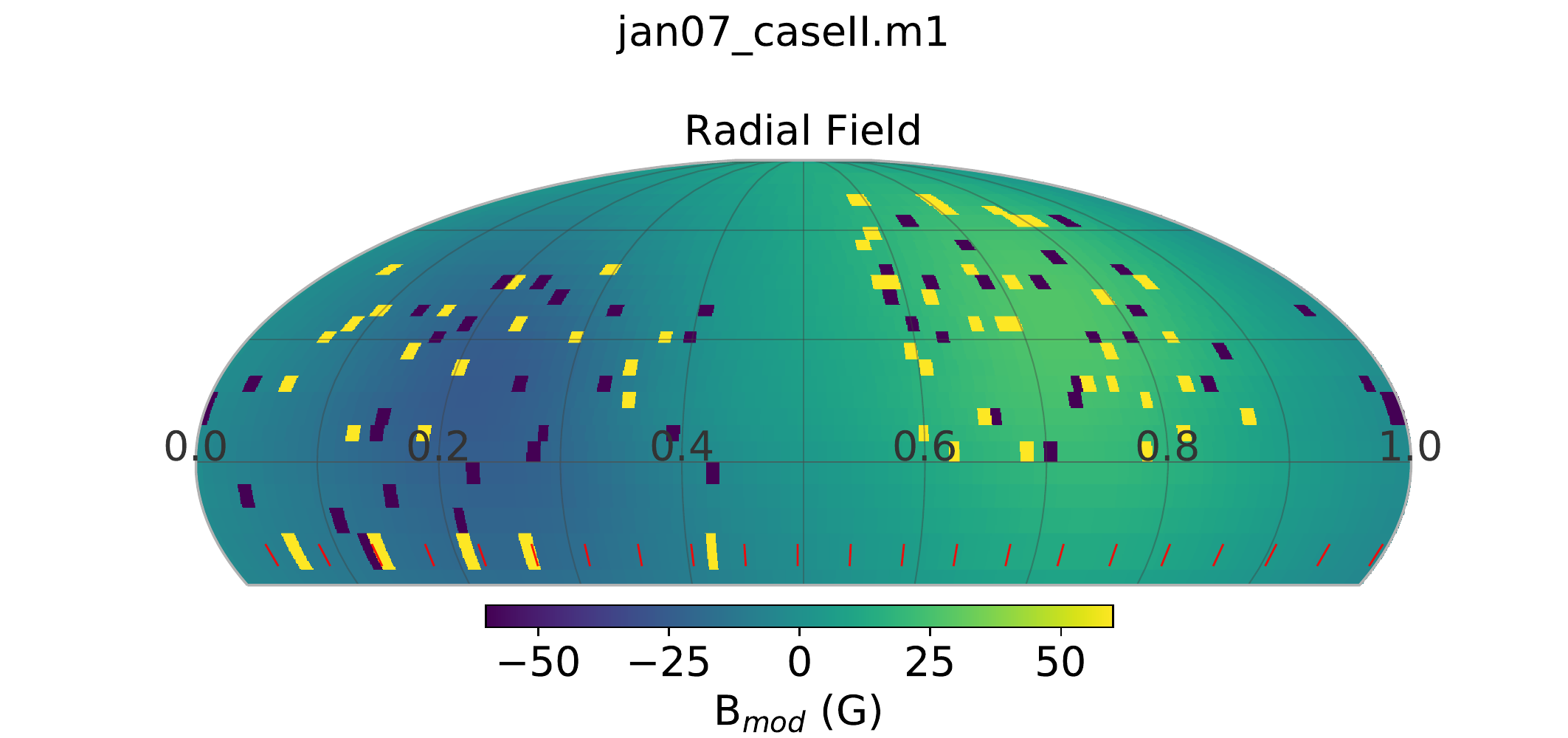}
        \label{fig:case_v_jan07}
        \vspace{0.5cm}
    \end{subfigure}
    \begin{subfigure}{0.49\textwidth}
        \vspace{0.5cm}
        \caption{2008.09}
        \centering
        \includegraphics[width=.95\linewidth, trim={1.8cm 0cm 1.7cm 1.5cm},clip]{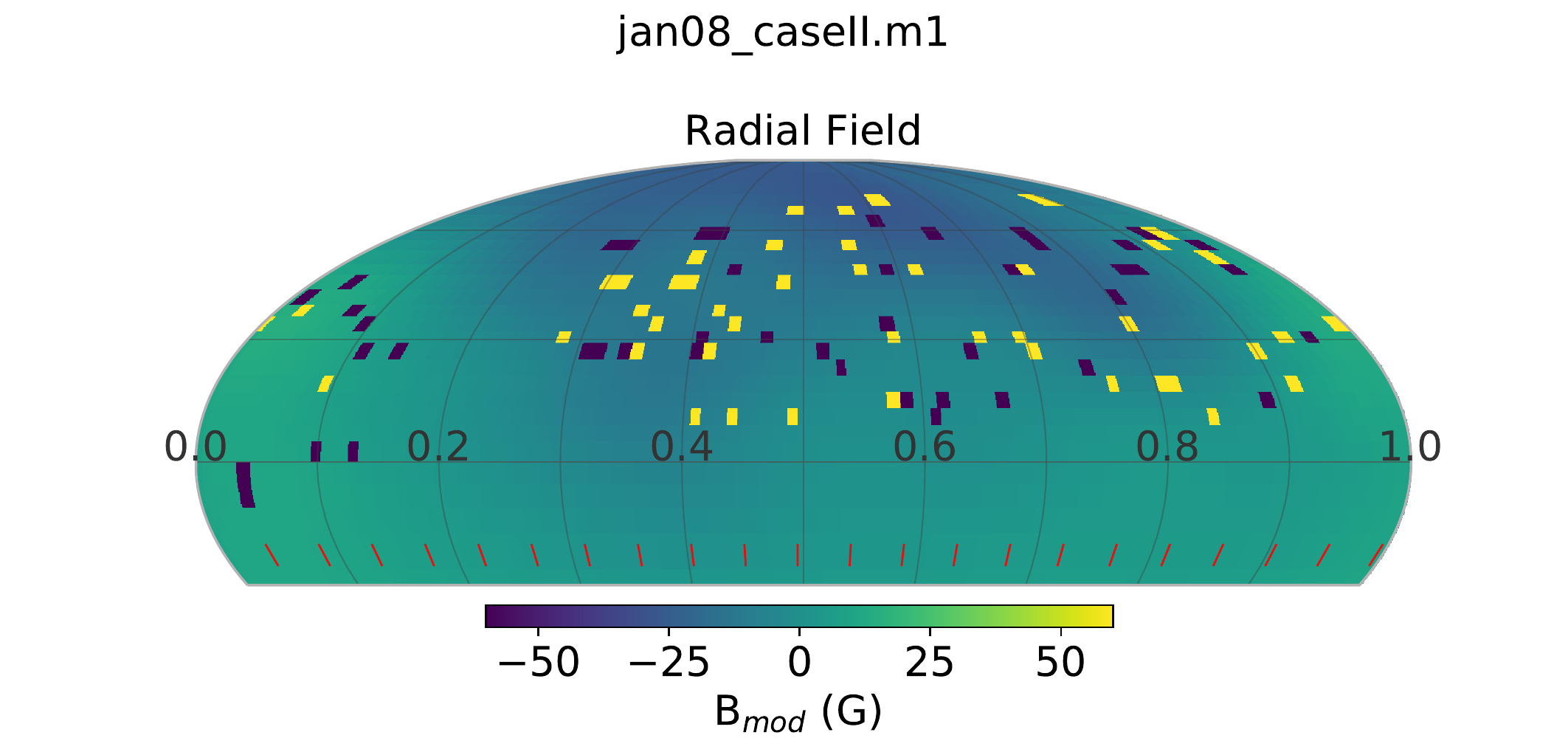}
        \label{fig:case_v_jan08}
        \vspace{0.5cm}
    \end{subfigure}
    \begin{subfigure}{0.49\textwidth}
        \vspace{0.5cm}
        \caption{2010.04}
        \centering
        \includegraphics[width=.95\linewidth, trim={1.8cm 0cm 1.7cm 1.5cm},clip]{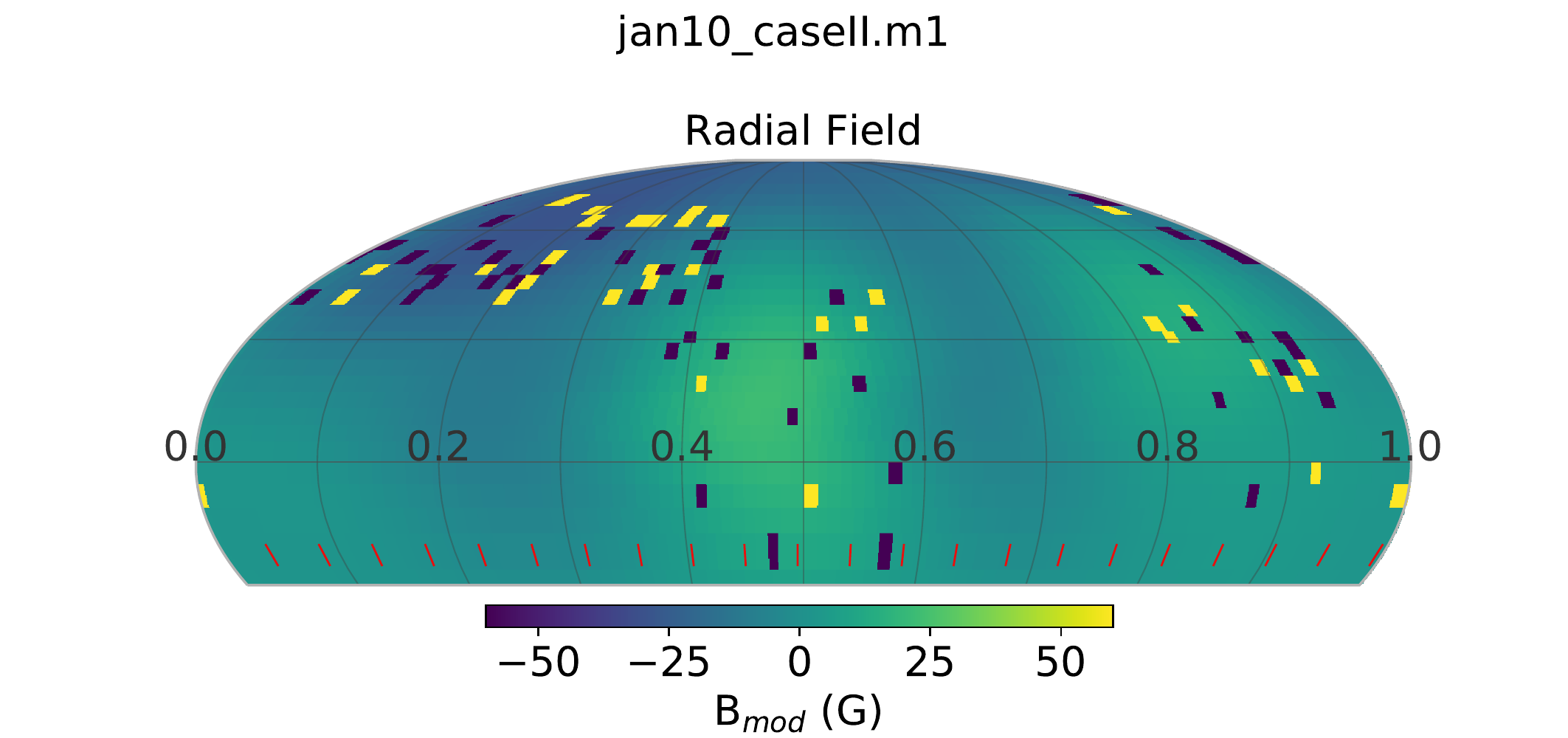}
        \label{fig:case_v_jan10}
        \vspace{0.5cm}
    \end{subfigure}
    \begin{subfigure}{0.49\textwidth}
        \vspace{0.5cm}
        \caption{2011.81}
        \centering
        \includegraphics[width=.95\linewidth, trim={1.8cm 0cm 1.7cm 1.5cm},clip]{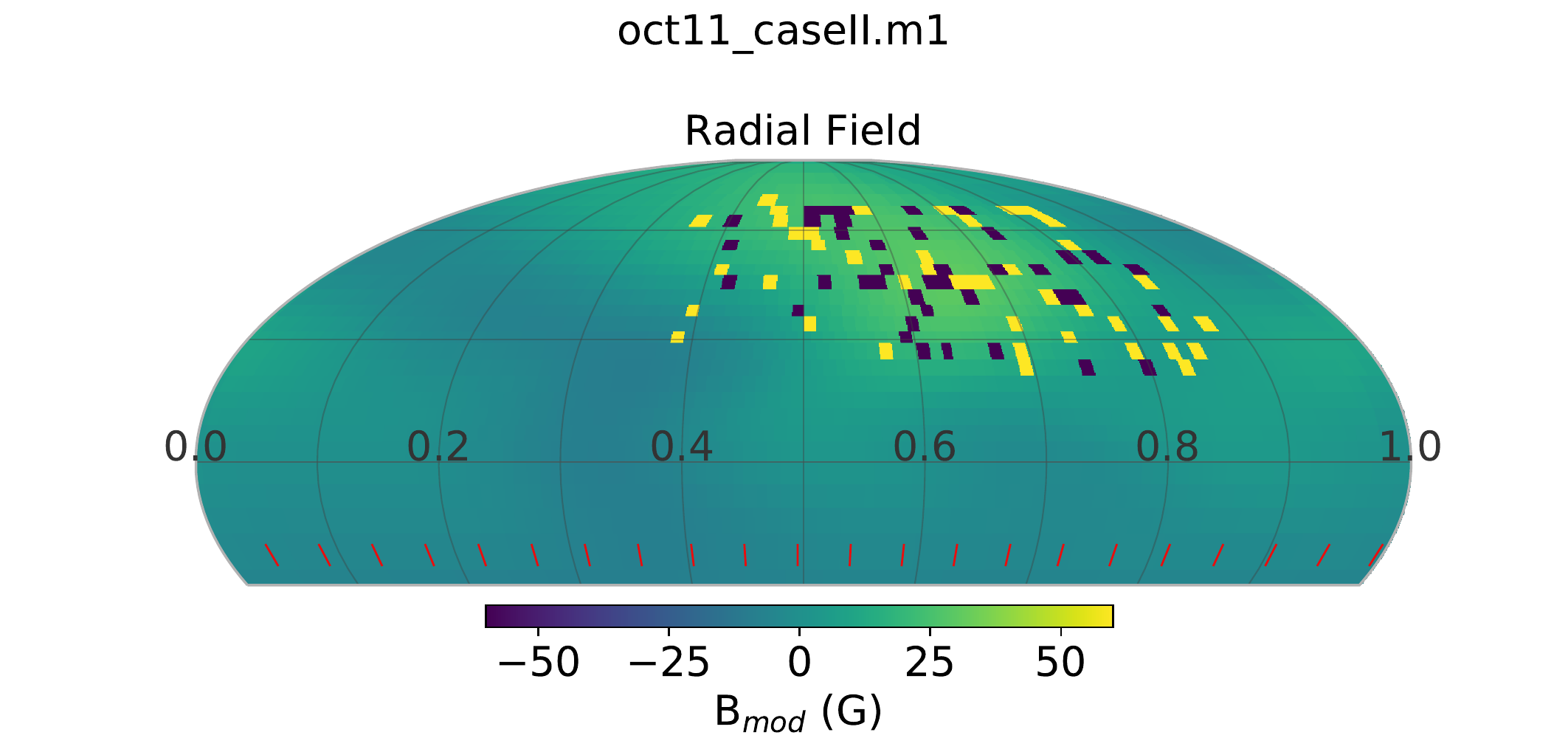}
        \label{fig:case_v_oct11}
        \vspace{0.5cm}
    \end{subfigure}
    \begin{subfigure}{0.49\textwidth}
        \vspace{0.5cm}
        \caption{2012.82}
        \centering
        \includegraphics[width=.95\linewidth, trim={1.8cm 0cm 1.7cm 1.5cm},clip]{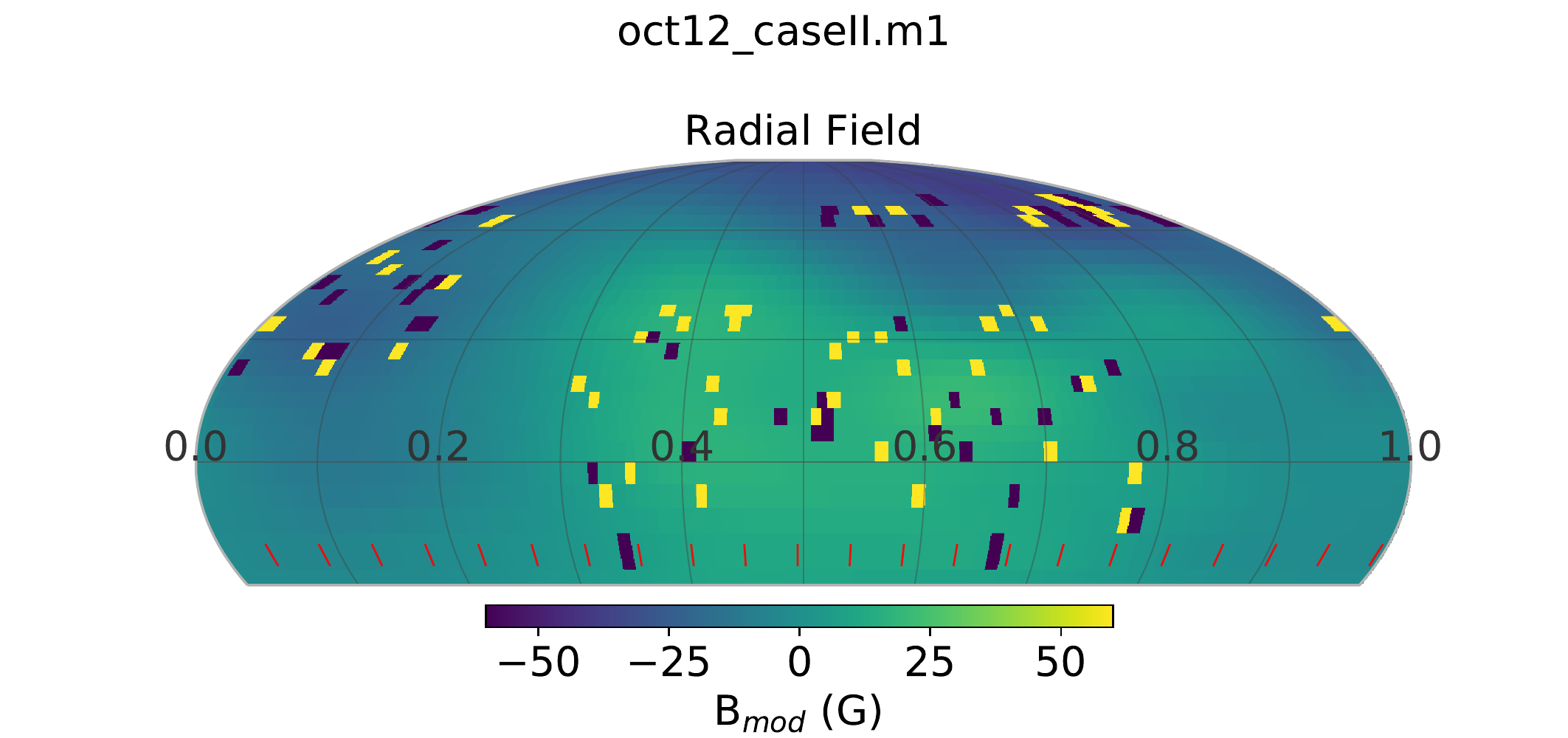}
        \label{fig:case_v_oct12}
        \vspace{0.5cm}
    \end{subfigure}
    \begin{subfigure}{0.49\textwidth}
        \vspace{0.5cm}
        \caption{2013.75}
        \centering
        \includegraphics[width=.95\linewidth, trim={1.8cm 0cm 1.7cm 1.5cm},clip]{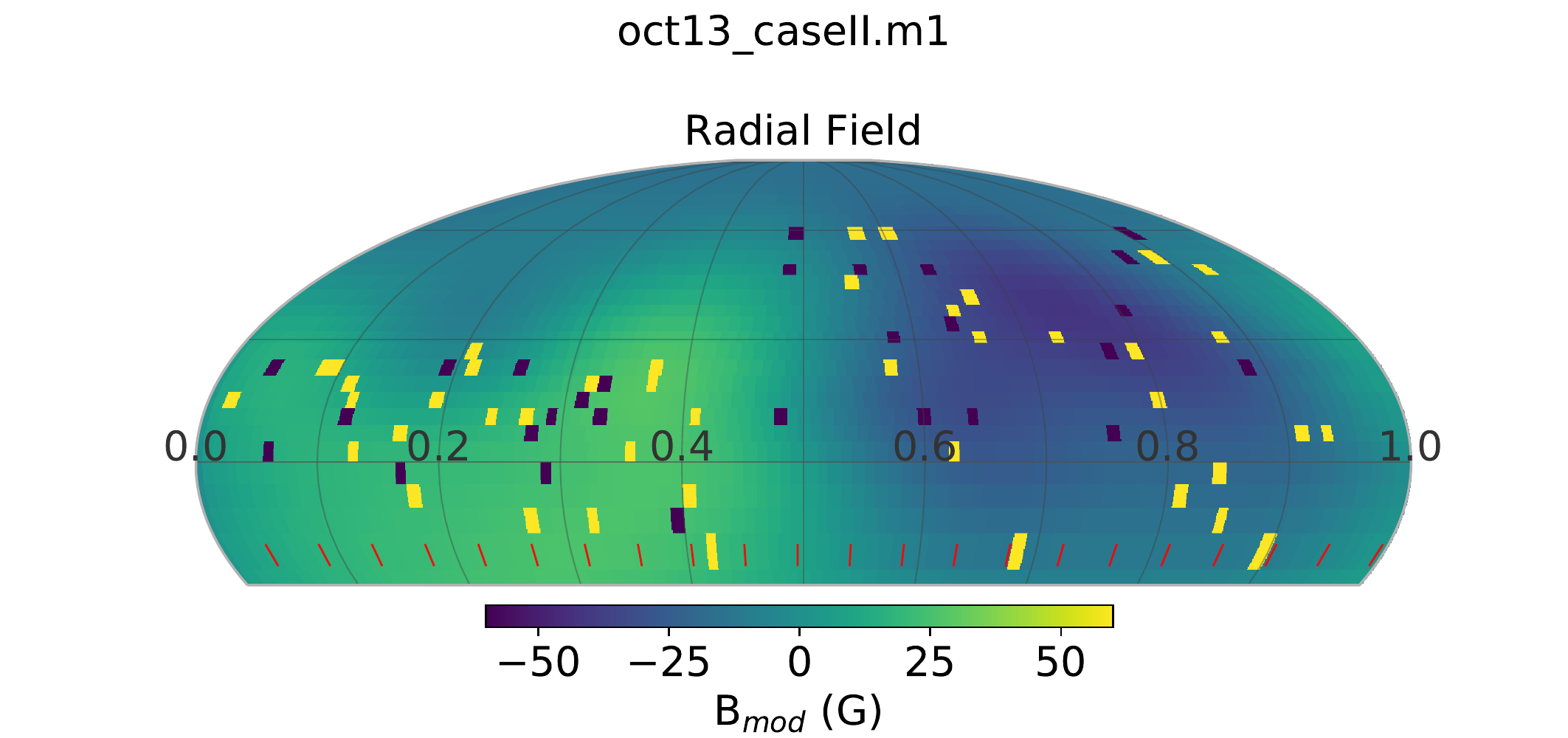}
        \label{fig:case_v_oct13}
        \vspace{0.5cm}
    \end{subfigure}
    \begin{subfigure}{0.49\textwidth}
        \vspace{0.5cm}
        \caption{2014.84}
        \centering
        \includegraphics[width=.95\linewidth, trim={1.8cm 0cm 1.7cm 1.5cm},clip]{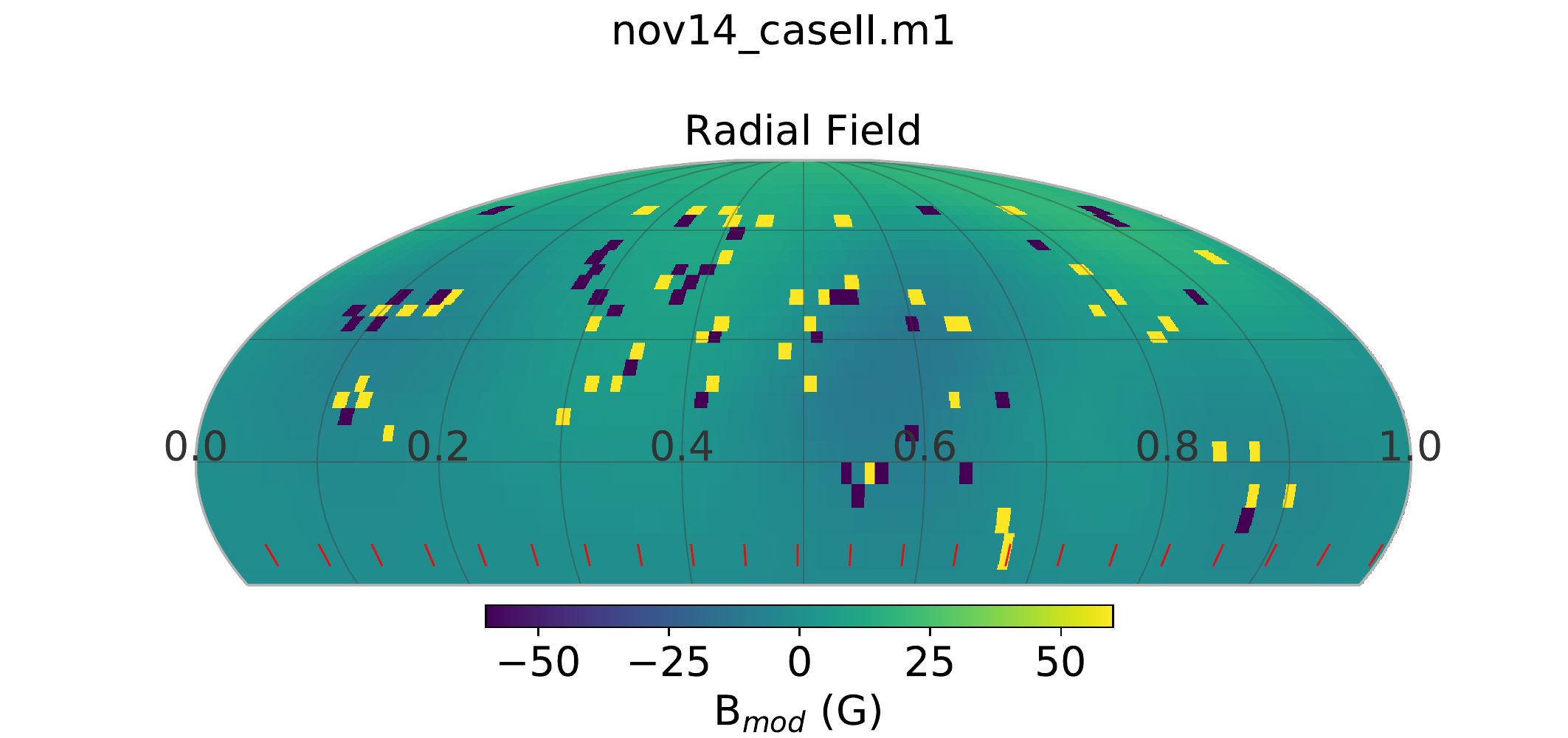}
        \label{fig:case_v_nov14}
        \vspace{0.5cm}
    \end{subfigure}
    \begin{subfigure}{0.49\textwidth}
        \vspace{0.5cm}
        \caption{2015.01}
        \centering
        \includegraphics[width=.95\linewidth, trim={1.8cm 0cm 1.7cm 1.5cm},clip]{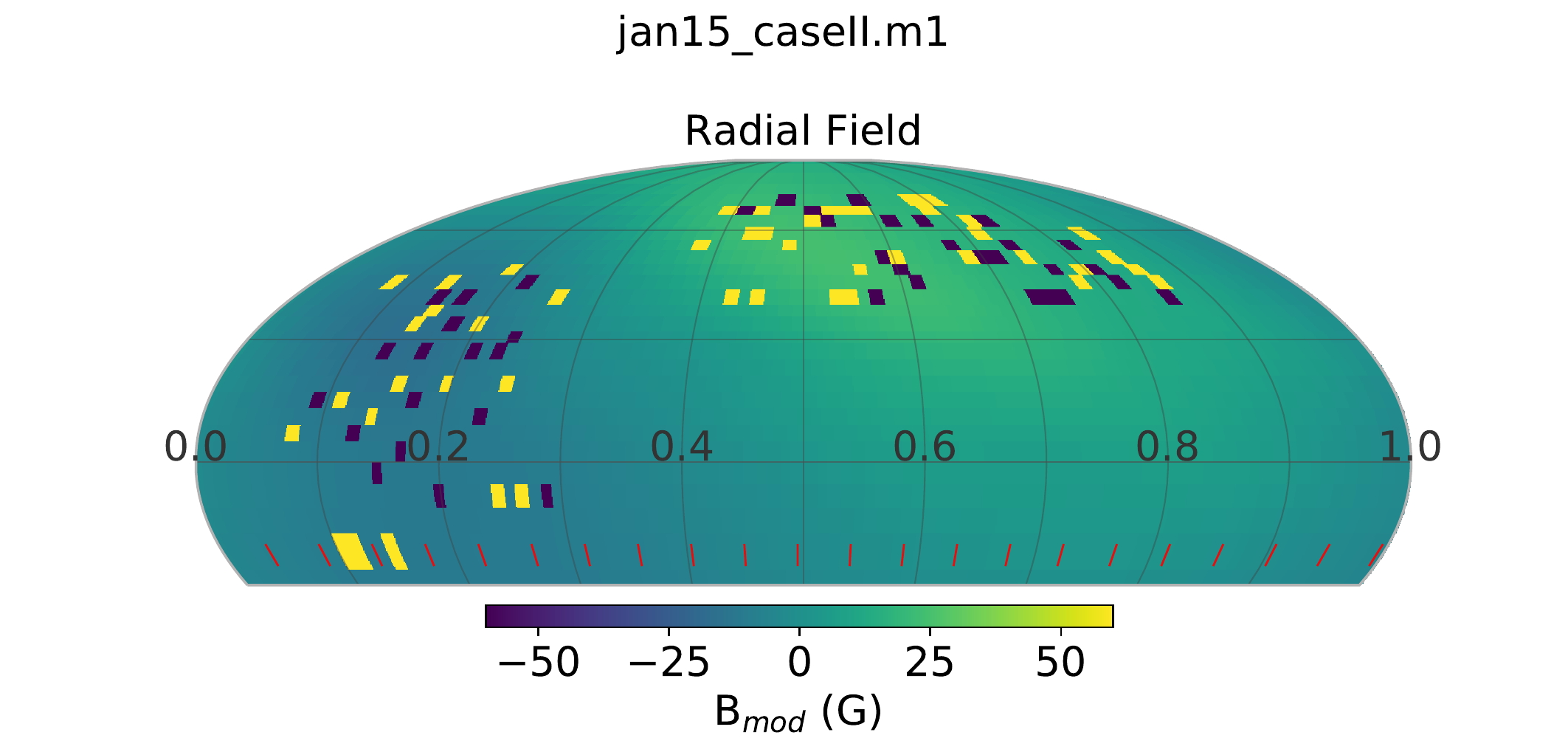}
        \label{fig:case_v_jan15}
        \vspace{0.5cm}
    \end{subfigure}
    \caption{
        Magnetic field maps of $\epsilon$ Eridani for eight epochs of observations (see sub-captions) and simulated magnetic spots, according to {\it case v}.
        Bright areas indicate positive polarity and dark -- negative polarity.
    }
    \label{fig:case_v_maps}
\end{figure*}

\begin{figure*}
    \centering
    
    \begin{subfigure}{0.49\textwidth}
        \vspace{0.5cm}
        \caption{2007.08}
        \centering
        \includegraphics[width=.95\linewidth, trim={1.8cm 0cm 1.7cm 1.5cm},clip]{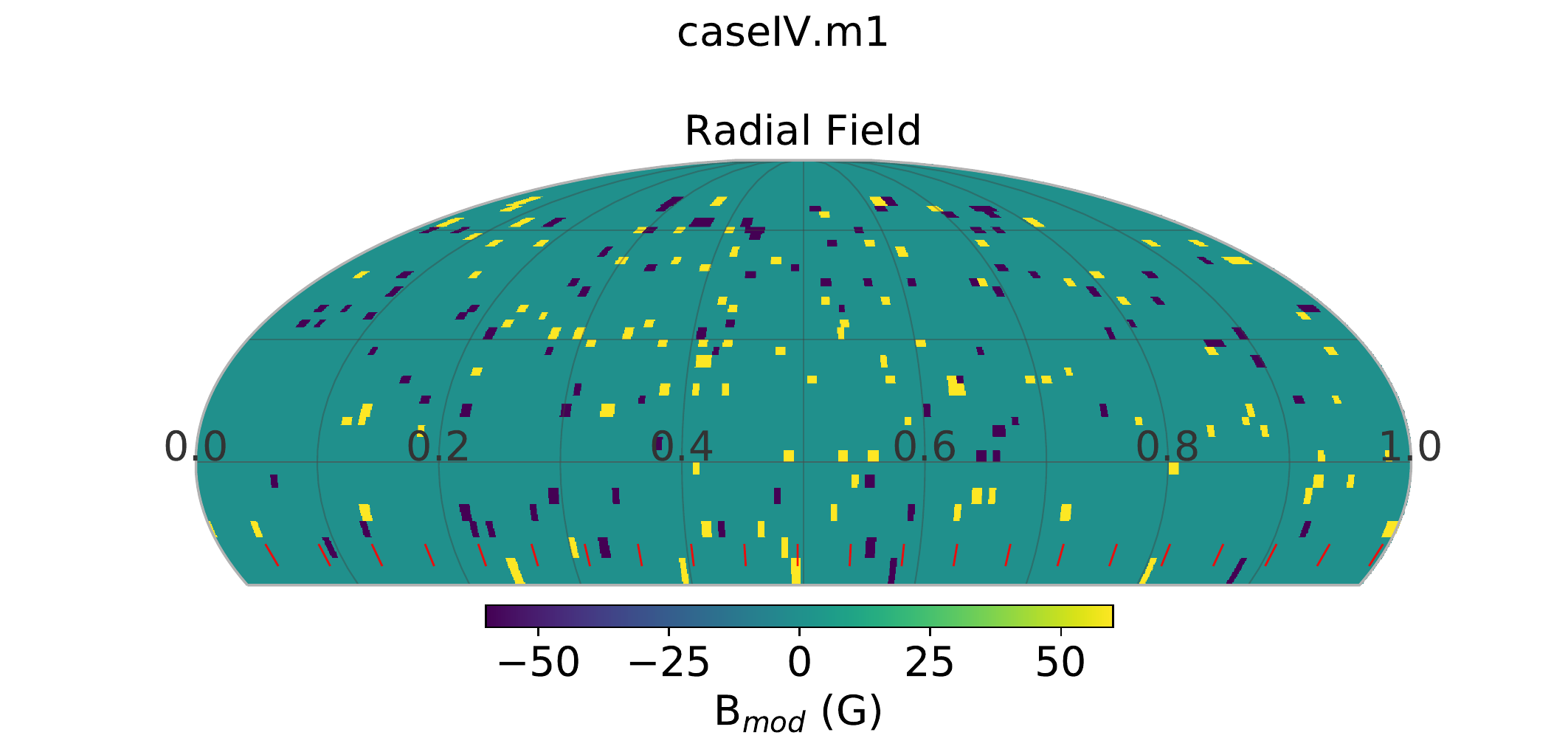}
        \label{fig:case_vi_jan07}
        \vspace{0.5cm}
    \end{subfigure}
    \begin{subfigure}{0.49\textwidth}
        \vspace{0.5cm}
        \caption{2008.09}
        \centering
        \includegraphics[width=.95\linewidth, trim={1.8cm 0cm 1.7cm 1.5cm},clip]{figures/maps/caseIV/caseIV_vir_60.pdf}
        \label{fig:case_vi_jan08}
        \vspace{0.5cm}
    \end{subfigure}
    \begin{subfigure}{0.49\textwidth}
        \vspace{0.5cm}
        \caption{2010.04}
        \centering
        \includegraphics[width=.95\linewidth, trim={1.8cm 0cm 1.7cm 1.5cm},clip]{figures/maps/caseIV/caseIV_vir_60.pdf}
        \label{fig:case_vi_jan10}
        \vspace{0.5cm}
    \end{subfigure}
    \begin{subfigure}{0.49\textwidth}
        \vspace{0.5cm}
        \caption{2011.81}
        \centering
        \includegraphics[width=.95\linewidth, trim={1.8cm 0cm 1.7cm 1.5cm},clip]{figures/maps/caseIV/caseIV_vir_60.pdf}
        \label{fig:case_vi_oct11}
        \vspace{0.5cm}
    \end{subfigure}
    \begin{subfigure}{0.49\textwidth}
        \vspace{0.5cm}
        \caption{2012.82}
        \centering
        \includegraphics[width=.95\linewidth, trim={1.8cm 0cm 1.7cm 1.5cm},clip]{figures/maps/caseIV/caseIV_vir_60.pdf}
        \label{fig:case_vi_oct12}
        \vspace{0.5cm}
    \end{subfigure}
    \begin{subfigure}{0.49\textwidth}
        \vspace{0.5cm}
        \caption{2013.75}
        \centering
        \includegraphics[width=.95\linewidth, trim={1.8cm 0cm 1.7cm 1.5cm},clip]{figures/maps/caseIV/caseIV_vir_60.pdf}
        \label{fig:case_vi_oct13}
        \vspace{0.5cm}
    \end{subfigure}
    \begin{subfigure}{0.49\textwidth}
        \vspace{0.5cm}
        \caption{2014.84}
        \centering
        \includegraphics[width=.95\linewidth, trim={1.8cm 0cm 1.7cm 1.5cm},clip]{figures/maps/caseIV/caseIV_vir_60.pdf}
        \label{fig:case_vi_nov14}
        \vspace{0.5cm}
    \end{subfigure}
    \begin{subfigure}{0.49\textwidth}
        \vspace{0.5cm}
        \caption{2015.01}
        \centering
        \includegraphics[width=.95\linewidth, trim={1.8cm 0cm 1.7cm 1.5cm},clip]{figures/maps/caseIV/caseIV_vir_60.pdf}
        \label{fig:case_vi_jan15}
        \vspace{0.5cm}
    \end{subfigure}
    \caption{
        Magnetic field maps of $\epsilon$ Eridani for eight epochs of observations (see sub-captions) and simulated magnetic spots, according to {\it case vi}.
        Bright areas indicate positive polarity and dark -- negative polarity.
    }
    \label{fig:case_vi_maps}
\end{figure*}

%% file: paper.bbl
\begin{thebibliography}{}
\makeatletter
\relax
\def\mn@urlcharsother{\let\do\@makeother \do\$\do\&\do\#\do\^\do\_\do\%\do\~}
\def\mn@doi{\begingroup\mn@urlcharsother \@ifnextchar [ {\mn@doi@}
  {\mn@doi@[]}}
\def\mn@doi@[#1]#2{\def\@tempa{#1}\ifx\@tempa\@empty \href
  {http://dx.doi.org/#2} {doi:#2}\else \href {http://dx.doi.org/#2} {#1}\fi
  \endgroup}
\def\mn@eprint#1#2{\mn@eprint@#1:#2::\@nil}
\def\mn@eprint@arXiv#1{\href {http://arxiv.org/abs/#1} {{\tt arXiv:#1}}}
\def\mn@eprint@dblp#1{\href {http://dblp.uni-trier.de/rec/bibtex/#1.xml}
  {dblp:#1}}
\def\mn@eprint@#1:#2:#3:#4\@nil{\def\@tempa {#1}\def\@tempb {#2}\def\@tempc
  {#3}\ifx \@tempc \@empty \let \@tempc \@tempb \let \@tempb \@tempa \fi \ifx
  \@tempb \@empty \def\@tempb {arXiv}\fi \@ifundefined
  {mn@eprint@\@tempb}{\@tempb:\@tempc}{\expandafter \expandafter \csname
  mn@eprint@\@tempb\endcsname \expandafter{\@tempc}}}

\bibitem[\protect\citeauthoryear{{Anglada-Escud{\'e}} \&
  {Butler}}{{Anglada-Escud{\'e}} \& {Butler}}{2012}]{2012ApJS..200...15A}
{Anglada-Escud{\'e}} G.,  {Butler} R.~P.,  2012, \mn@doi [\apjs]
  {10.1088/0067-0049/200/2/15}, \href
  {http://adsabs.harvard.edu/abs/2012ApJS..200...15A} {200, 15}

\bibitem[\protect\citeauthoryear{{Anglada-Escud{\'e}}
  et~al.,}{{Anglada-Escud{\'e}} et~al.}{2016}]{2016Natur.536..437A}
{Anglada-Escud{\'e}} G.,  et~al., 2016, \mn@doi [\nat] {10.1038/nature19106},
  \href {https://ui.adsabs.harvard.edu/abs/2016Natur.536..437A} {536, 437}

\bibitem[\protect\citeauthoryear{{Astropy Collaboration}}{{Astropy
  Collaboration}}{2013}]{2013A&A...558A..33A}
{Astropy Collaboration} 2013, \mn@doi [\aap] {10.1051/0004-6361/201322068},
  \href {https://ui.adsabs.harvard.edu/#abs/2013A&A...558A..33A} {558, A33}

\bibitem[\protect\citeauthoryear{{Barnes}}{{Barnes}}{2007}]{2007ApJ...669.1167B}
{Barnes} S.~A.,  2007, \mn@doi [\apj] {10.1086/519295}, \href
  {http://adsabs.harvard.edu/abs/2007ApJ...669.1167B} {669, 1167}

\bibitem[\protect\citeauthoryear{{Barnes} et~al.,}{{Barnes}
  et~al.}{2017}]{2017MNRAS.466.1733B}
{Barnes} J.~R.,  et~al., 2017, \mn@doi [\mnras] {10.1093/mnras/stw3170}, \href
  {https://ui.adsabs.harvard.edu/abs/2017MNRAS.466.1733B} {466, 1733}

\bibitem[\protect\citeauthoryear{{Berdyugina}}{{Berdyugina}}{2005}]{2005LRSP....2....8B}
{Berdyugina} S.~V.,  2005, \mn@doi [Living Reviews in Solar Physics]
  {10.12942/lrsp-2005-8}, \href
  {http://adsabs.harvard.edu/abs/2005LRSP....2....8B} {2}

\bibitem[\protect\citeauthoryear{{Brogi}, {Marzari}  \& {Paolicchi}}{{Brogi}
  et~al.}{2009}]{2009A&A...499L..13B}
{Brogi} M.,  {Marzari} F.,   {Paolicchi} P.,  2009, \mn@doi [\aap]
  {10.1051/0004-6361/200811609}, \href
  {https://ui.adsabs.harvard.edu/abs/2009A%26A...499L..13B} {499, L13}

\bibitem[\protect\citeauthoryear{{Campbell}, {Walker}  \& {Yang}}{{Campbell}
  et~al.}{1988}]{1988ApJ...331..902C}
{Campbell} B.,  {Walker} G.~A.~H.,   {Yang} S.,  1988, \mn@doi [\apj]
  {10.1086/166608}, \href {http://adsabs.harvard.edu/abs/1988ApJ...331..902C}
  {331, 902}

\bibitem[\protect\citeauthoryear{{Cumming}, {Marcy}  \& {Butler}}{{Cumming}
  et~al.}{1999}]{1999ApJ...526..890C}
{Cumming} A.,  {Marcy} G.~W.,   {Butler} R.~P.,  1999, \mn@doi [\apj]
  {10.1086/308020}, \href {http://adsabs.harvard.edu/abs/1999ApJ...526..890C}
  {526, 890}

\bibitem[\protect\citeauthoryear{{Donati} \& {Landstreet}}{{Donati} \&
  {Landstreet}}{2009}]{2009ARA&A..47..333D}
{Donati} J.~F.,  {Landstreet} J.~D.,  2009, \mn@doi [\araa]
  {10.1146/annurev-astro-082708-101833}, \href
  {https://ui.adsabs.harvard.edu/abs/2009ARA&A..47..333D} {47, 333}

\bibitem[\protect\citeauthoryear{{Donati} et~al.,}{{Donati}
  et~al.}{2017}]{2017MNRAS.465.3343D}
{Donati} J.-F.,  et~al., 2017, \mn@doi [\mnras] {10.1093/mnras/stw2904}, \href
  {http://adsabs.harvard.edu/abs/2017MNRAS.465.3343D} {465, 3343}

\bibitem[\protect\citeauthoryear{{Giguere}, {Fischer}, {Zhang}, {Matthews},
  {Cameron}  \& {Henry}}{{Giguere} et~al.}{2016}]{2016ApJ...824..150G}
{Giguere} M.~J.,  {Fischer} D.~A.,  {Zhang} C.~X.~Y.,  {Matthews} J.~M.,
  {Cameron} C.,   {Henry} G.~W.,  2016, \mn@doi [\apj]
  {10.3847/0004-637X/824/2/150}, \href
  {http://adsabs.harvard.edu/abs/2016ApJ...824..150G} {824, 150}

\bibitem[\protect\citeauthoryear{{Gonz{\'a}lez Hern{\'a}ndez}, {Pepe}, {Molaro}
   \& {Santos}}{{Gonz{\'a}lez Hern{\'a}ndez}
  et~al.}{2018}]{2018haex.bookE.157G}
{Gonz{\'a}lez Hern{\'a}ndez} J.~I.,  {Pepe} F.,  {Molaro} P.,   {Santos} N.~C.,
   2018, {ESPRESSO on VLT: An Instrument for Exoplanet Research}.
p.~157, \mn@doi{10.1007/978-3-319-55333-7_157}

\bibitem[\protect\citeauthoryear{{Grandjean} et~al.,}{{Grandjean}
  et~al.}{2020}]{2020A&A...633A..44G}
{Grandjean} A.,  et~al., 2020, \mn@doi [\aap] {10.1051/0004-6361/201936038},
  \href {https://ui.adsabs.harvard.edu/abs/2020A&A...633A..44G} {633, A44}

\bibitem[\protect\citeauthoryear{{Hatzes}}{{Hatzes}}{2002}]{2002AN....323..392H}
{Hatzes} A.~P.,  2002, \mn@doi [Astronomische Nachrichten]
  {10.1002/1521-3994(200208)323:3/4<392::AID-ASNA392>3.0.CO;2-M}, \href
  {https://ui.adsabs.harvard.edu/abs/2002AN....323..392H} {323, 392}

\bibitem[\protect\citeauthoryear{{Hatzes} et~al.,}{{Hatzes}
  et~al.}{2000}]{2000ApJ...544L.145H}
{Hatzes} A.~P.,  et~al., 2000, \mn@doi [\apjl] {10.1086/317319}, \href
  {http://adsabs.harvard.edu/abs/2000ApJ...544L.145H} {544, L145}

\bibitem[\protect\citeauthoryear{{Haywood} et~al.,}{{Haywood}
  et~al.}{2020}]{2020arXiv200513386H}
{Haywood} R.~D.,  et~al., 2020, arXiv e-prints, \href
  {https://ui.adsabs.harvard.edu/abs/2020arXiv200513386H} {p. arXiv:2005.13386}

\bibitem[\protect\citeauthoryear{{H{\'e}brard}, {Donati}, {Delfosse}, {Morin},
  {Boisse}, {Moutou}  \& {H{\'e}brard}}{{H{\'e}brard}
  et~al.}{2014}]{2014MNRAS.443.2599H}
{H{\'e}brard} {\'E}.~M.,  {Donati} J.-F.,  {Delfosse} X.,  {Morin} J.,
  {Boisse} I.,  {Moutou} C.,   {H{\'e}brard} G.,  2014, \mn@doi [\mnras]
  {10.1093/mnras/stu1285}, \href
  {http://adsabs.harvard.edu/abs/2014MNRAS.443.2599H} {443, 2599}

\bibitem[\protect\citeauthoryear{{H{\'e}brard}, {Donati}, {Delfosse}, {Morin},
  {Moutou}  \& {Boisse}}{{H{\'e}brard} et~al.}{2016}]{2016MNRAS.461.1465H}
{H{\'e}brard} {\'E}.~M.,  {Donati} J.-F.,  {Delfosse} X.,  {Morin} J.,
  {Moutou} C.,   {Boisse} I.,  2016, \mn@doi [\mnras] {10.1093/mnras/stw1346},
  \href {http://adsabs.harvard.edu/abs/2016MNRAS.461.1465H} {461, 1465}

\bibitem[\protect\citeauthoryear{Hunter}{Hunter}{2007}]{Hunter:2007}
Hunter J.~D.,  2007, Computing In Science \& Engineering, 9, 90

\bibitem[\protect\citeauthoryear{{Jeffers}, {Barnes}, {Jones}, {Reiners},
  {Pinfield}  \& {Marsden}}{{Jeffers} et~al.}{2014a}]{2014MNRAS.438.2717J}
{Jeffers} S.~V.,  {Barnes} J.~R.,  {Jones} H.~R.~A.,  {Reiners} A.,  {Pinfield}
  D.~J.,   {Marsden} S.~C.,  2014a, \mn@doi [\mnras] {10.1093/mnras/stt1950},
  \href {https://ui.adsabs.harvard.edu/abs/2014MNRAS.438.2717J} {438, 2717}

\bibitem[\protect\citeauthoryear{{Jeffers}, {Petit}, {Marsden}, {Morin},
  {Donati}  \& {Folsom}}{{Jeffers} et~al.}{2014b}]{2014AA...569A..79J}
{Jeffers} S.~V.,  {Petit} P.,  {Marsden} S.~C.,  {Morin} J.,  {Donati} J.-F.,
  {Folsom} C.~P.,  2014b, \mn@doi [\aap] {10.1051/0004-6361/201423725}, \href
  {http://adsabs.harvard.edu/abs/2014A%26A...569A..79J} {569, A79}

\bibitem[\protect\citeauthoryear{{Jeffers}, {Boro Saikia}, {Barnes}, {Petit},
  {Marsden}, {Jardine}, {Vidotto}  \& {BCool Collaboration}}{{Jeffers}
  et~al.}{2017}]{2017MNRAS.471L..96J}
{Jeffers} S.~V.,  {Boro Saikia} S.,  {Barnes} J.~R.,  {Petit} P.,  {Marsden}
  S.~C.,  {Jardine} M.~M.,  {Vidotto} A.~A.,   {BCool Collaboration} 2017,
  \mn@doi [\mnras] {10.1093/mnrasl/slx097}, \href
  {https://ui.adsabs.harvard.edu/abs/2017MNRAS.471L..96J} {471, L96}

\bibitem[\protect\citeauthoryear{Jones, Oliphant, Peterson  \& Others}{Jones
  et~al.}{2001}]{jones_scipy_2001}
Jones E.,  Oliphant T.,  Peterson P.,   Others 2001, {SciPy}: Open source
  scientific tools for Python, \url {http://www.scipy.org/}

\bibitem[\protect\citeauthoryear{{K{\H{o}}v{\'a}ri} et~al.,}{{K{\H{o}}v{\'a}ri}
  et~al.}{2019}]{2019A&A...624A..83K}
{K{\H{o}}v{\'a}ri} Z.,  et~al., 2019, \mn@doi [\aap]
  {10.1051/0004-6361/201834810}, \href
  {https://ui.adsabs.harvard.edu/abs/2019A&A...624A..83K} {624, A83}

\bibitem[\protect\citeauthoryear{{Keenan} \& {McNeil}}{{Keenan} \&
  {McNeil}}{1989}]{1989ApJS...71..245K}
{Keenan} P.~C.,  {McNeil} R.~C.,  1989, \mn@doi [\apjs] {10.1086/191373}, \href
  {https://ui.adsabs.harvard.edu/abs/1989ApJS...71..245K} {71, 245}

\bibitem[\protect\citeauthoryear{{Kochukhov}, {Hackman}, {Lehtinen}  \&
  {Wehrhahn}}{{Kochukhov} et~al.}{2020}]{2020A&A...635A.142K}
{Kochukhov} O.,  {Hackman} T.,  {Lehtinen} J.~J.,   {Wehrhahn} A.,  2020,
  \mn@doi [\aap] {10.1051/0004-6361/201937185}, \href
  {https://ui.adsabs.harvard.edu/abs/2020A&A...635A.142K} {635, A142}

\bibitem[\protect\citeauthoryear{{Lagrange}, {Desort}  \& {Meunier}}{{Lagrange}
  et~al.}{2010}]{2010A&A...512A..38L}
{Lagrange} A.~M.,  {Desort} M.,   {Meunier} N.,  2010, \mn@doi [\aap]
  {10.1051/0004-6361/200913071}, \href
  {https://ui.adsabs.harvard.edu/abs/2010A&A...512A..38L} {512, A38}

\bibitem[\protect\citeauthoryear{{Lagrange}, {Meunier}, {Chauvin}, {Sterzik},
  {Galland}, {Lo Curto}, {Rameau}  \& {Sosnowska}}{{Lagrange}
  et~al.}{2013}]{2013A&A...559A..83L}
{Lagrange} A.~M.,  {Meunier} N.,  {Chauvin} G.,  {Sterzik} M.,  {Galland} F.,
  {Lo Curto} G.,  {Rameau} J.,   {Sosnowska} D.,  2013, \mn@doi [\aap]
  {10.1051/0004-6361/201220770}, \href
  {https://ui.adsabs.harvard.edu/abs/2013A&A...559A..83L} {559, A83}

\bibitem[\protect\citeauthoryear{{Lang}, {Jardine}, {Morin}, {Donati},
  {Jeffers}, {Vidotto}  \& {Fares}}{{Lang} et~al.}{2014}]{2014MNRAS.439.2122L}
{Lang} P.,  {Jardine} M.,  {Morin} J.,  {Donati} J.~F.,  {Jeffers} S.,
  {Vidotto} A.~A.,   {Fares} R.,  2014, \mn@doi [\mnras]
  {10.1093/mnras/stu091}, \href
  {https://ui.adsabs.harvard.edu/abs/2014MNRAS.439.2122L} {439, 2122}

\bibitem[\protect\citeauthoryear{Lehmann, Hussain, Jardine, Mackay  \&
  Vidotto}{Lehmann et~al.}{2018}]{Lehmann2018}
Lehmann L.~T.,  Hussain G. A.~J.,  Jardine M.~M.,  Mackay D.~H.,   Vidotto
  A.~A.,  2018, \mn@doi [Monthly Notices of the Royal Astronomical Society]
  {10.1093/mnras/sty3362}, 483, 5246

\bibitem[\protect\citeauthoryear{{Lomb}}{{Lomb}}{1976}]{1976Ap&SS..39..447L}
{Lomb} N.~R.,  1976, \mn@doi [\apss] {10.1007/BF00648343}, \href
  {http://adsabs.harvard.edu/abs/1976Ap%26SS..39..447L} {39, 447}

\bibitem[\protect\citeauthoryear{{L{\"o}ptien}, {Lagg}, {van Noort}  \&
  {Solanki}}{{L{\"o}ptien} et~al.}{2020}]{2020A&A...635A.202L}
{L{\"o}ptien} B.,  {Lagg} A.,  {van Noort} M.,   {Solanki} S.~K.,  2020,
  \mn@doi [\aap] {10.1051/0004-6361/201936975}, \href
  {https://ui.adsabs.harvard.edu/abs/2020A&A...635A.202L} {635, A202}

\bibitem[\protect\citeauthoryear{{Mandal} \& {Banerjee}}{{Mandal} \&
  {Banerjee}}{2016}]{2016ApJ...830L..33M}
{Mandal} S.,  {Banerjee} D.,  2016, \mn@doi [\apjl]
  {10.3847/2041-8205/830/2/L33}, \href
  {https://ui.adsabs.harvard.edu/abs/2016ApJ...830L..33M} {830, L33}

\bibitem[\protect\citeauthoryear{{Mawet} et~al.,}{{Mawet}
  et~al.}{2019}]{2019AJ....157...33M}
{Mawet} D.,  et~al., 2019, \mn@doi [\aj] {10.3847/1538-3881/aaef8a}, \href
  {https://ui.adsabs.harvard.edu/abs/2019AJ....157...33M} {157, 33}

\bibitem[\protect\citeauthoryear{McKinney}{McKinney}{2010}]{mckinney}
McKinney W.,  2010, in van~der Walt S.,  Millman J.,  eds, Proceedings of the
  9th Python in Science Conference. pp 51 -- 56

\bibitem[\protect\citeauthoryear{{Metcalfe} et~al.,}{{Metcalfe}
  et~al.}{2013}]{2013ApJ...763L..26M}
{Metcalfe} T.~S.,  et~al., 2013, \mn@doi [\apjl] {10.1088/2041-8205/763/2/L26},
  \href {http://adsabs.harvard.edu/abs/2013ApJ...763L..26M} {763, L26}

\bibitem[\protect\citeauthoryear{{Meunier}, {Mignon}  \& {Lagrange}}{{Meunier}
  et~al.}{2017}]{2017A&A...607A.124M}
{Meunier} N.,  {Mignon} L.,   {Lagrange} A.~M.,  2017, \mn@doi [\aap]
  {10.1051/0004-6361/201731017}, \href
  {https://ui.adsabs.harvard.edu/abs/2017A&A...607A.124M} {607, A124}

\bibitem[\protect\citeauthoryear{{Mortier}}{{Mortier}}{2016}]{2016csss.confE.134M}
{Mortier} A.,  2016, in 19th Cambridge Workshop on Cool Stars, Stellar Systems,
  and the Sun (CS19). Cambridge Workshop on Cool Stars, Stellar Systems, and
  the Sun.
p.~134, \mn@doi{10.5281/zenodo.59214}

\bibitem[\protect\citeauthoryear{{Pepe} et~al.,}{{Pepe}
  et~al.}{2010}]{2010SPIE.7735E..0FP}
{Pepe} F.~A.,  et~al., 2010, in Ground-based and Airborne Instrumentation for
  Astronomy III. p. 77350F, \mn@doi{10.1117/12.857122}

\bibitem[\protect\citeauthoryear{{Petit} et~al.,}{{Petit}
  et~al.}{2008}]{2008MNRAS.388...80P}
{Petit} P.,  et~al., 2008, \mn@doi [\mnras] {10.1111/j.1365-2966.2008.13411.x},
  \href {https://ui.adsabs.harvard.edu/abs/2008MNRAS.388...80P} {388, 80}

\bibitem[\protect\citeauthoryear{{Petit}, {Dintrans}, {Morgenthaler}, {Van
  Grootel}, {Morin}, {Lanoux}, {Auri{\`e}re}  \&
  {Konstantinova-Antova}}{{Petit} et~al.}{2009}]{2009A&A...508L...9P}
{Petit} P.,  {Dintrans} B.,  {Morgenthaler} A.,  {Van Grootel} V.,  {Morin} J.,
   {Lanoux} J.,  {Auri{\`e}re} M.,   {Konstantinova-Antova} R.,  2009, \mn@doi
  [\aap] {10.1051/0004-6361/200913285}, \href
  {https://ui.adsabs.harvard.edu/abs/2009A&A...508L...9P} {508, L9}

\bibitem[\protect\citeauthoryear{{Reiners}}{{Reiners}}{2012}]{2012LRSP....9....1R}
{Reiners} A.,  2012, \mn@doi [Living Reviews in Solar Physics]
  {10.12942/lrsp-2012-1}, \href
  {https://ui.adsabs.harvard.edu/abs/2012LRSP....9....1R} {9, 1}

\bibitem[\protect\citeauthoryear{{Reiners}, {Shulyak}, {Anglada-Escud{\'e}},
  {Jeffers}, {Morin}, {Zechmeister}, {Kochukhov}  \& {Piskunov}}{{Reiners}
  et~al.}{2013}]{2013A&A...552A.103R}
{Reiners} A.,  {Shulyak} D.,  {Anglada-Escud{\'e}} G.,  {Jeffers} S.~V.,
  {Morin} J.,  {Zechmeister} M.,  {Kochukhov} O.,   {Piskunov} N.,  2013,
  \mn@doi [\aap] {10.1051/0004-6361/201220437}, \href
  {https://ui.adsabs.harvard.edu/abs/2013A&A...552A.103R} {552, A103}

\bibitem[\protect\citeauthoryear{{Rueedi}, {Solanki}, {Mathys}  \&
  {Saar}}{{Rueedi} et~al.}{1997}]{1997AA...318..429R}
{Rueedi} I.,  {Solanki} S.~K.,  {Mathys} G.,   {Saar} S.~H.,  1997, \aap, \href
  {http://cdsads.u-strasbg.fr/abs/1997A%26A...318..429R} {318, 429}

\bibitem[\protect\citeauthoryear{{Saar}}{{Saar}}{1990}]{1990IAUS..138..427S}
{Saar} S.~H.,  1990, in {Stenflo} J.~O.,  ed.,  IAU Symposium Vol. 138, Solar
  Photosphere: Structure, Convection, and Magnetic Fields. pp 427--441

\bibitem[\protect\citeauthoryear{{Saar} \& {Donahue}}{{Saar} \&
  {Donahue}}{1997}]{1997ApJ...485..319S}
{Saar} S.~H.,  {Donahue} R.~A.,  1997, \mn@doi [\apj] {10.1086/304392}, \href
  {https://ui.adsabs.harvard.edu/abs/1997ApJ...485..319S} {485, 319}

\bibitem[\protect\citeauthoryear{{Scargle}}{{Scargle}}{1982}]{1982ApJ...263..835S}
{Scargle} J.~D.,  1982, \mn@doi [\apj] {10.1086/160554}, \href
  {http://adsabs.harvard.edu/abs/1982ApJ...263..835S} {263, 835}

\bibitem[\protect\citeauthoryear{{Shulyak} et~al.,}{{Shulyak}
  et~al.}{2019}]{2019A&A...626A..86S}
{Shulyak} D.,  et~al., 2019, \mn@doi [\aap] {10.1051/0004-6361/201935315},
  \href {https://ui.adsabs.harvard.edu/abs/2019A&A...626A..86S} {626, A86}

\bibitem[\protect\citeauthoryear{{Strassmeier}}{{Strassmeier}}{2009}]{2009A&ARv..17..251S}
{Strassmeier} K.~G.,  2009, \mn@doi [\aapr] {10.1007/s00159-009-0020-6}, \href
  {https://ui.adsabs.harvard.edu/abs/2009A&ARv..17..251S} {17, 251}

\bibitem[\protect\citeauthoryear{Van Der~Walt, Colbert  \& Varoquaux}{Van
  Der~Walt et~al.}{2011}]{van2011numpy}
Van Der~Walt S.,  Colbert S.~C.,   Varoquaux G.,  2011, Computing in Science \&
  Engineering, 13, 22

\bibitem[\protect\citeauthoryear{{Wade}, {Donati}, {Landstreet}  \&
  {Shorlin}}{{Wade} et~al.}{2000}]{2000MNRAS.313..823W}
{Wade} G.~A.,  {Donati} J.-F.,  {Landstreet} J.~D.,   {Shorlin} S.~L.~S.,
  2000, \mn@doi [\mnras] {10.1046/j.1365-8711.2000.03273.x}, \href
  {http://adsabs.harvard.edu/abs/2000MNRAS.313..823W} {313, 823}

\makeatother
\end{thebibliography}
